# Adjudicating Conduction Mechanisms in High Performance Carbon Nanotube Fibers

John Bulmer*
Aerospace Systems Directorate, Air Force Research Laboratory, Wright-Patterson Air Force Base, Ohio 45433
University of Cincinnati, 598 Rhodes Hall, P.O. Box 210072, Cincinnati, OH, USA

Chris Kovacs, Thomas Bullard, Charlie Ebbing, Timothy Haugan
Aerospace Systems Directorate, Air Force Research Laboratory, Wright-Patterson Air Force Base, Ohio 45433

Ganesh Pokharel
Perry College of Mathematics, Computing and Sciences, University of West Georgia, Carrollton, GA 30118, USA
Materials Department, University of California, Santa Barbara, CA 93106, USA

Stephen D. Wilson
Materials Department, University of California, Santa Barbara, CA 93106, USA

Fedor F. Balakirev, Oscar A. Valenzuela
National High Magnetic Field Laboratory, Los Alamos National Laboratory, Los Alamos, New Mexico 87545, USA

Michael A. Susner, David Turner
Materials and Manufacturing Directorate, Air Force Research Laboratory, 2179 12th Street, Wright-Patterson Air Force Base, Ohio 45433, USA

Pengyu Fu
The Center for High Performance Power Electronics, The Ohio State University, Columbus, OH 43210 USA

Teresa Kulka
Institute of Theoretical Physics, Faculty of Physics, University of Warsaw, Pasteura 5, PL-02093 Warsaw, Poland

Jacek Majewski
Institute of Theoretical Physics, Faculty of Physics, University of Warsaw, Pasteura 5, PL-02-093 Warszawa, Poland
Terahertz Research and Application Center CENTERA2, Center of Advanced Materials and Technologies CEZAMAT, Warsaw University of Technology, Poland

Irina Lebedeva
CIC nanoGUNE, Donostia-San Sebastián 20018, Spain
Simune Atomistics, Donostia-San Sebastián 20018, Spain

Karolina Z. Milowska
CIC nanoGUNE, Donostia-San Sebastián 20018, Spain
Ikerbasque, Basque Foundation for Science, Bilbao 48013, Spain

Agnieszka Lekawa-Raus
Centre for Advanced Materials and Technologies (CEZAMAT) , Warsaw University of Technology , Warsaw , Poland

Magdalena Marganska
Institute for Theoretical Physics, University of Regensburg, 93040 Regensburg, Germany
Institute of Theoretical Physics, Wrocław Tech, Wybrzeże Wyspiańskiego 27, 50-370 Wrocław, Poland

The performance of carbon nanotube (CNT) cables, a contender for copper-wire replacement, is tied to its metallic and semi-conducting-like conductivity responses with temperature; the origin of the semi-conducting-like response however is an underappreciated incongruity in literature. With controlled aspect-ratio and doping-degree, over 61 unique cryogenic experiments including anisotropy and Hall measurements, CNT cable performance is explored at extreme temperatures (65 mK) and magnetic-fields (60 T). A semi-conducting-like conductivity response with temperature becomes temperature-independent approaching absolute-zero, uniquely demonstrating the necessity of heterogeneous fluctuation induced tunneling; complete de-doping leads to localized hopping, contrasting graphite's pure metallic-like response. High-field magneto-resistance (including +22% longitudinal magneto-resistance near room-temperature) is analyzed with hopping and classical two-band models, both similarly yielding a parameter useful for conductor development. Varying field-orientation angle uncovers two- and four-fold symmetries from Aharonov-Bohm-like corrections to curvature-induced bandgap. Tight-binding calculations using Green's Function formalism model the largest, coherent transport to-date in commensurate CNT bundles in magnetic-field, revealing non-uniform transmission across bundle cross-sections with doping restoring uniformity; independent of doping, transport in bundle-junction-bundle systems are predominantly from CNTs adjacent to the other bundle. The final impact is predicting the ultimate conductivity of heterogeneous CNT cables using temperature and field-dependent transport, surpassing conductivity of traditional metals.

*corresponding author bulmerjn@ucmail.uc.edu

Keywords: carbon nanotube, bundles, junctions, electrical conductivity, magneto-resistance, heterogeneous transport, Aharonov-Bohm effect

## Introduction

The performance of carbon nanotube (CNT) conductors is steadily improving and now surpasses copper's conductivity on a weight-basis[1] [2] [3] [4] [5] [6] [7] and conventional carbon-fiber's tensile-strength[8], with continued development expected with greater aspect-ratio[9] [10] and incorporating dopants with large-scale order[11]. Older, more fragile, carbon-based conductors however, such as graphitic-intercalation-compounds (GICs)[12], have higher conductivities (50% greater than copper in absolute terms[13] [14]) and, along with its highly-ordered graphite host, have a completely metallic-like resistance *R* temperature *T* response ($dR/dT$>0) down to liquid-helium temperatures[12]. Despite individual metallic CNTs[15] and some CNT bundles[16] also having this full metallic-like $dR/dT$>0, it is established[10] [17] [18] [19] [20] [21] that CNT materials with aligned microstructure typically have both a semi-conducting-like response ($dR/dT$<0) at colder temperatures and a metallic-like response at higher temperatures ($dR/dT$>0), forming a u-shaped *R* vs *T* plot. Moderately ordered graphite[12] [22] [23] and conductive-polymers[17] [24] [25] also have this u-shaped response, where the semi-conducting component is from extrinsic factors (junctions, voids and misalignment); when minimized, higher room-temperature conductivity is achieved[17] [25]. When finally a conductive-polymer was fabricated with full $dR/dT$>0, its conductivity was regarded as a pinnacle limit[26] [27]. Despite 25 years of CNT-conductor development, a CNT-conductor with a full $dR/dT$>0 is not yet realized.

The semi-conducting-like $dR/dT$<0 contribution in CNT-conductors is an underappreciated controversy in literature, where studies pick from either homogeneous-transport (variable-range-hopping, insulator-to metal-transition, or weak-localization) or heterogenous-transport (fluctuation-induced-tunneling combined with metallic conduction) without possibly full consideration of other mechanisms. Homogeneous-transport, prevalently applied to disordered materials, assumes disordered uniformity with one governing characteristic-length. With variable-range-hopping (VRH) for example, widely applied for lower conductivity CNT materials, charge-carriers are localized over the characteristic-length *a* and tunnel to distant sites with matching energies (opposed to nearest neighbors with phonon exchange) resulting in exponentially diverging *R* as *T*->0 according to[28] [29] [30] [31] [32] [33]

$$R[T] = R_c \, exp \left[\left(\frac{T_M}{T}\right)^{\frac{1}{D+1}}\right] \tag{1}$$

where $R_C$ is a fitting prefactor, *D* is the dimensionality, and $T_M$ is the Mott temperature (figure 1a). $T_M$ is the maximum temperature where VRH applies with $T_M = C/(k_b N(\varepsilon_F) a^D)$ and *C*= 18.2 for *D*= 3[34] [35], *C* = 13.8 for *D* = 2 [36], $k_b$ is Boltzmann's constant, and $N(\varepsilon_F)$ is the density of states at the Fermi level. Above $T_M$, the semi-conducting-like component becomes Arrhenius thermal-activation. As material order improves (say, by increasing chemical doping[19] or CNT metallicity[37]), $a \to \infty$ and the material undergoes an insulator to metal transition. Now, the homogeneous-transport is weak-localization where the phase-coherence length ($L_{Phase}$) becomes the limiting characteristic-length. Here, electron backscatter from crystal defects adds coherently over $L_{Phase}$ and leads to a small resistance increase (<1% for 8 nm thin Cu films at 4 K[38], <3% for pre-graphitic carbon-fibers at 4 K[39]). Temperature impedes the phase-coherence leading to a gradual $dR/dT$ <0. Weak-localization originally explained the anomalous resistance correction and -MR in metal thin-films[38] [40], and later carbon-fiber[39] [41], graphitic-intercalation-compounds[42], and CNTs[19] [21] [37] [43] [44] [45] [46] [47] [48] [49].

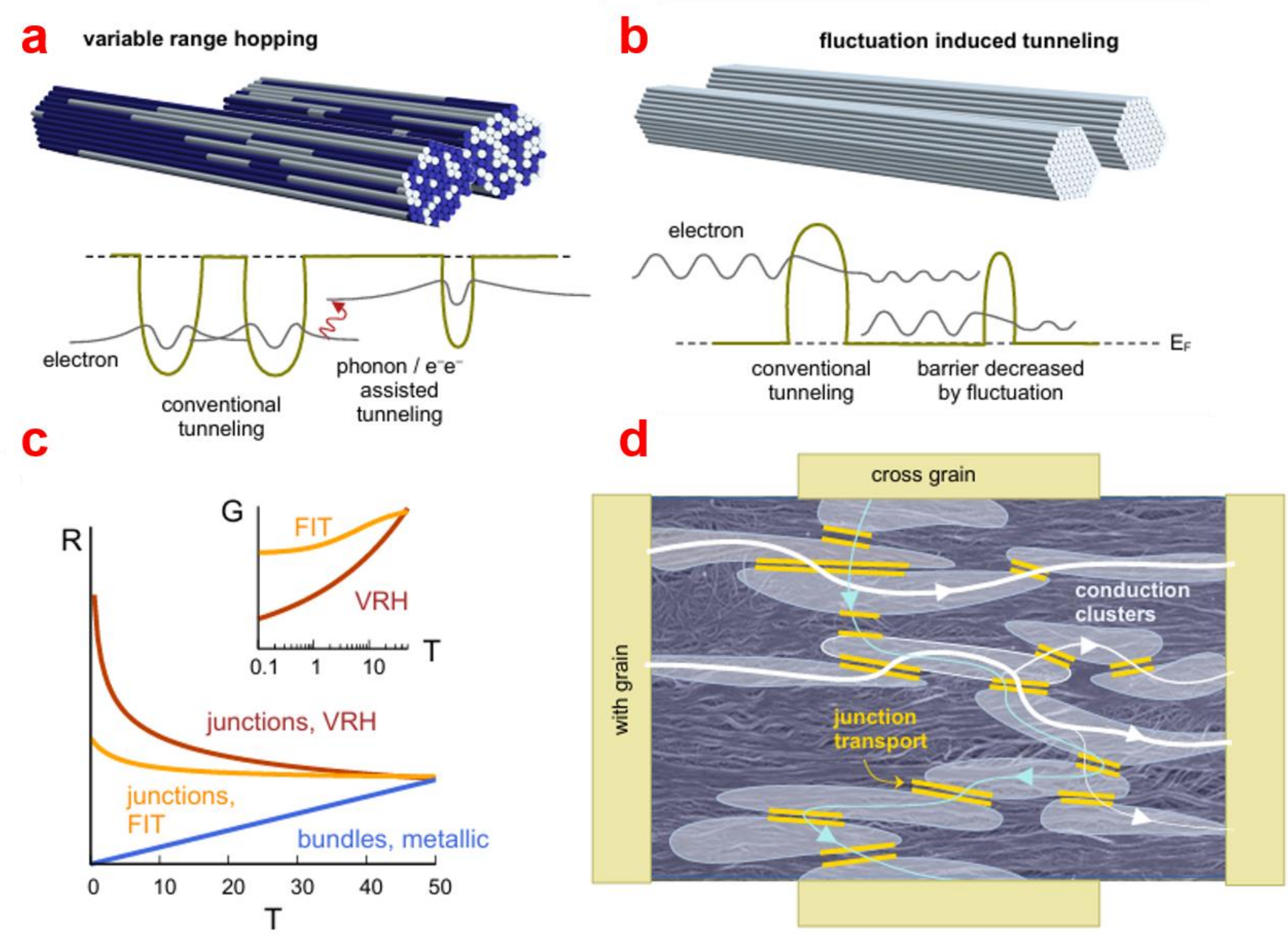

**Figure 1| Proposed heterogeneous CNT network. a, de-doped bundles of mixed semi-conducting (dead-weight, dark blue) and metallic CNTs (light blue); junctions are modeled as variable-range-hopping (VRH) where localized charge-carriers tunnel to distant locations with matching energy. b, With as-is (doped), all CNTs participate and junctions are modeled with fluctuation-induced-tunneling (FIT), where long delocalized conduction paths are separated by small insulating gaps. Thermal fluctuations enhance transmission, but are not necessary. c, cartoon of the models' temperature-dependent resistances, with inset showing conductivity. d, SEM photograph of CNT ribbon superimposed by metallic conduction paths (with and against microstructure-alignment) with interrupting yellow semi-conducting junctions.**

Other CNT studies, somewhat separately, use heterogeneous-transport models, comprised of independent mechanisms that combine in some way, for more complex $R/T$ responses. Substantiated by thermo-power measurements[50], the $R/T$ u-shape is extrinsically-tunable based on heterogeneous composition, opposed to an intrinsically-fixed material property. The room-temperature resistance of transparent, unaligned CNT films, for example, was estimated by series-addition of CNT resistance with CNT junction resistance[51] [52] [53] [54] [55]. Temperature-dependent resistance was modeled for unaligned armchair CNT films[56] by a phonon-assisted tunneling term added in series with a hopping term. Fluctuation induced tunneling (FIT)[57] is another empirical-based heterogeneous-transport mechanism typically used for u-shape $R/T$ responses where small insulating junctions separate long conductive structures (figure 1b). If junction capacitance is sufficiently small, relative to the long conductive structure capacitance, thermal noise voltage fluctuations become sufficient to modify the tunneling potential. This results in the temperature-dependent resistance[17] [57]:

$$R(T) = R_c \, exp \left[\frac{T_1}{T+T_2}\right] + R_{metallic}[T] \tag{2}$$

with fitting constant $R_c$, $k_bT_1$ the approximate energy barrier of the junction, $T_2$ is the temperature above which thermal noise becomes important, and $R_{metallic}[T]$ is the system-specific long conductive structure resistance. Note that FIT has the simplest mathematical form to generate constant resistance when $T<<T_2$ (signaling temperature-independent tunneling), while generating Arrhenius-style thermal-activation when $T>>T_2$. Equation (2) has been applied to conductive-polymers[58], CNT-conductors[18] [21] [31] [49] [59], and individualized metal-nanowires (where controlled metal-nanowires form a single junction for FIT model validation[60] [61]). Magnetic-field dependence is not yet established[31] and, anecdotally, FIT is not widely utilized outside of the polymer community. A sketch of $R(T)$ for the metallic (bundle transport) and junction (VRH and FIT for de-doped and as-is CNTs, respectively) is shown in figure 1c, with an inset showing the junctions' conductance $G(T)$. A schematic illustration of the CNT fiber with the conducting regions separated by junctions is shown in figure 1d.

Here, we study the effects of aspect-ratio and doping-degree in the latest CNT-conductors (fiber and ribbon) and explore their homogeneous and heterogeneous-transport in uniquely extreme temperatures (<65 mK -300 K), magnetic-fields (60 T), and magnetic-field orientation. These results are compared with coherent tight-binding transport models for an individual CNT rotating in magnetic-field and for spatially-large CNT bundle. This is noteworthy because, for the first time, homogeneous and heterogeneous-transport models were systematically compared in high-performance CNT-conductors with well controlled variation in CNT parameters in extreme environmental conditions. We show that heterogeneous-transport out of necessity must be used and, in particular, FIT best describes the coldest temperature response. This enables determination of an ultimate CNT-conductor conductivity, as well as demonstrates new quality-control techniques for development. In the supplemental, a well-narrated database exhaustively provides the complete cryogenic and room-temperature property data, as well as model-fitting, correlation tables, and literature-based compendium of MR mechanisms.

**Results**

We obtained ribbons and fiber composed of high-quality few-wall CNTs with aligned microstructures from Dexmat using the established acid-solution spinning-process[1] [9] with different molecular aspect-ratios (AR: 1200, 3100, 4800, and 5600). These CNT materials are p-doped from their acid-based production process (labeled “as-is”), which we can remove with a 1000 °C 1-hour bakeout in flowing $H_2$. Afterwards laboratory air and moisture re-exposure lightly p-dopes the de-doped CNT-conductor[62] [63], so we also implemented a 100 °C vacuum bake-out within every electrical characterization apparatus to remove the physisorbed species[37] [64] (Labeled “de-doped”, further details in supplemental-section 1.0). Four-wire resistance, either with or perpendicular to the microstructure-alignment, was measured in Quantum Design Physical Properties Measurement System (PPMS) from room-temperature down to 1.9 K, with magnetic-field up to 9 T and the ability to change sample orientation in field. In two cases, a different PPMS system enabled measurement down to <65 mK and fields up to 14 T. Resistance was converted to a specific-conductivity for easier comparisons across samples, which is a more useful metric than standard conductivity because a cross-sectional area, which is variable in porous materials, is not required for calculation[10].

**Temperature-dependent conductivity.** Compiling all 28 zero-field cryogenic resistance measurements, figure 2a plots the room-temperature specific-conductivity $\sigma$ across all samples (ribbon, fibers, as-is, and de-doped) against their cryogenic-resistance-ratio (resistance at 300 K divided by resistance at 10 K). Specific-conductivity measured either parallel or perpendicular microstructure-alignment have the same linear relationship, although parallel is on average 20x higher (Supplemental-figure 2.2-4). A CNT meta-analysis also demonstrated similar correlation across many studies[10]. These correlations demonstrate that understanding and controlling the metallic-like and semi-conducting-like temperature responses leads to improved performance. Figure 2b plots aspect-ratio versus cryogenic-resistance-ratio, showing positive correlation for as-is (r>90%). These aspect-ratios correspond to CNT lengths 1.8, 4.65, 6.75, and 8.4 µm (assuming a 1.5 nm diameter CNT), which is greater than the characteristic-lengths in homogeneous-transport models: an individual CNT's phonon-limited mean-free-path (1 µm at room-temperature[65]); phase-coherence-length (10 to 40 nm from 100 to 2 K[66]) in weak-localization; localization lengths (< 6 nm [35] [66]) in VRH. This supports the thesis that the cryogenic-resistance-ratio is dictated by a heterogenous network of CNT structures and junctions, opposed to homogeneous-transport mechanisms. De-doping categorically decreases the cryogenic-resistance-ratio (on average, by a factor of 3, figure 2b and Supplemental-figure 2.2-3), while any upward trends with aspect-ratio are no longer statistically significant. This correlation loss indicates a different transport regime primarily controlled just by junctions.

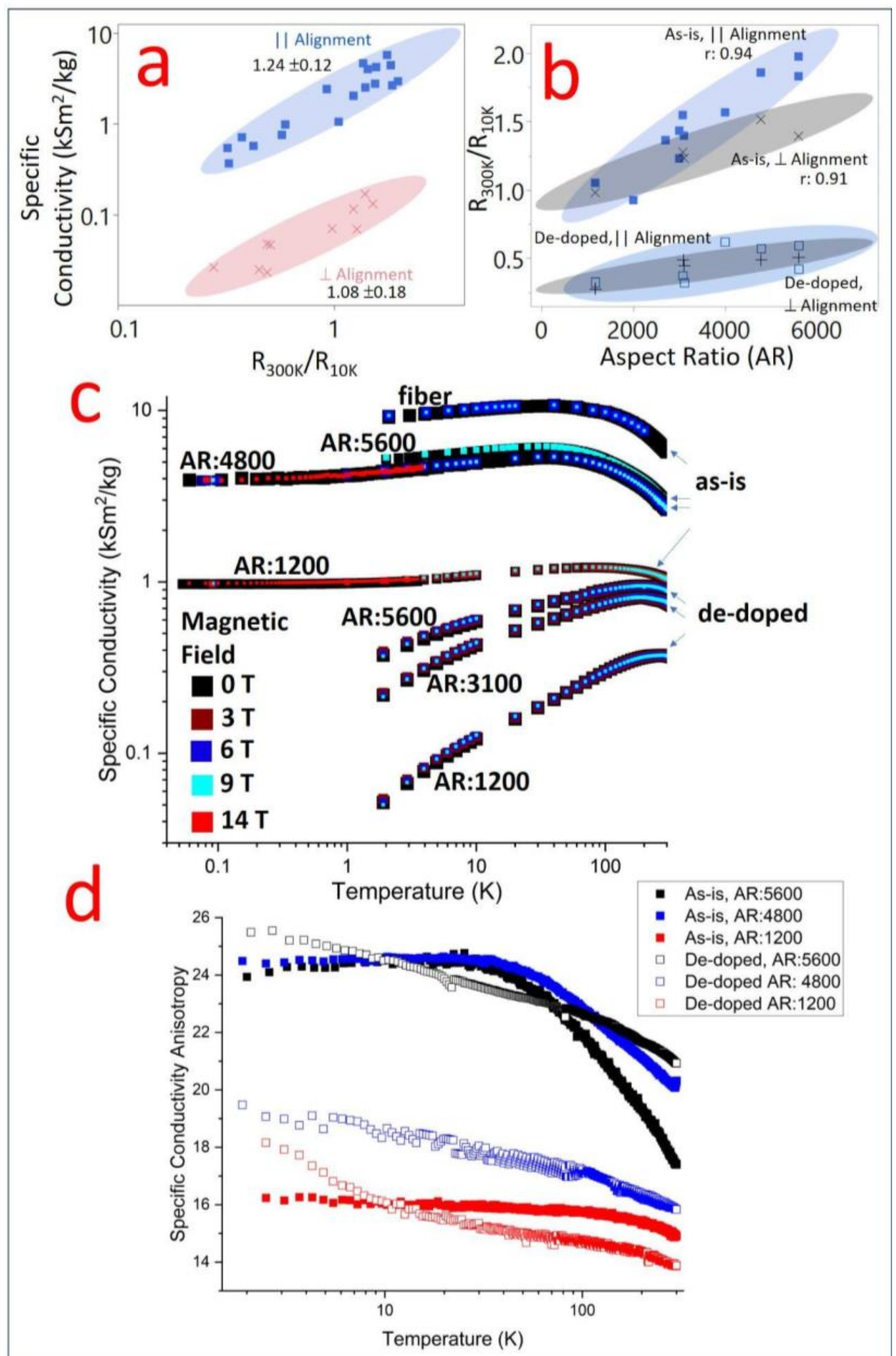

**Figure 2| Conductivity temperature dependence. a, zero-field room-temperature specific-conductivity versus cryogenic-resistance-ratio (*R* at 300 K/ *R* at 10 K) across all materials, with partitioning according to microstructure-alignment, with power-law exponent. b, cryogenic-resistance-ratio versus aspect-ratio (AR) partitioned by doping-status and microstructure-alignment, with correlation indicated for as-is. c, Specific examples of parallel-alignment specific-conductivity vs temperature for various AR and doping-status.** Colors represent the transverse magnetic-field *H*. **d, temperature-dependence of the specific-conductivity anisotropy.**

Figure 2c shows specific examples of specific-conductivity (parallel microstructure-alignment) versus temperature with aspect-ratio and doping-status indicated, all showing the upside-down "u" profile familiar from literature[10] [17] [18] [19] [20] [21]. The as-is has a pronounced metallic-like component ($d\sigma/dT<0$) compared to its semi-conducting-like component ($d\sigma/dT>0$), where the semi-conducting component notably levels-off to a constant value approaching absolute-zero. This conclusive level-off, shown for the first time for these high-performance CNT-conductors at particularly low temperatures (<65 mK), validates their fundamentally metallic nature. Transverse DC magnetic-field (*H*, depicted up to 14 T with different colors) shows a small positive conductivity contribution. The semi-conducting component under field still remains, however, implying an origin not from weak-localization (which is suppressed by field).

The dependence on aspect-ratio, with CNT length far greater than typical homogeneous characteristic-lengths, implies heterogeneous-transport with independent terms for CNT structures and their junctions; further, the upside-down u-shape with one maximum implies a total resistance that is the series-sum of one metallic-like resistance term and one semi-conducting-like resistance term. The heterogeneous-transport model FIT (equation 2) is consistent with the conductivity level-off approaching absolute-zero. The best FIT fits were with adding a metallic power-law resistivity term ($R_{metallic}[T] \propto T^x$ with exponent $x$ between 1.4 and 2.5), opposed to a more standard linear or quasi-1D metallic term[67]. Correlations with the heterogeneous fits are shown in supplemental-section 2.3.

De-doped CNT-conductors are less conductive (on average, 6x lower, Supplemental-figure 2.2-2) and have a greater semi-conducting component that, rather than level-off, rapidly approaches zero conductivity approaching absolute-zero. Reduced-activation-energy analysis and fitting to specific VRH models (Supplemental section 2.1) indicates the insulator side of the metal/insulator transition with typically 3D VRH. If the de-doped sample was not vacuum-baked within the measurement apparatus, it would be on the metal/insulator transition (indicated by power-law dependence between conductivity and temperature approaching absolute-zero), rather than hopping conduction[19]. This is similar to earlier reports on de-doped CNT fibers[19] and contrasts the fully metallic temperature response of high-quality graphite[68] [69] [70]. Note that the de-doping post-process does not negatively impact the CNT structure itself (supplemental-section 1.0). Despite VRH at colder temperatures, above ≈200 K a small metallic-like component appears ($d\sigma/dT<0$). Similar to as-is, the u-shaped mixed metallic/semi-conducting components indicate a heterogeneous mixture of junctions and CNT structures, which is expected considering the CNT lengths and typical VRH localization lengths (4.6 to 9 nm[28] [30] [33] [35]). Dissimilar to as-is, we saw that the cryogenic-resistance-ratio was not significantly correlated to aspect-ratio and this indicates a difference in junction mechanisms. As suggested in [37], the completely de-doped semi-conducting CNTs are essentially always insulating (for $d$= 1.5 nm, semi-conducting bandgap ≈233 meV, corresponding to a thermal-activation of 2700 K); therefore, the de-doped metallic CNTs are always responsible for electrical conductivity over the temperature range of interest (65 mK to 300 K) and the network now more sparse and percolative relative to the as-is/highly-doped case where all CNTs can contribute.

Figure 2d shows the specific-conductivity anisotropy versus temperature for CNT ribbons, defined as the ratio of specific conductivity with the current flowing with-grain to that with the current flowing across-grain. With decreasing temperature for as-is, the specific-conductivity anisotropy first increases and then becomes constant below ≈100 K. This level-off is also reflected in the fitted parameters for FIT, where $T_1$ and $T_2$ do not differ significantly between parallel and perpendicular cases (Supplemental-section 2.3). De-doped changed gradually with temperature without this level-off. The change of anisotropy all-together however was never large (from 300 K to 2 K, the change was <33%). Temperature independence of conductivity anisotropy has been observed before in aligned CNT materials[71]; it was concluded there was no fundamental difference in junction or CNT transport in different directions and that anisotropy manifested by the extrinsic differences in the parallel and perpendicular geometries of the larger percolating network. This contrasts graphite where conductivity anisotropy, in one case, was 12.5 at room-temperature and increased to 54 at cryogenic temperatures[72]; this also contrasts the quasi-1D metal tetrathiofulvalinium tetracyanoquinodimethan (TTF)(TCNQ) going from 500 at room-temperature to 10,000 at cryogenic temperatures[73]. Considering our anisotropy changes little with temperature implies the current distribution through the CNT structure-junction network is also changing little with temperature. In other words, if tributary parallel connections became relevant at higher temperature, less tortuous paths against-the-grain would be taken at higher temperature; then, the anisotropy would be closer to 1 at higher temperature and more extreme at cold temperature. Indeed, this behavior is observed on a small scale, relative to the overall resistance change, and demonstrates the impact of tributary parallel connections is small. This smallness in anisotropy change supports heterogeneous-transport models where CNT-conductor resistance is a series-sum of CNT structure resistors and junction resistors, without tributary parallel connections (e.g., equation (2)). See supplemental-section 2.2 for additional correlations, reduced-activation-energy-analysis, VRH and FIT fitting.

**High-field positive (+) MR.** High-field magnetoresistance ($MR(H)$=100%*($R(H)$-$R(0)$)/$R(0)$, sweeping magnetic-field $H$ up to 60 T and measuring four-wire resistance $R$) was accomplished with a pulse magnet at National High Magnetic Field Laboratory; figure 3 shows $MR$ versus $H$ for select aspect-ratios (AR) and temperature $T$. De-doped CNT-conductors under transverse-field (figure 3a) have an initially negative (-) MR, followed by +MR to an inflection point at ≈10 T, where +MR continues and is broadly quadratic (MR: 48% at 1.5 K, 60 T) without saturation. Near room-temperature, the +MR is smaller though still present (MR: 0.8% at 270 K, 50 T). In contrast, as-is (figure 3b) lacks positive quadratic MR. Further, for as-is and de-doped, there is no obvious qualitative difference between the 4800 and 1200 aspect-ratios. These results resemble our previous high-field MR of CNT-conductors from direct spinning, where CNTs are considerably longer[21] [74], although are more heterogeneous and porous; this highlights that our MR response applies to varieties of CNT-conductors. Notably, large longitudinal MR (figure 3c, $H$ now parallel to current and CNT alignment, MR goes up to 41% at 1.5 K and 22% at 270 K) matches or exceeds the transverse MR ($H$ perpendicular to current). While this particular longitudinal sample was de-doped by our standard 1000°C hydrogen treatment, because of logistical constraints, our 100°C vacuum bake-out within the magnet did not occur. This means there was some light-doping from atmosphere physisorption and the +MR could be even greater if fully de-doped. Compiling all de-doped results onto one graph, figure 3d plots temperature against the fitted straight-line slope of +MR vs $H^2$. This fitted slope monotonically decreases with temperature where transverse MR subsides faster than longitudinal (unabridged high-field MR, supplemental-section 3.0).

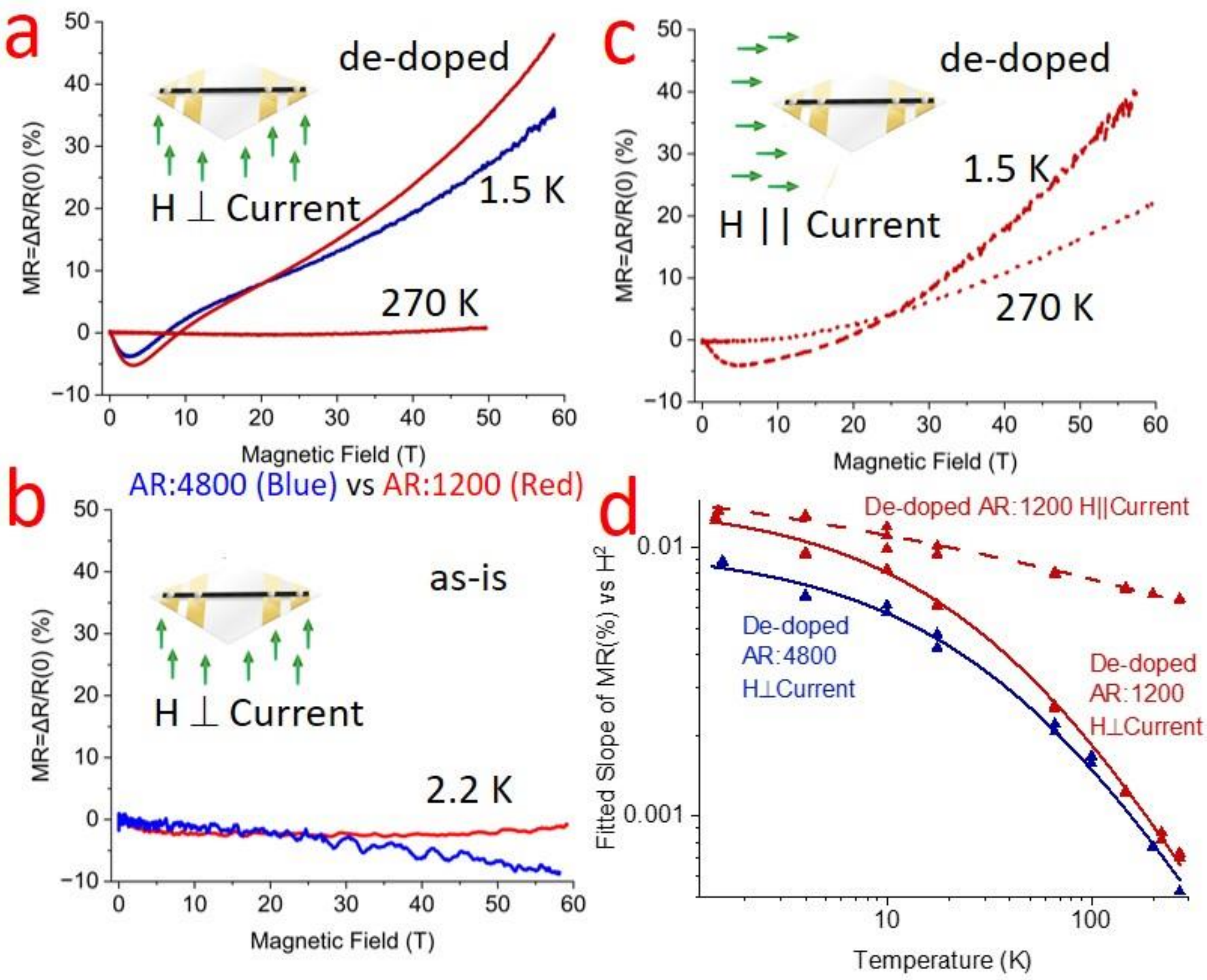

**Figure 3| High-field magneto-resistance (MR(%)=100% (*R*(*H*)-*R*(0))/*R*(0)) versus magnetic-field *H*, for select temperatures *T* for short (AR: 1200, red) and long (AR: 4800, blue) aspect-ratios (AR) and probe-current always || CNT alignment. a, de-doped with *H* ⊥ CNTs. b, as-is with *H* ⊥ CNTs. c, de-doped now with *H* || CNTs. d, just for de-doped, plotting the fitted slope of MR vs $H^2$ coefficient against *T*, which is useful for VRH and classical two-band analysis.**

**High-field +MR under VRH.** The high-field +MR increases quadratically without saturation and this discourages explanations using anisotropic magneto-resistance (AMR) dealing with magnetic impurities[75] [76] [77] or field-induced spin polarization mechanisms[21] [78]. Zeeman splitting becomes relevant when $gu_B H > k_b T$ , where $g \approx 2$ is the g-factor, and $u_B$ is the Bohr magneton; increased charge-carrier interaction from Zeeman splitting causes +MR for other carbon-based conductors[48] [79] [80] [81], although is regarded as a liquid-helium phenomena and does not explain the +MR near room-temperature. With transverse MR, for our CNT diameters ($d \approx 1.5$ nm), *H* is too small for Landau levels where ≈1000 T is required for significant +MR [82] [83] [84] [85] [86]. Note that only CNT-conductors with high-field +MR, the de-doped ones, also follow VRH. A well-known[34] [87] [88] [89] VRH MR mechanism predicts approximately quadratic +MR via H-induced constriction of the localized wavefunction; this leads to reduced hopping and lower conductivity, with a power-law temperature-dependence[34] [36]. :

$$MR(\%) = 100\% \frac{t\, e^2 a^4}{\hbar^2} \left(\frac{T_M}{T}\right)^S H^2 \quad (3)$$

where $a$ is the charge-carrier localization-length, $e$ electron charge, $\hbar$ is reduced Plank's constant, $t$ = 0.00248, and $s$ = 3/4 for 3D VRH [34] and 1 for 2D VRH [36]. In this picture, for the de-doped heterogeneous mixture of CNT junctions and CNT structures adding in series, the de-doped CNT junctions contribute to the VRH behavior while the de-doped metallic CNTs are not significantly contributing to MR. Best fits of VRH hoping models leads to a temperature-independent localization-length between 5 and 9 nm, which concurs with some CNT reports (4.6 to 9 nm [28] [30] [33] [35]). However, the MR's temperature dependence does not have the predicted power-law form (figure 3d does not follow a power-law), nor do the most charitable fits for exponent $s$ (0.25-0.27 for $T<T_M$ or 0.33-0.36 for all $T$) agree with the temperature power-law value in (equation 3). Further, for VRH, MR should decrease going from transverse to longitudinal $H$ orientations[87] [88] [89] (where field and electron-propagation are parallel, the resultant electron wave constriction does not hinder forward propagation); this decrease in longitudinal MR is not observed. While other mechanisms could be obfuscating the pure VRH response, our +MR also occurs near room-temperature, well above $T_M$ where VRH no longer applies; both transverse and longitudinal +MR is present when there is a metallic-like $dR/dT > 0$ response.

**Classical High-field +MR.** A classical two-band model accounts for the curved trajectory of free-carriers from the Lorentz force and models the +MR response of graphite[90], carbon-fiber[91] [92], and CNT materials[21] [45] according to:

$$\mathrm{MR}(\%) = 100\%\ \mu^2 \mathrm{H}^2 \quad (4)$$

where $\mu$ is temperature-dependent electronic mobility, with equal numbers of holes and electrons, with equal $\mu$, and with $\mu H \ll 1$. The cyclotron radius $r_c$ can be calculated with conservative values[21] of effective mass $m^*$ ($7.8 \times 10^{-32}$ kg) and Fermi velocity $v_f$ ($8 \times 10^5$ m/s) according to $r_c = m^*\, v_f\, e^{-1} H^{-1}$, yielding $r_c$ values 6.5 nm to 39 nm for magnetic-fields 60 to 10 T, respectively, and is small enough to fit within CNT bundles. This opens the possibility that anisotropic CNT bundles could support two-band MR between junctions. More complicated classical two-band variations account for unbalanced carriers (with electron $n_e$ and hole $n_h$ carrier densities) and angle $\theta$ between the graphitic-plane and $H$[91] [92] [93]:

$$MR(\%) = 100\% \frac{\frac{4 n_e n_h}{(n_e+n_h)^2} H^2 Cos^2[\theta]}{\left(\frac{1}{\mu^2} + \left(\frac{n_e-n_h}{n_e+n_h}\right)^2 H^2 Cos^2[\theta]\right)} \quad (5)$$

Here, the MR is initially quadratic with $H$, although saturates depending on hole/electron imbalances; as shown in (5), domination of one carrier leads to no MR (because the Hall voltage exactly balances the Lorentz force[94]) and this explains the lack of +MR in as-is CNT-conductors. Supporting this, Hall measurements on as-is (supplemental-section 4.4) demonstrate holes are dominant with a relatively temperature-independent carrier-density. Alternatively, the +MR of de-doped CNT-conductors is quadratic without saturation, so it fits best to the simpler (equation 4), supporting that electron/hole carrier densities are practically equal/compensated. Hall measurements for de-doped (supplemental-section 4.4) had a Hall voltage that fit less reliably with field; further, the Hall sign would sporadically flip between hole and electrons, followed by a dramatic and consistent change to electron conduction over 300 K. A Hall sign change indicates competition between electron and hole conduction (weighted by mobility and number density). Similar results were observed with graphitic-intercalation-compounds, where the pristine graphite has large quadratic +MR (≈$10^5$ at 4.2 K and 2.3 T)[90], which can be partly [12] [95] or fully [96] [97] suppressed after doping.

Applying (4) on de-doped MR, fitted $\mu$ ranged from ≈20 to 110 cm²/V s for 270 to 1.5 K respectively (figure 3d plots $\mu^2$ versus $T$); this fitted $\mu$ is a composite mobility affected by both junctions, as well as CNT structures interactions. $\mu$ is on-par with unaligned CNT networks (2-200 cm²/V s[51] [96]) and direct-spun CNT fibers measured similarly (≈120 to 210 cm²/V s [21]), though is well lower than the best individual CNTs (10,000 cm²/V s[98]), graphitized carbon-fiber ( ≈7,000 to 50,000 cm²/V s[91]), and graphite (15,000 to 1,300,000 cm²/V s[90]). Using our room-temperature $\mu$ from the transverse field (20 cm²/V s) and typical SWCNT film carrier densities from literature (0.5 to 1.3 x$10^{20}$ $cm^{-3}$)[99] [100], this arrives at a conductivity of ≈0.04 $MSm^{-1}$ and is closer to our de-doped ribbon conductivities measured perpendicular to the microstructure-alignment (0.01 to 0.02 $MSm^{-1}$). According to Matthiessen's rule, mobilities from different scattering mechanisms add in parallel: one mobility for elastic interaction ($T$ independent) and, in parallel, another mobility for phonon interaction (T-power-law with exponent x). We get[21] [91]:

$$\mu(T) = (\mu_{ELASTIC}^{-1} + \mu_{PHONON} T^{-x})^{-1} \qquad (6)$$

Fits to this model (which are superior to our VRH fits) are shown in figure 3d with fitted $x$ of 0.70. This $x$ is greater than direct spun CNT fibers (x= 0.5)[21], although less than graphitized carbon-fiber[91] and individual CNTs[15] [51] [101] [102] [103] [104] ($x$=1, from thermal phonon scattering). Using MR for $\mu$-characterization has proven useful for graphitized carbon-fiber where differences in graphitic perfection can no longer be resolved by Raman spectroscopy or X-ray diffraction[105]. Further, *quantitative* Hall measurements are difficult in heterogeneous systems like CNT films (as demonstrated by others[100] and us, supplemental-section 4.4 and 4.5), where the obscuration of the intrinsic Hall voltage amongst the heterogeneous conduction system over-estimates the carrier-density by several orders of magnitude.

**Longitudinal High-field MR.** Our de-doped CNT fibers also had significant *longitudinal* MR and, near room-temperature, this was greater than the transverse MR. While Lorentz force is zero with $H$ parallel to current, classical two-band anisotropic semimetals can still have longitudinal MR from off-diagonals in the mobility matrix[106] [107] [108], notably such as graphite's large longitudinal MR when $H$ and probe-current are normal the graphite plane (MR= 1200% at 8 T and 4.2 K[109]). Beyond classical MR, another contender for longitudinal +MR can be bandgap modulation from the Aharonov-Bohm effect (discussed next section). It is likely both quantum and classical effects simultaneously contribute to longitudinal MR in varying degrees, which obfuscates longitudinal high-field analysis compared to the simpler transverse orientation.

It is interesting that both *transverse* two-band and VRH models predict similar characteristic-lengths: distance between elastic scatter (4-4.7 nm, approximately calculated by $\mu_{ELASTIC}\ v_f\ m^*/e)$ or VRH's localization length (5- 9 nm). While our MR results largely favor classical two-band over VRH, independent of the selected mechanism, the quadratic +MR response in transverse high-field is correlated to improved conductivity of pristine CNT materials.

**Angular MR.** Our PPMS more precisely explored angle-dependent MR across a wider sample set by continuously rotating the sample in static field, albeit lower strength (≤9T). Starting with typical *transverse* MR, where the angle is fixed with low-field perpendicular to the sample, our results concur with literature[19] [21] [35] [66] [76] [79] ; that is, a -MR that becomes more negative with lower temperature (for T<10 K, MR: -1.5 -> -3% (as-is) and -4 -> -6% (de-doped), supplemental-figure 4.1-1--4). Then, just for de-doped below ≈5 K, a positive +MR component appears at higher field. Agreeing with multiple CNT studies[21] [37] [43] [44] [46] [47] [79] , our -MR for all CNT-conductors (as-is/de-doped; ||/⊥ alignment; various aspect-ratios, supplemental-section 4.2) fits best to 2D weak-localization with electron-electron dephasing; a somewhat worse fit is obtained from 1D weak-localization (used in InAs nanowires[110] [111] [112] [113] and suggested for CNTs in[66]) and no significant fit with 3D weak-localization[19] [43] [48] [114] [115]. For all categories, our fitted 2D phase-coherence length ranges from ≈51 to 10 nm for temperatures 1.9 to 100 K respectively (supplemental-figure 4.2-1), similar to other CNT reports[21] [79]. De-doped's closely-matching 2D weak-localization response was not expected; typically, -MR in VRH is from a conceptually similar interference mechanism for localized carriers, the so-called Sivan Entin-Wohlman Imry model (SEI)[34] [116] [117] [118] used in multiple CNT studies [19] [28] [29] [30] [31] [32] [33] [35] [119]. Note that, because of the low-field +MR present in de-doped for *T*≤5 K, we only considered de-doped weak-localization when T≥10 K (where +MR influence is assumed small). All-together, 2D weak-localization suggests charge-carrier confinement. This confinement is possibly on CNT bundles surface;[43] conductive atomic force microscope (AFM) studies demonstrated thinner bundles are intrinsically more conductive than wider ones[52] [54] [55] [103] [120]. Alternatively, contiguous sheets of metallic CNTs embedded in the bundle, together with non-participating semi-conducting ones, also form a crumpled-sheet where 2D-confinement could apply. The distribution of current within a CNT bundle is explored in the tight-binding simulations in the next section. Supplemental-section 4.2 has the complete fitted weak-localization parameters and phase-coherence plots.

To now explore angle-dependent MR, which is less addressed in carbon-conductor literature (with exceptions[76] [105]), we mechanically rotated the CNT-conductor orientation continuously in a steady DC magnetic-field *H*. Figure 4a and supplemental-figure 4.3-3--4 shows MR modulation of a de-doped CNT fiber at 4.5 T: sweeping from angle 0° (*H*⊥CNT), through 90° (*H*||CNTs), all the way around to 360°. In this particular run, the probe-current is fixed to parallel the microstructure-alignment. Multiple temperatures are shown. At higher temperatures (≈60 K), MR at 0 ° (H⊥CNTs) starts negative and, moving away from 0°, plateaus to more negative MR, then repeats at 180°. At colder temperatures (≈3.5 K), a +MR component now develops approaching 90 ° (H||CNTs). With higher field (9 T, figure 4b), the +MR component at 90 ° now dominates across all temperatures. As-is has a similar response, although the MR modulation is smaller (MR <1.5%, compared to de-dope's 6 to 20%, supplemental-figure 4.3-5). Fourier analysis of the periodic traces (supplemental-section 4.3-6) shows the 180° Fourier component is largest, corresponds to two-fold symmetry as expected from fiber morphology. 90° Fourier components are the next largest, corresponding to four-fold symmetry. Note that a constant and higher-order components are present, although are notably smaller. Supplemental-section 4.3-7--8 shows another angle-dependent

MR with probe-current now perpendicular to the microstructure-alignment; the MR modulation is again similar to the original, supporting no intrinsic transport difference between with or against-the-grain.

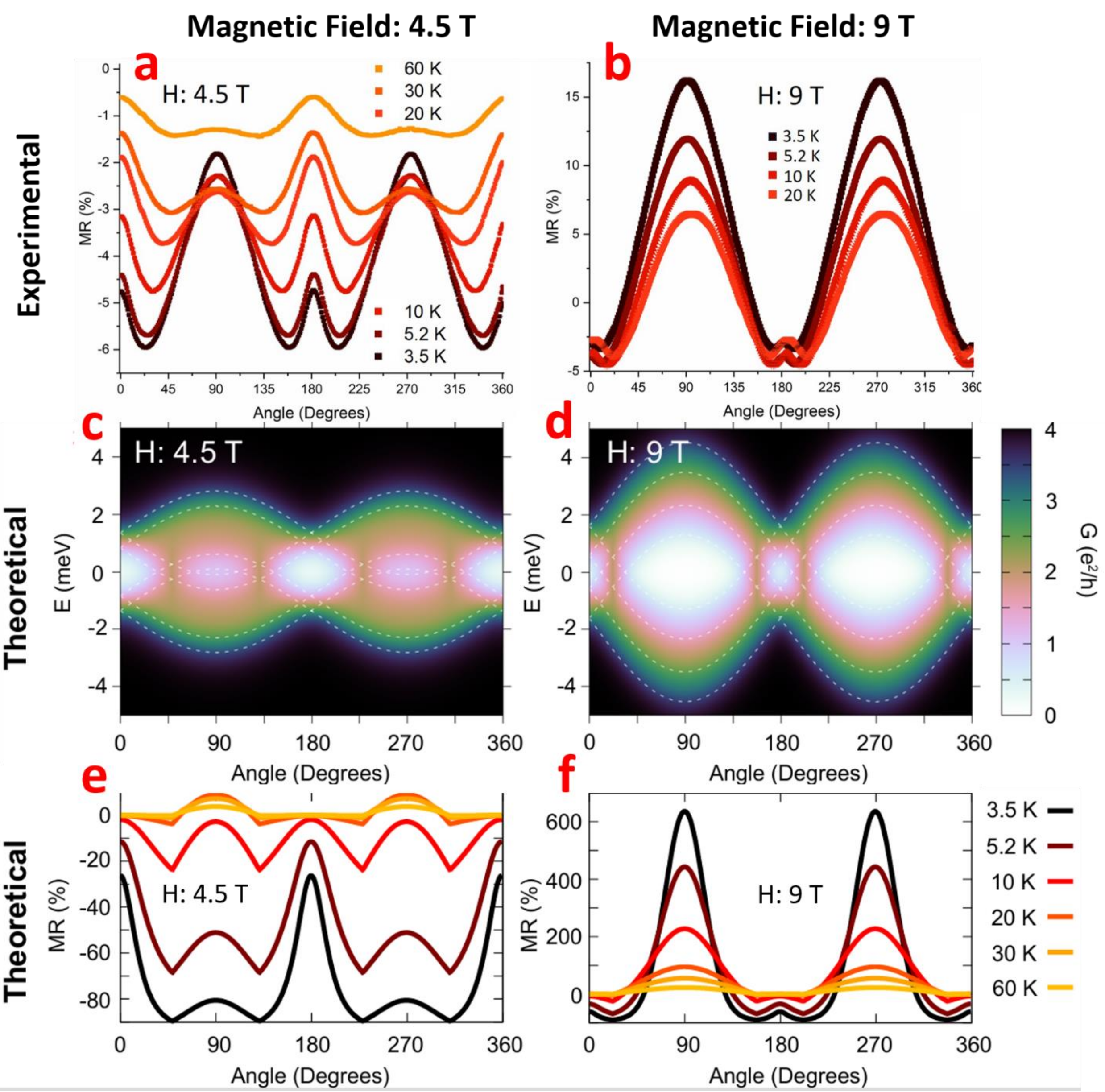

**Figure 4| a, MR modulation of a de-doped CNT fiber as it continuously rotates in steady field *H*=4.5 T, with probe-current always || CNT microstructure-alignment. 0° represents H⊥CNTs and 90° represents ||. b, the same except at 9 T. c, tight-binding simulation of the AB-modified curvature-induced bandgap versus field-angle, for an individual non-armchair metallic single-wall CNT, with *H*=4.5 T and *T*=3.5 K. Color represents transmission in units of quantum conductance ($G= e^2/h$). d, the same except *H*=9 T. e, Slicing the band-diagram at the Fermi level ($E_f$=0) allows MR (%) calculation, qualitatively reproducing the experimental results.**

Similar two and four-fold symmetry was also observed in the earlier angular-dependent MR study[76], which was attributed to significant residual iron catalyst; our case has only small traces of residual iron catalyst (supplemental-figure 1.3-1--2). Alternatively, classical two-band MR could contribute at low-field too; considering the greatest fitted high-field mobilities (1.5 K, transverse MR $\mu$= 94-112 cm²/Vs and longitudinal MR $\mu$= 117 cm²/Vs), this yielded MR≈50% at 60 T, although at 9 T this is an insignificant MR≈1%. Within weak-localization theory, with only inelastic dephasing (electron–electron/phonon) and no spin terms, MR is negative. When 3D, it is angle-independent. When 2D, it is most negative at 0° (H⊥confining-2D-plane) and MR should approach 0% for 90° (H||confining-2D-plane). Clearly, the angle-dependent MR differs from this simple weak-localization picture.

Due to CNT's cylindrical topology, their band-structure is sensitive to magnetic-flux threading the CNT diameter $\phi = H\pi(d/2)^2$, with their 1D subbands shifting with field via the Aharonov-Bohm (AB) effect. This phenomenon is caused by coupling of the orbital motion to the magnetic-field, and is periodic with period $\varphi_0 = h/e$, where $\varphi_0$ is the flux quantum with Planck's constant $h$. In gapless CNTs at low-fields this response results in the gradual opening of a bandgap, and in initially gapped CNTs in its closing[86] [121] [122] [123], which we propose as the main source of our -MR in parallel magnetic-field. The closing of semi-conducting bandgaps requires >100 T fields, but it is established that in metallic non-armchair CNTs the curvature of the atomic lattice opens a smaller bandgap, proportional to $\cos(3\vartheta)/d^2$ where $\vartheta$ is chiral angle[124]. This gap ranges from a few to a few tens of meV and is easily closed by experimentally available magnetic-fields. In[125], for example, an individual CNT (with small bandgap and $d$= 1.5 nm) is brought near the charge-neutrality point with a gate-electrode and then subjected to pulsed, parallel-oriented 60 T magnetic-field. At 82 K, the conductance increases by 14% from 0 to 5.9 T and, from there, the conductance plummets to ≈0 by 40 T. This modulation is a possible quantum contribution to our angle-dependent MR and classical longitudinal-MR in our high-field (hinted previously, figure 3c), and motivated the following effective model simulation.

**Calculation of the orbital response in magnetic-field.** The bandgap modulation via the Aharonov-Bohm mechanism is primarily caused by the coupling of the parallel component of the field to the orbital degree of freedom, which in CNTs corresponds to the valley, inherited from the CNTs parent graphene. This coupling has opposite sign for the K and K' valley, corresponding to clockwise and counterclockwise motion of the electron around the CNTs perimeter. The spin in turn couples to the full magnetic-field through the Zeeman effect, but the orbital response is usually several times stronger. Model details are found in methods and supplemental-section 5.1. The linear conductance through the infinite CNT with field-modulated band-structure is shown in figure 4 c,d as a function of the field-orientation, with the dashed white lines marking the position of the band edges and the color background encoding the linear conductance smeared by the temperature at 3.5 K (other temperatures are explored in supplemental-figure 5.1-1--2). In this calculation the leads are formed by the left and right semi-infinite parts of the CNT, while the central part is finite. The bandgap varies periodically with a dominant 180° period, but it has also a weaker 90°-periodic component (more visible for $H_{//}$ = 4.5 T). This subdominant component is caused by the fact that the gap closes at four positions: at $H_{//} = \pm H_{//,0}$ and at $H_{//} = \pm H_{//,0}$+180°. At 4.5 T the gap closes near 45°, which leads to the apparent 90° periodicity. At $H$ = 9 T the opened gap is so wide that $H_{//}$ reaches $H_{//,0}$ only close to 22°.

Slicing these data at $E_F$=0 at different temperatures yields the MR of a single CNT in the de-doped state. Similar to our experimental angle-dependent MR, in figure 4 e,f we see that at 4.5 T the MR at 0° is negative and decreasing with temperature; it features four dips spaced by ≈90° and flattens out towards higher temperatures. At 9 T the distinct +MR at 90° dominates the response. Other doping-levels are explored in supplemental-figure 5.1-3; this lowers the magnitude of the MR modulation, which was another experimental observation.

While the qualitative agreement between the experimental and theoretical result is quite good, there are some quantitative discrepancies. Firstly, the values for MR are unrealistically high. This is to be expected, since the calculation deals with a single CNT while the fiber is a composite of many CNTs with varying diameters and helicities, connected in parallel and in series – proper ensemble averaging would yield much lower MR magnitudes. Secondly, some features of the MR behave differently in the experiment and in theory. For instance, at $H$ = 4.5 T the value of MR near 90° changes very little with temperature, but the theoretical result varies very strongly, to the point of the MR becoming positive at high temperatures. There again the collective nature of the CNT fiber certainly comes into play, and with it likely weak-localization, missing in the single CNT calculation.

**CNT Bundle Simulations.** The CNT bundle is fundamental to the heterogeneous conductive network yet its distributed behavior is not fully understood. To investigate the electronic-transport distribution within a CNT bundle, with magnetic-field, we performed large-scale quantum transport calculations using tight-binding non-equilibrium Green's function (TB–NEGF) formalism. We modelled bundles composed of seven CNTs arranged in a flower-like, hexagonal configuration with the same or mixed helicity (either (9,9), (21,0), (12,3) or (20,0) or a mixture, see supplemental table 5.2-1 for geometry and selected transmission distribution). To match the experimental semi-conducting fraction (≈2/3), mixed bundles were carefully constructed with comparable composition and symmetry. We also modeled a large bundle of nineteen (9,9) CNTs arranged with hexagonal close-packed geometry, as well two CNT bundles separated by a junction (forcing inter-bundle tunnelling for conduction)**.**

Our approach leverages TBtrans and the sisl Python library, both extended and customized for this work. To capture magneto-transport at high magnetic fields—where perturbative approaches break down—we implemented the Peierls substitution directly into both the Hamiltonian and the overlap matrices across the entire system, including electrodes. This modification allows a fully gauge-consistent treatment of external magnetic-fields up to 60 T, oriented perpendicular to the CNT bundle (Supplemental Table 5.2-1, Supplementary Figure 5.2-4). Importantly, all systems span more than 10 nm in length and contain thousands of atoms, making this study among the first to achieve such transport simulations at this scale and level of theory, while retaining full quantum coherence.

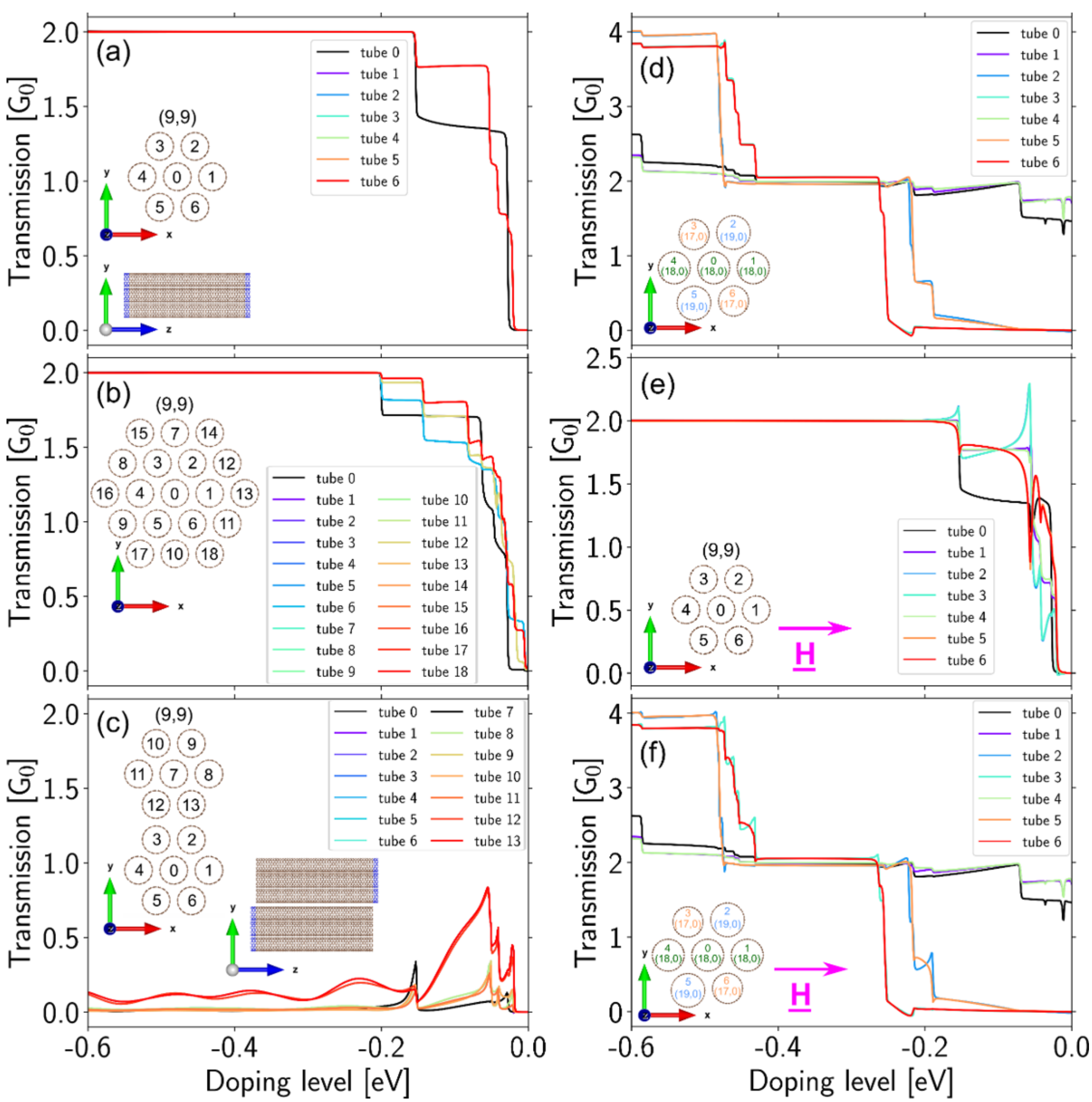

**Figure 5 | Tube-resolved, individualized transmission for various CNT bundles versus doping-level; the insets show side and cross-sectional views of the hexagonal CNT bundle arrangement where all CNTs are connected to semi-infinite electrodes (blue). a, Metallic-bundle composed of seven (9,9) CNTs at *H*=0 T; b, Larger metallic-bundle composed of nineteen (9,9) CNTs at *H*=0 T; inset shows a central CNT is surrounded by six in a middle layer and twelve in an outer layer, forming a finite hexagonal close-packed (HCP) arrangement. c, Junction composed of two metallic-bundles of seven (9,9) CNTs each at *H*=0 T. The Inset demonstrates how electron tunnelling between bundles is required for conduction. d, Mixed bundle consisting of three metallic (18,0) CNTs (CNTs: 0, 1, and 4), two semi-conducting (19,0) CNTs (CNTs: 2 and 5), and two semi-conducting (17,0) CNTs (CNTs: 3 and 6) at *H*=0 T. e, Metallic (9,9) CNT bundle with *H*=9 T perpendicular to the bundle (along the x-axis, as indicated). f, Mixed CNT bundle**

**with *H*=9 T perpendicular to the bundle (along the x-axis as indicated). Semi-conducting CNTs are labelled with lighter-colored numbers and metallic with darker-colored numbers. CNTs are plotted to scale, so differences in cross-sectional size reflect actual diameter variations.**

We first consider individual CNT bundles composed of all-metallic, identical CNTs with each CNT end uniformly connected to electrodes, without magnetic-field. When evaluated near the Fermi level $E_f$ =0 (un-doped), the transmission is frequently lower in the bundle core; this is shown in a seven-armchair bundle (figure 5a), particularly very close to Fermi level $E_f$ =0 at the pseudo-gap formed by armchair interaction (5.6% core vs 15.7% outer, supplemental table 5.2-1). Lower core transmission is also for the seven-(12,3) bundle (0.33% core vs 16.5 % outer, supplemental table 5.2-1). Near-zero transmission is also present in the center CNT of an un-doped, large 19-armchair bundle (0.25% core vs 3-7% outer, figure 5b and supplemental table 5.2-1). Structural relaxation of the bundle slightly deforms the central CNT from circular to off-circular, but largely does not alter the overall transmission profile. This demonstrates that our primary conclusions, even if drawn from mostly unrelaxed systems, remain valid and physically meaningful.

There are exceptions to lower core transmission in de-doped CNTs. A de-doped seven-(21,0) bundle has instead near uniform bundle transmission (supplemental table 5.2-1). This was a relatively large diameter and, in this isolated case, incorporating relaxation resulted in return to lower core transmission (6.5% core vs 15.5% outer, supplemental table 5.2-1 and figure 5.2-2). Further, when in a seven-(9,9)-armchair bundle, the center CNT was individually axially rotated by 20° (Supplemental-figure 5.2-2–3), the transmission through the rotated central CNT greatly exceeded the outer CNTs at $E_f$ =0 (75.3% core vs 4.1% outer). However, geometry optimization reveals that this rotated configuration is energetically less favorable (Supplemental Table 5.2-2), suggesting that the outer-conducting configuration is more likely in realistic conditions. For the large 19-armchair bundle, while the central CNT has near-zero de-doped transmission (0.25%), the adjacent CNT middle layer carries a larger de-doped transmission density (7.2%) than the surface layer CNT (3.3- 6.2%). Critically however, after even small doping (>0.1 eV), the transmission across any metallic-bundle becomes near uniform and overall greatly increases; the doped bundle conductivity converges to the sum of ballistic metallic CNT conductivity (each contributing two $G_0 = 4e^2/h$) or greater (for seven-(21,0) bundle, 50% greater from sub-band doping). This underscores doping's importance, even for all-metallic CNT bundles.

Now consider bundles with semi-conducting CNTs. For un-doped bundles of mixed metallicities, metal CNTs carry the transmission at $E_f$ =0. For a mixed seven-CNT bundle with three (18,0) metallic CNTs in a row (figure 5d), the bundle transport is concentrated across the metallic CNTs with less transmission in the middle (29.89% core vs 34.88% and 35.63% for outer metallic CNTs). Although the outer values are not identical, due to local atomic arrangements and inter-tube interactions, the trend remains: outer CNTs have higher transmission. Exploring this further, we simulated a 7-mixed bundle composed of two outer and opposing metallic CNTs with five interchangeable semi-conducting CNTs (Supplementary Figure 5.2-1). These semi-conducting CNTs were composed of two different semi-conducting helicities ((19,0) and (20,0)), where their bundle position was swapped without changing composition. In the de-doped case, the metallic CNTs always dominated the transport as expected, although metallic CNT-4 had a lion-share of the metallic transmission over metallic CNT-1 due to the inter-tube interactions. Swapping the positions of the semi-conducting helicities did quantitatively change the overall bundle transmission and somewhat modify the transmission partitioning across the metallic CNTs; qualitatively however, CNT-4 still captured most of the transport independent of adjacent semi-conducting helicities. This swapping observation, along with our other core bundle results, supports our previous case of the mixed bundle where transmission is less in the middle metallic CNT; therefore, the lower core transmission is better explained by its central location opposed to interaction from particular adjacent semiconducting CNTs. A de-doped fully semi-conducting seven-(20,0) bundle (Supplementary Figure 5.2-3d), due to its bandgap, exhibits zero transmission and the first non-zero contributions appear only at higher energies. As such, the CNT-resolved transmission profile at low energies is not directly representative of conducting behavior and should not be interpreted in the same way as metallic systems. Contrasting all-metallic-bundles, greater doping (0.6 eV) is required to make uniform non-zero transmission. Once activated however, the mixed or all-semi-conducting bundles (supplemental table 5.2-1) can have transmission greater than the equivalent doped metallic bundle of similar diameter (compare the results for a seven-(21,0) versus seven-(20,0) zigzag bundles); this illustrates that sufficiently doped semi-conducting CNTs can be better conductors then metallic CNTs, doped or un-doped.

Now consider two parallel seven-armchair CNT bundles separated by a junction; the bundles' far-ends uniformly connect to electrodes and the bundles' close-ends overlap (10.33 nm overlap with 0.34 nm separation, figure 5c), forcing transmission across the junction. Despite uniform electrode connection, and independent of the doping level, only CNTs directly adjacent to the other bundle participate in conduction (four out of 14, figure 5c). This is the only studied situation resulting in low bundle cross-section utilization, independent of doping-level. This key result implies that, for real-world CNT fibers with bundle junctions, thinner CNT bundles always have greater cross-sectional utilization.

Next, magnetic-field is considered for CNT bundle transport. Figure 5e and 5f show transmission under a 9 T perpendicular field for a metallic (9,9) bundle and a mixed bundle, respectively. A direct comparison with the zero-field configurations (Figure 5a and 5d) reveals the spatial transmission distribution remains largely unchanged from field. In Figure 5e and Supplementary Figures 5.2-4 a–c however, we observe spikes in the individual transmission curves for outer tubes when field is applied, particularly within the energy range −0.2 eV to 0 eV. These spikes highlight the sensitivity of coherent transport in metallic systems to magnetic phase effects and inter-tube interference. Despite these local doping spikes for individual CNTs, the overall bundle transmission is not significantly modulated for perpendicular magnetic-field (from 4.5 to 60 T), the seven-armchair or mixed bundle cases, across the doping energy continuum (supplemental table 5.2-1, figure 5.2-4).

**Discussion**

Heterogeneous FIT has long been proposed[50] for CNT-conductors, although is often overlooked in authoritative CNT transport papers[19], and here we uniquely show its unavoidable relevance with the semi-conducting-like temperature response leveling-off to a constant value approaching absolute-zero. Further support includes 1) the well-established mixed metallic-like and semi-conducting-like temperature responses; 2) the dependence of cryogenic-resistance-ratio on CNT length, which is longer than any homogeneous characteristic-length; and 3) Drude-conduction signatures with high-field MR and Hall measurements. Homogeneous mechanisms still, however, contribute: 1) weak-localization causes transverse -MR and 2) de-doping leads to VRH. Further, considering the conductivity-anisotropy does not change significantly temperature implies that: 1) the anisotropy is created by the extrinsic difference in path-lengths between "with-grain" and "against-the-grain"; 2) that current distribution across the various network paths does not change much with temperature. 3) Thus, a simple series-sum of CNT structure and junction resistances are sufficient for network modeling and tributary pathways that open up at higher temperature, while present to some degree, are not overall significant. Similar extrinsic network geometry narratives [71] [126] and simple resistance series-sums [18] [21] [31] [49] [59] have been found in other CNT materials.

What are the consequences of accepting a heterogeneous system where resistance components simply sum in series? At sufficiently warm temperatures, both VRH and FIT become Arrhenius style thermal-activation $R_{SemiCond}(T) = R_c exp\left(\frac{K}{T}\right)$ with fitting factors $R_c$ and $K$. This Arrhenius region may be identified as a straight segment on an Arrhenius plot (ln$R$ vs $T^{-1}$) before the upswing from metallic resistance. Using the heterogeneous model $R(T) = R_{SemiCond}(T) + R_{Metal}(T)$ and just considering the warmer region where Arrhenius style thermal-activation applies, the semi-conducting contribution may be subtracted-out after fitting. This leaves the intrinsic metallic resistance without specifying the semi-conducting form and is shown in figure 6 below. Figure 6a shows the relative metallic contribution to the total resistance at room-temperature $\left(100\% \frac{R_{Metal}(300K)}{R(300K)}\right)$ versus aspect-ratio. For as-is with parallel microstructure-alignment, this relative metallic contribution (≈16 to 55%) increases almost linearly with aspect-ratio (power-law 0.85 +/- 0.11), because junctions take less share of network resistance with longer CNTs; extrapolating, 100% metallic conduction requires an aspect-ratio of at least ≈12000. De-doped have lower metallic fractions (≈4 to 12%) without correlation to aspect-ratio, because junctions here dominate network resistance independent of CNT length. Cross-grain samples have similar metallic fractions and correlations as their counterparts, despite a ≈20x difference in conductivity; its network path has the same junction/structure relationship as for the parallel-alignment, although the cumulative network path is meandering and much longer (in this framework, 20x longer).

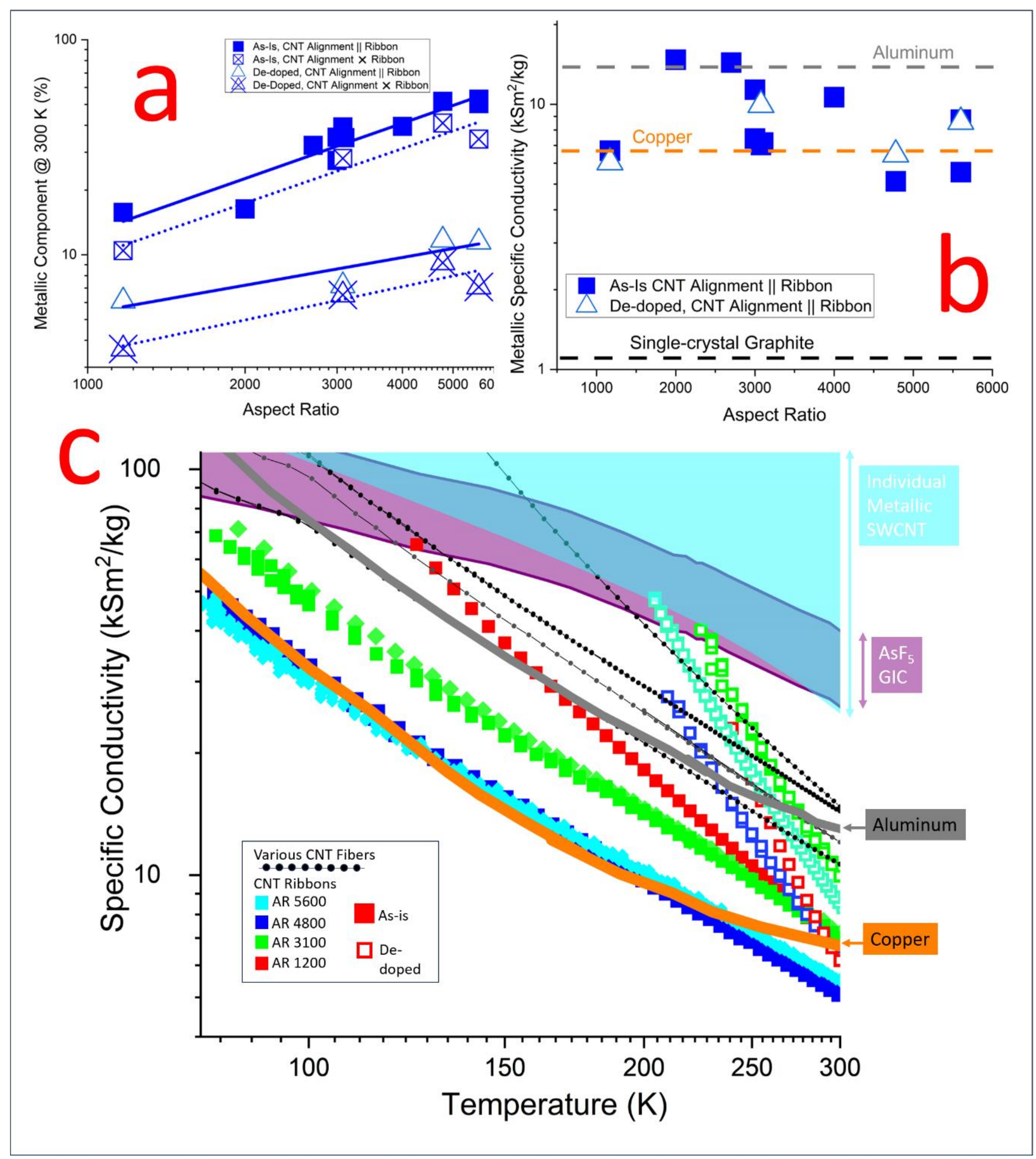

**Figure 6| Estimating metallic-component conductivity. Subtracting the fitted component of Arrhenius thermal-activation leaves the metallic component with relatively few assumptions on transport mechanisms. a, aspect-ratio (AR) versus the metallic percentage of total room-temperature resistance. b, The room-temperature specific-conductivity of the metallic-component compared to copper and aluminum. c, temperature-dependent specific-conductivity of metallic-components with different aspect-ratios (AR) and doping-status, alongside relevant benchmarks[127] [128].**

Figure 6b shows the room-temperature, metallic-component specific-conductivity of as-is and de-doped CNT-conductors if extrinsic junctions were eliminated, compared against relevant benchmarks. This calculation assumes that the network path-length is minimized; for parallel microstructure-alignment this is approximately true, while perpendicular are not considered here. As shown, metallic-component specific-conductivity are well above copper and some as-is fibers are above aluminum (the highest practical specific-conductivity metal). There is no correlation with aspect-ratio, reflecting its intrinsic nature. The metallic-component specific-conductivity of de-doped and as-is are comparable at room-temperature. Conductive AFM studies on sparse CNT networks show that doping both enhances transmission across junctions as well as increases carrier-density[55] [129], contrasting graphitic-intercalation-compounds (GICs) where doping just increases carrier-density[12]. In our case, the doping-enhanced carrier-density and improved junction transmission may be partially countered by increased scattering within bundles. Figure 6c shows the temperature-dependence of the metallic-component specific-conductivity with different aspect-ratio and doping-status. Other advanced carbon conductors (GICs, individual metal CNTs) and metal benchmarks (copper, aluminum) are shown based on their reported resistance vs temperature plots and room-temperature specific-conductivity. As-is is between copper and aluminum. De-doped traces have a steeper temperature-dependence, likely from a decrease in the proposed doping induced scattering.

In conclusion, the quantitative understanding and controlling of the semi-conducting-like and metallic-like conductivity responses to temperature is practical and necessary for overall conductivity improvement. Were it not for the junctions, the conductivity of CNT-conductors already exceeds copper and aluminum. For both as-is and de-doped, increasing aspect-ratio decreases the junctions' impact; a 12000 aspect-ratio should then lead to significant junction suppression at room-temperature. Increasing aspect ratio further will have diminishing returns at room-temperature, although could lead to a completely metallic-like conductivity response down to liquid-helium temperatures ($d\sigma/dT<0$ for all $T$). From that point, continued conductivity improvement is now reliant on improving the intrinsic metallic structure (thinner bundles, less defects, and optimally doped semi-conducting CNTs). Fully de-doped CNT-conductors become localized at cryogenic temperatures, despite the relatively favorable degree of microstructure-alignment, graphitic quality, and aspect-ratio. This contrasts the behavior of its older analog single-crystal graphite which already has the full metallic-like temperature response. A fundamental difference is that single-crystal graphite has planes in registry and this order does not easily exist for CNTs within a bundle; further, fully de-doped semi-conducting CNTs are simply deadweight. Notably large tight-binding calculations showed complex current distributions within a de-doped bundle, although transmission tended to be lower in the core of metallic CNTs; doping universally increased the uniformity and overall transmission of bundles of mixed, fully-semi-conducting, and fully-metallic CNT composition. Critically, they demonstrated that when one bundle is connected to another in series, only the CNTs adjacent to the other bundle participate in the transport, independent of the doping-level. This favors thinner bundles with more connections for higher conductivity. The impact of the magnetic field in on transmission distribution within a bundle is small.

The low-field negative MR across all samples best follows 2D weak-localization; we conclude that the main components of the network are not individual CNTs but their bundles, and that bundle transport is at least partially coherent. As found by others, the 2D character hints that the transport occurs either at the bundle surface or in contiguous crumpled-sheets of metallic CNTs embedded in the bundle. The magnetic-field orientation study, including substantial longitudinal MR near room-temperature, indicates that the transport response carries a strong signature due to the individual components, in the form of the gap modulation via the Aharonov-Bohm effect due to the parallel component of the field. We found that the positive quadratic MR at high-field for de-doped CNT materials is governed by a characteristic parameter related to higher conductivity, likely classical in nature. As conductivity and the characteristic-length increases, this +MR signal will become more accessible at lower field; this is particularly useful as Raman and X-Ray diffraction lose resolution for highly graphitic systems and Hall measurements are less effective for heterogeneous systems.

## Methods

**Database.** All data is organized and provided in the supplemental database. Also included are material property correlation tables (p-values <0.05 were statistically significant) and a compilation of the transport models fits. Linear regression and non-linear curve fitting was accomplished in OriginPro where error was propagated to calculate a reduced $X^2$ for model comparison.

**Materials.** Provided from Dexmat, aligned CNT ribbons and fibers were made from the established[1] [9] method of wet-spinning solutions of chlorosulfuric acid (CSA) and CNTs through a spinneret into a coagulant, such as acetone or water. The CNTs were few-walled and of high graphitic quality; this was verified by Raman spectroscopy showing radial breathing modes (RBMs) and a high G:D ratio (supplemental-figure 1.2-1). After the 1000 °C $H_2$ bakeout, Raman spectra became sharper and there was no degradation of the G:D ratio. Scanning electron microscopy shows the degree of microstructure-alignment (supplemental-figure 1.1-1). Thermogravimetric analysis (TGA) showed residual iron content <0.6 % and X-ray fluorescence showed surface iron contamination was < 0.1 % (Supplemental-figure 1.3-1--2). CNT ribbons and fibers were composed of CNTs with company-controlled molecular aspect-ratios, to include 5600, 4800, 3100, and 1200. These aspect-ratios were verified by dissolving the CNTs back into CSA solution and conducting extensional rheometry[130].

**Physical Property Measurement System.** CNT-conductors were measured in a Quantum Design Physical Property Measurement System (PPMS) using a standard four-wire resistance technique with probe-wires mounted with silver paint (ribbons were typically 1 cm long, 1-2 mm wide, 4- 5 mm between inner leads). Sample resistances ranged from 0.02 to 0.16 Ω (as-is ribbons), 0.1 to 2 Ω (de-doped ribbons), 0.5 to 37 Ω (CNT fibers, as-is and de-doped) with temperatures ranging from 1.9 to 373 K. Different probe-currents were used for a sample run (20 uA, 100 uA and 1000 uA) where trace overlap indicated sample heating was not an issue; the DC current switched directions to remove thermoelectric effects. When used, the DC magnetic-field (9 T max) was applied normal to the surface of the CNT sample and consequentially perpendicular to both the probe-current and CNT microstructure-alignment. In some circumstances a specialized probe holder was used to continuously rotate the CNT sample in the static magnetic-field. 0° indicates that the magnetic-field is perpendicular to the sample surface, 90° indicates the magnetic-field is parallel to the sample surface. The sample first rotated without any magnetic-field to generate a base line (supplemental-figure 4.3-10). For resistance measurement for temperatures between 0.05 to 3.8 K, select CNT ribbons were measured using a separate Quantum Design Dynacool PPMS equipped with a dilution refrigerator (DR) insert and Electrical Transport Option (ETO) module. An in-situ cryo-clean was made for optimal performance of DR, especially to stabilize the temperature at ultra-low temperatures. Probe-currents ranged from 10 to 50 mA without signs of heating. Temperature ramp rate was 0.025 K/min to ensure sample thermo-equilibrium. Both 0 T and 14 T were applied with magnetic-field normal to the CNT ribbon surface.

**High-Pulsed Field.** Four-wire probe magneto-resistance measurements up to 60 T (≈9 ms rise, 30 ms fall) were accomplished with a pulse magnet at the National High Magnetic Field Laboratory (NHMFL), Los Alamos National Laboratory (LANL). Two samples of each CNT ribbon category were measured simultaneously for redundancy. Samples were hooked up to fine magnet wires with silver paint, with care taken to minimize wire loops and anchor wires to the supporting structure with GE varnish, to minimize vibration. Probe-current was always in the direction of microstructure-alignment and the magnetic-field was typically perpendicular to the sample surface (transverse MR) and one sample run had magnetic-field parallel to the probe-current and microstructure-alignment (longitudinal MR). The magnetic-field was measured by a calibrated B-dot probe inductor, which was also used to precisely align the sample in the magnet's center when installing the sample. Supplemental-figure 3.1-1 shows the measurement circuit. An AC signal with frequency above the magnet noise (typically either 32 kHz or 175 kHz) drove current through the sample, with sample voltage sent through a Stanford Instruments SR560 Preamp (typically 1000x amplification, 10- 300 kHz bandpass), before digitization recording (3.2 MS/s). A 10.2 Ω was put in series with the sample to probe the sample current, where its voltage drop was measured in a similar way. A post-process lock in amplifier technique isolated the sample resistance from electromagnetic noise. A magnetic pulse measurement was accomplished for multiple temperature setpoints from 1.45 to 270 K. For the de-doped CNT ribbons, before measurement, a bake-out for ≈100 °C, 1 hour under vacuum was also accomplished within their measurement housing.

**Model of orbital response in magnetic-field.** The construction of the appropriate effective Hamiltonian, describing the physics of the bands close to the charge-neutrality point, is discussed in the supplemental-section 5.1. Its main parameters are the initial small bandgap (2.2 meV) and the magnitude of the orbital magnetic moment (0.323 meV/T); they were estimated from the fact that at high temperature the strongest -MR in de-doped fibers occurs at $H_{||,0} \approx 3.4$ T (figure 4 a,b) and from the value of d ≈1.5 nm measured in Raman spectroscopy. The reason why we used high temperature data for parameter extraction is that the experimental situation may be complicated by an additional contribution to MR from weak-localization; but between 30 K and 60 K the strongest -MR remains at 45°, so the weak-localization is likely already suppressed.

**Tight-binding bundle modelling.** The spin unpolarized tight-binding calculations of SWCNT bundles and SWCNT bundle junctions were performed using sisl python library[131] and TBtrans[132], a tight-binding code integrating the non-equilibrium Green's function formalism, with magnetic field included in the Hamiltonian by Peierls substitution. A complete description is provided in supplemental-section 5.2.

## Acknowledgements

This work was performed at the National High Magnetic Field Laboratory and is supported by the National Science Foundation, the Department of Energy, and the State of Florida through NSF Cooperative Grant No. DMR-1157490 and by U.S. DOE BES Science at 100 T project. Some of this work was supported by the Air Force Office of Scientific Research (LRIR #18RQCOR100 and #24RQCOR004). This research was performed while J.B. held an NRC Research Associateship award at the U.S. Air Force Research Laboratory (AFRL), Aerospace System Directorate (AFRL/RQ). We wish to thank the Carbon Hub and Dexmat for material support. SDW and GP gratefully acknowledge support via the UC Santa Barbara NSF Quantum Foundry funded via the Q-AMASE-i program under award DMR-1906325. Special thank you to Rober Waelder (AFRL) and Matthew Foster, Geoff Wehmeyer, Junichiro Kono (Rice University) and Vasili Perebeinos (University of Buffalo) for valuable discussions on carbon nanotube transport.

T.K. and K.Z.M. gratefully acknowledge the Interdisciplinary Centre for Mathematical and Computational Modelling at University of Warsaw, Poland (Grant No. G47-5) for providing computer facilities and technical support. T.K. and I.L. also acknowledge the technical and human support provided by the DIPC Supercomputing Center, Spain. T.K., K.Z.M. and I.L. are grateful to the Agencia Estatal de Investigación, Ministerio de Ciencia e Innovación, Spain for funding this research under Proyectos de Generación de Conocimiento 2022 program, PID2022-139776NB-C65. K.Z.M also would like to thank the European Commission (Marie Skłodowska-Curie Cofund Programme; grant no. H2020-MSCA-COFUND-2020-101034228-WOLFRAM2) for supporting this research. T.K. acknowledges the 3rd edition of Microgrants in Action IV.4.1 - 'A complex programme of support for UW PhD students', implemented as part of the 'Excellence Initiative - Research University' (IDUB) Programme. JAM acknowledges the support from Centera2 project (FENG.02.02-IP.02.01-IP.05-T0004/23) funded with IRA FENG program of Foundation for Polish Science, and co-financed by the EU FENG Programme. A.L.-R. would like to thank Warsaw University of Technology, Poland—Excellence Initiative (Materials Technologies–3 ADVANCED, grant agreement no 1820/359/Z01/POB5/2021) for funding this research. M.M. gratefully acknowledges helpful discussions with Leonid Golub.

## Conflict of Interest

The authors declare no conflict of interest.

Supplemental Section for

Adjudicating Conduction Mechanisms in High Performance Carbon Nanotube Fibers

**<u>Supplemental section 1.0-- Materials</u>**

Below is a materials property survey of the CNT ribbons and fibers (with the various aspect ratios (AR), in either as-is or de-doped status) with scanning electron microcopy (SEM), Raman spectroscopy, thermogravimetric analysis (TGA) and X-ray florescence (XRF).

**Supplemental Section 1.1 SEM.** Below are selected representative SEM images of the as-is and de-doped CNT ribbon with different aspect ratios (AR), accomplished with a Zeiss Gemini SEM with relevant SEM parameters provided in the photographs.

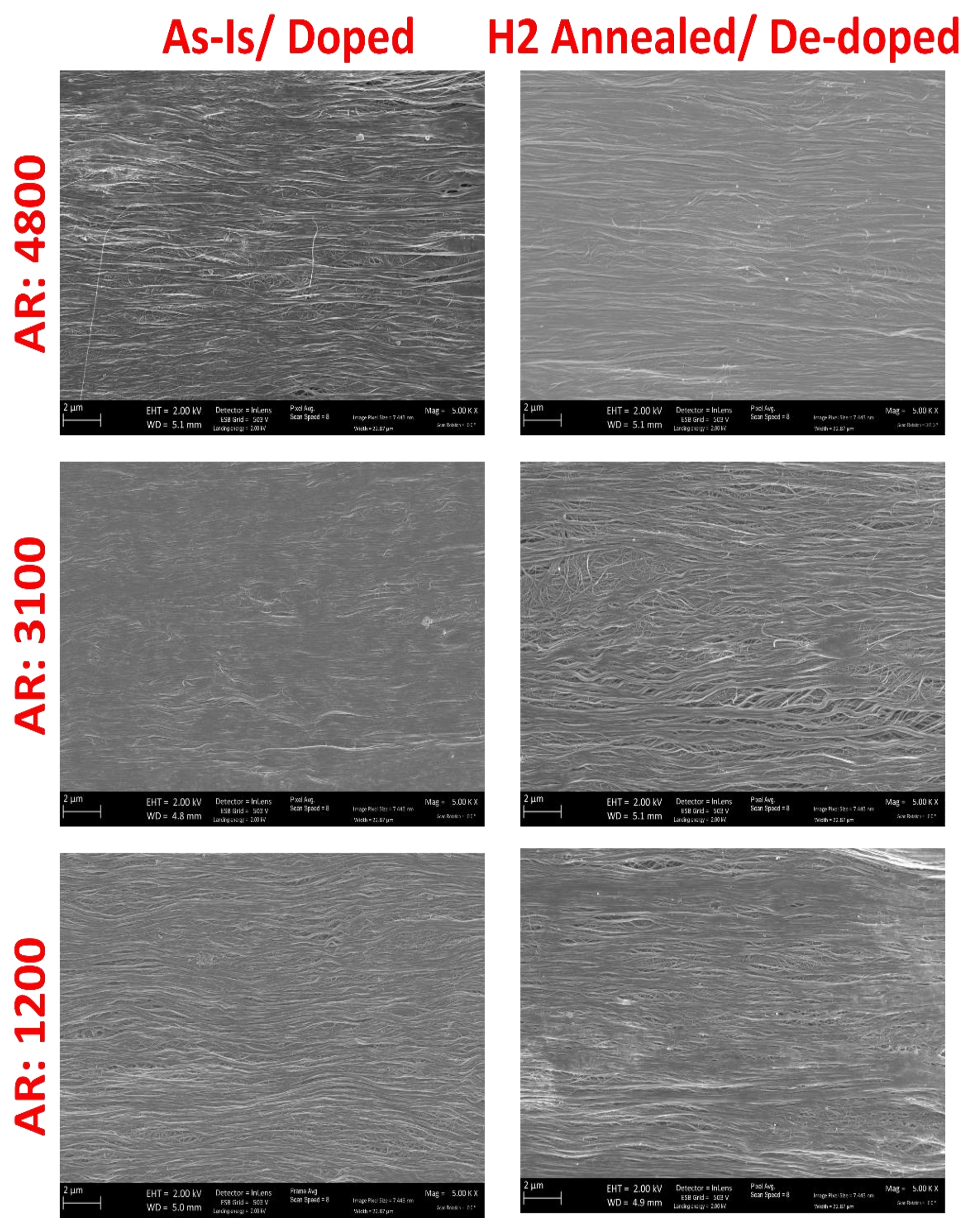

**Supplemental figure 1.1-1. SEM of the CNT ribbons, depending on doping status and aspect ratio (AR).**

**Supplemental Section 1.2 Raman**. Representative Raman spectra of the as-is and de-doped CNT ribbons with different aspect ratios. A silicon-calibrated Renshaw Raman spectrometer was used (shown here with 50x objective) and 785 nm laser excitation, with laser polarization parallel to the CNT microstructure alignment. These spectra show the existence of radial breathing modes (RBMs) that indicate the existence of few-walled CNTs (FWCNTs). 785 nm average Raman G:D ratios for as-is ranged from 54 (AR 1200) to 72 (AR 3100) to 60 (AR 4800); for de-doped this was 4.75 (AR 1200) to 36.9 (AR 3100) to 28.0 (AR 4800), all indicating a high degree of graphitic perfection. Complete Raman spectra for 785 nm, 633 nm, 514 nm, and 488 nm, as well as various calculated Raman metrics, are contained in the supplemental database.

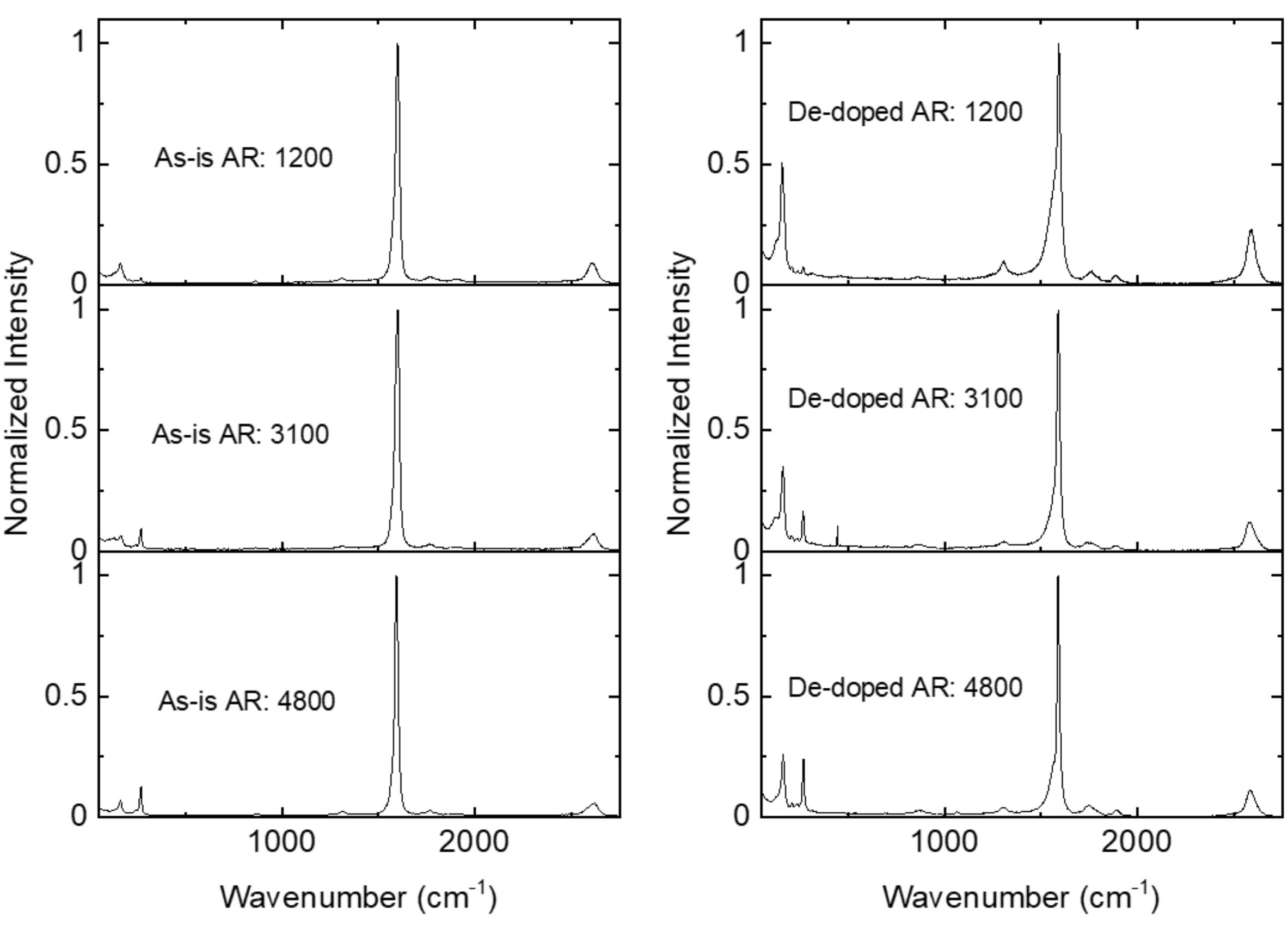

**Supplemental figure 1.2-1. 785 nm Raman of the as-is and de-doped CNT ribbons made from three different aspect ratios (AR).**

**Supplemental Section 1.3 TGA and XRF**. Thermogravimetric analysis (TGA) was accomplished with a TA TGA5500 with dynamic heating either in air or $N_2$ atmospheres to 1000 °C. X-Ray Florescence (XRF) was accomplished with a Bruker M4 Tornado and 25 sample points were collected to generate an average. Samples were stacked together in multiple layers to ensure only the CNT ribbons were characterized and not material underneath the sample.

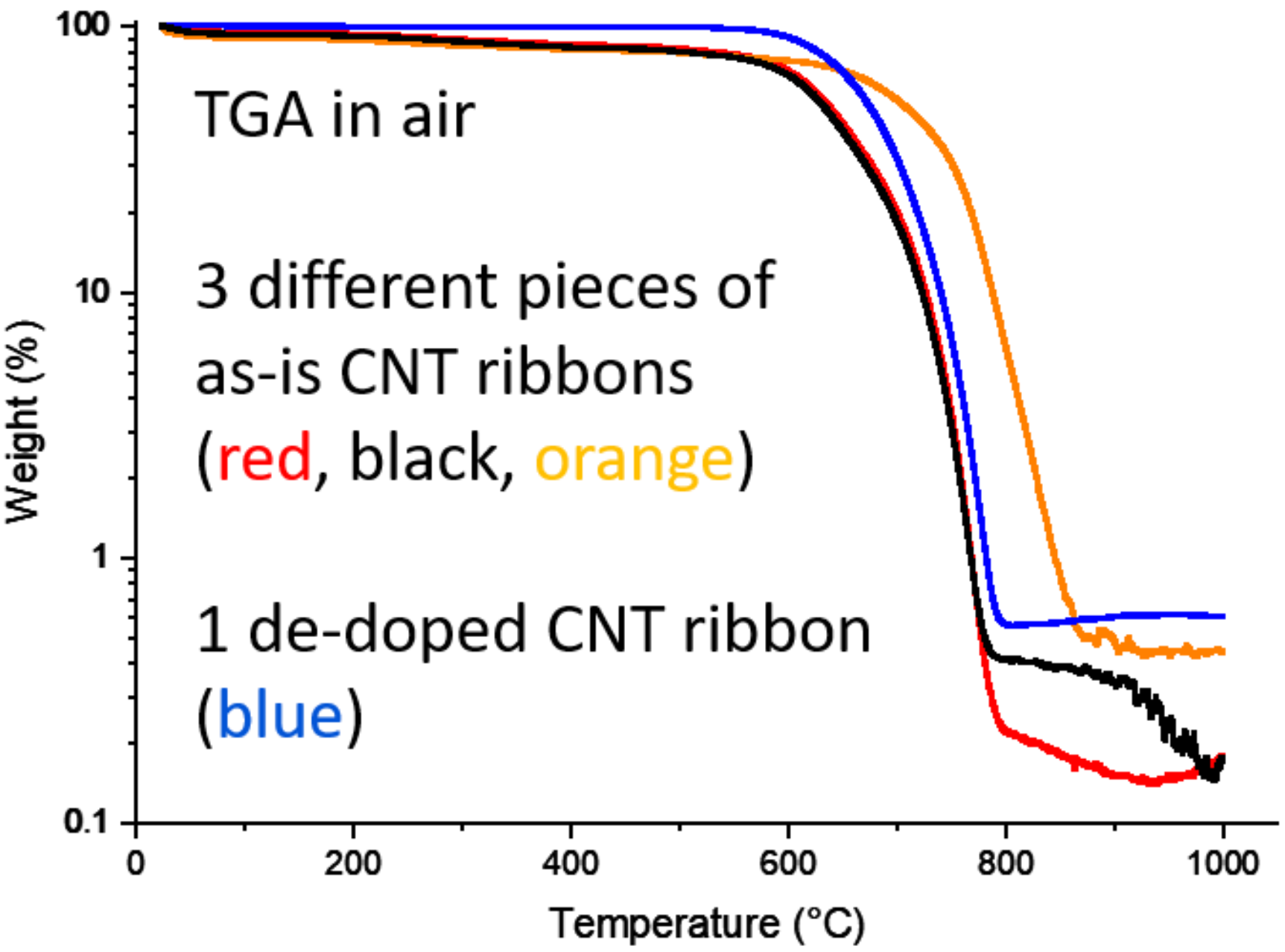

**Supplemental figure 1.3-1. TGA in air of three different pieces of as-is CNT ribbon (red, black, orange) and a de-doped CNT ribbon (blue). We see that the ash weight, and hence the residual iron content, is less than 0.6%.**

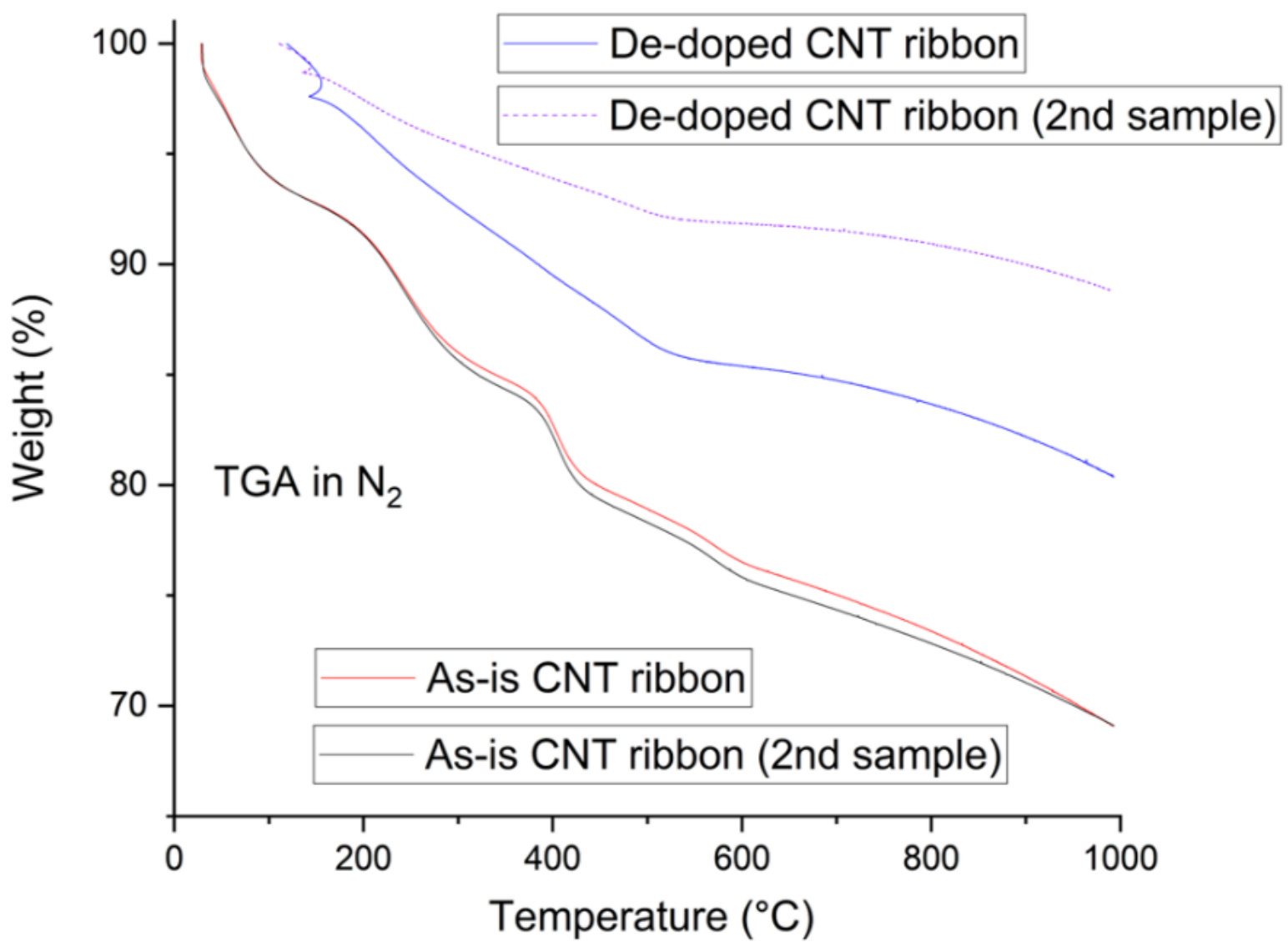

**Supplemental figure 1.3-2. TGA in an inert $N_2$ atmosphere for as-is CNT ribbons and de-doped CNT ribbons that were previously de-doped by a bake-out in pure $H_2$ at 1000 °C for at least one hour.**

**Supplemental Table 1.3-1. XRF elemental compositions**

| | As-is CNT Ribbon (5 layers) Mass % | De-doped CNT Ribbon (3 layers) Mass % |
|---|---|---|
| C | 94.758 | 99.787 |
| Si | 0.102 | 0.096 |
| S | 5.067 | 0.092 |
| Ca | 0.013 | 0.007 |
| Fe | 0.06 | 0.017 |

## Supplemental section 2.0-- Zero-field Conductivity versus Temperature

**Supplemental section 2.1 Reduced Activation Energy**. Reduced activation energy $W$ is a metric calculated from the zero magnetic field resistance $R$ versus temperature $T$ data according to $W = - d\ln[R]/d\ln[T]$ and is used to better determine conduction mechanisms with less ambiguity. For variable range hopping (VRH), approaching absolute zero, $W \propto -T^{S}$ where 3D VRH $S$= -1/4; 2D VRH $S$= -1/3; and 1D or ES VRH $S$= -1/2. For the insulator to metal transition, $W$ approaches a constant value approaching absolute zero. For metal systems with delocalized charge carriers approaching absolute zero, $W$ approaches zero[1]. Below are some selected graphed examples of the reduced activation energy function; the database contains transport assignment of each material variety based on smoothing the reduced activation energy function.

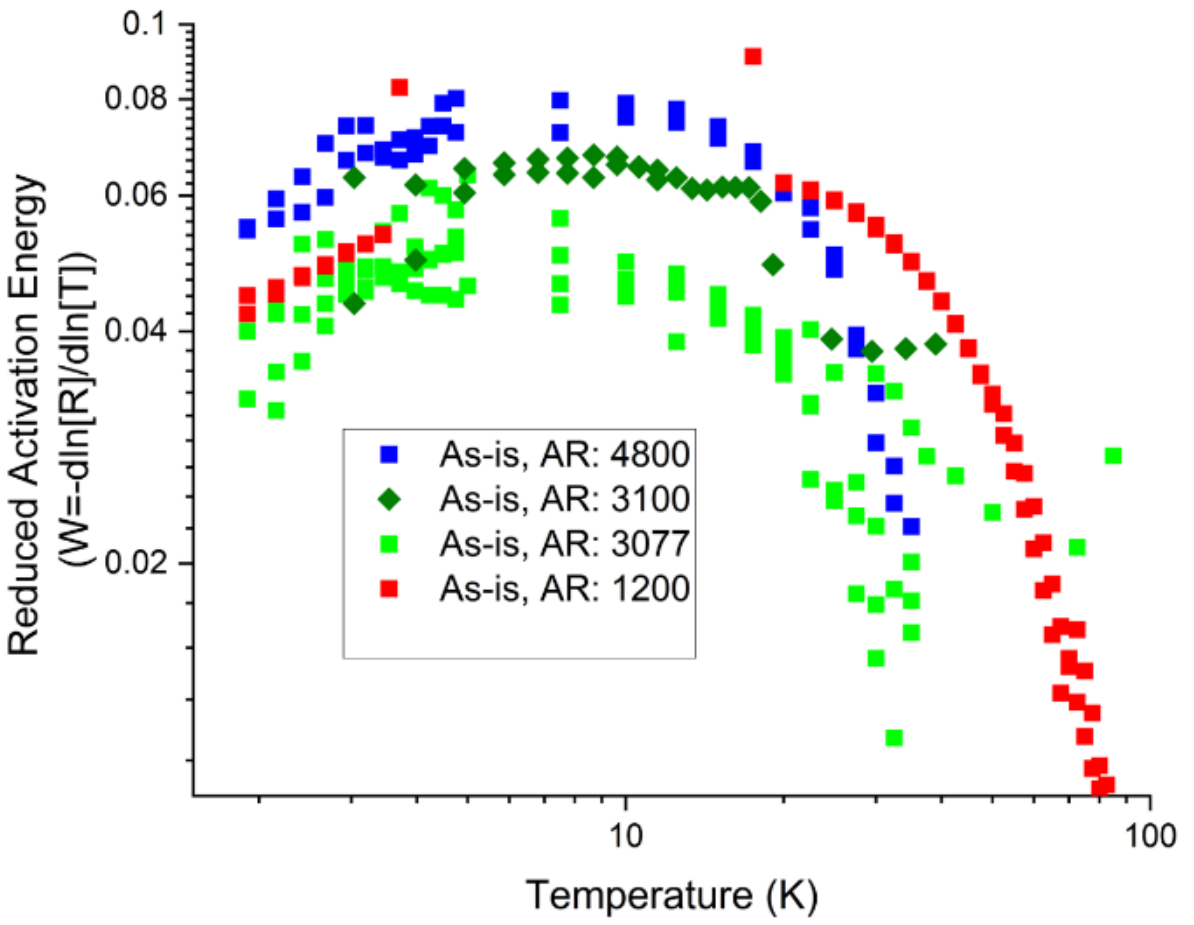

**Supplemental figure 2.1-1. Reduced activation energy for the as-is CNT ribbons, showing a decreasing trend approaching absolute zero on the log log plot, consistent with a metallic system.**

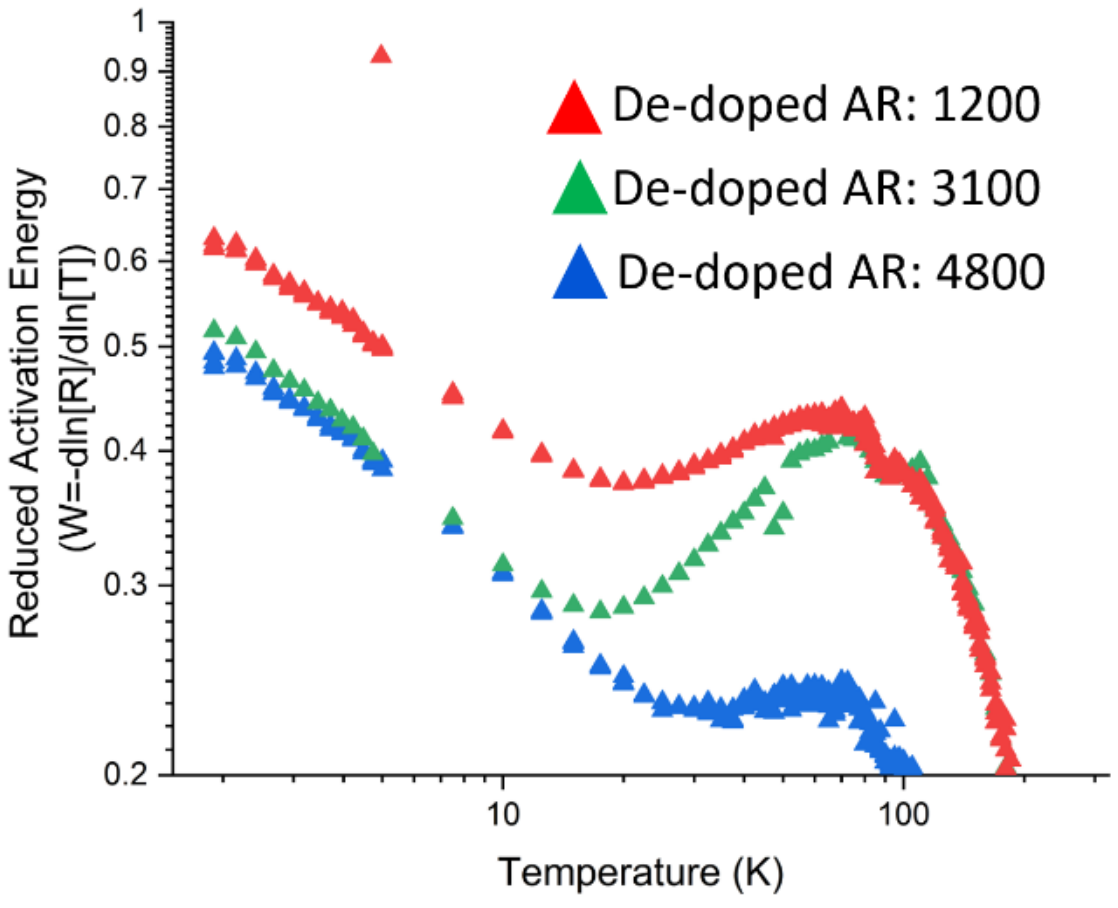

**Supplemental figure 2.1-2. Reduced activation energy for de-doped CNT ribbons, showing a power law approaching absolute zero that is consistent with variable range hopping.**

**Supplemental section 2.2 Property Correlations.** In this section we discuss electrical transport correlations and power-laws that exist for the CNT fibers and ribbons in the as-is and de-doped state. When we just consider ribbons, their morphology allows us to measure the conductivity and cryogenic ratio across the grain of the microstructure alignment. In general, we measure each material multiple times to generate a standard deviation, with full material property calculations found in the data base.

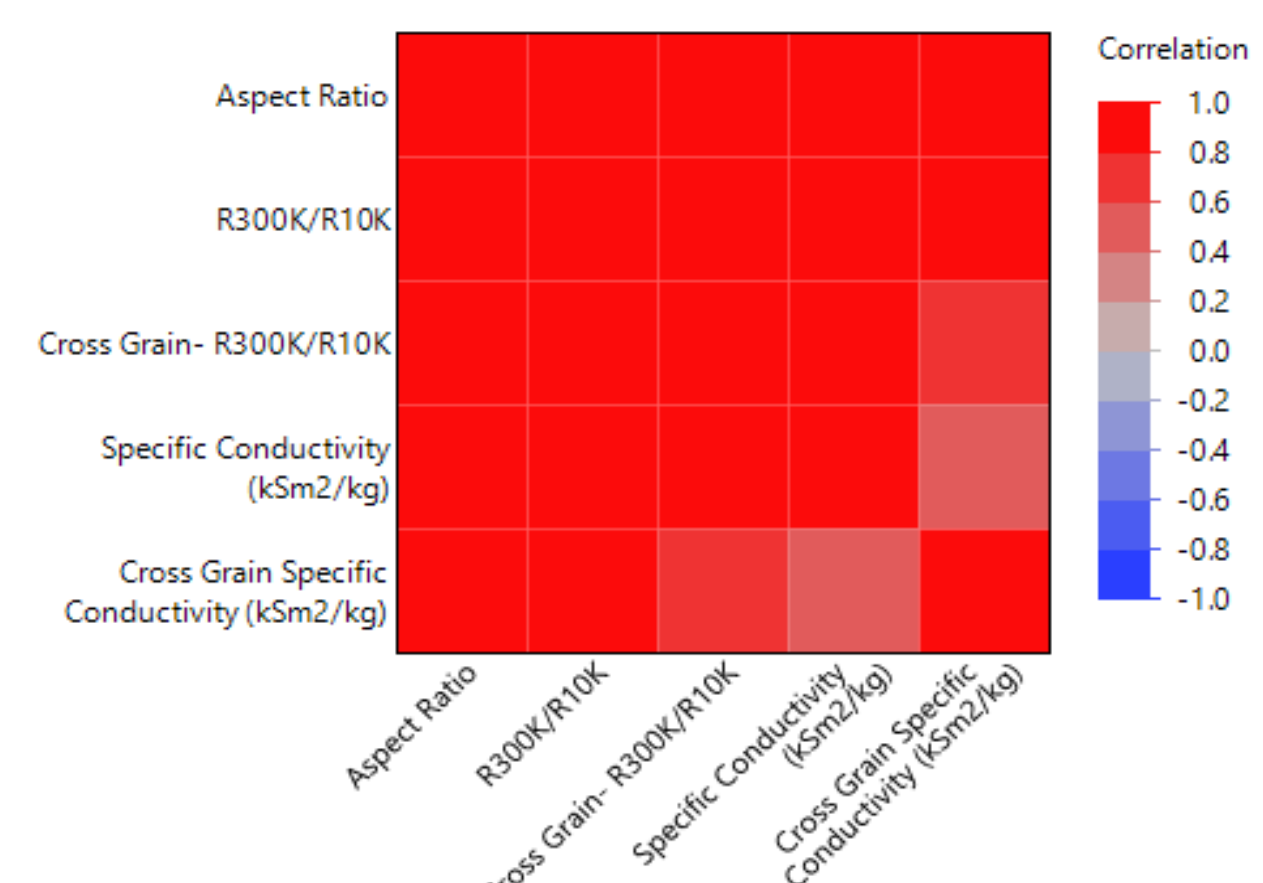

Correlations-- Restricted to **As-Is/Doped CNT Ribbons**, over 5 samples with different aspect ratios

**Supplemental Table 2.2-1. Correlation table for As-is/ doped CNT ribbons, for 5 CNT ribbons.**

| | **Aspect Ratio** | **Specific Conductivity (kSm2/kg)** | **X-Grain Specific Conductivity (kSm2/kg)** | **R300K/R10K** | **X-Grain R300K/R10K** |
|---|---|---|---|---|---|
| **Aspect Ratio** | 1 | 0.842214207 | **0.8873676** | **0.987067651** | **0.914230329** |
| **Specific Conductivity (kSm2/kg)** | 0.842214207 | 1 | 0.593728665 | 0.854928067 | 0.849008765 |
| **Cross Grain Specific Conductivity (kSm2/kg)** | **0.8873676** | 0.593728665 | 1 | 0.807704251 | 0.693614822 |
| **R300K/R10K** | **0.987067651** | 0.854928067 | 0.807704251 | 1 | **0.937058685** |
| **Cross Grain- R300K/R10K** | **0.914230329** | 0.849008765 | 0.693614822 | **0.937058685** | 1 |

**Supplemental Table 2.2-2. Associated p-values for correlation table for As-is/ doped CNT ribbons.**

| | **Aspect Ratio** | **Specific Conductivity (kSm2/kg)** | **X-Grain Specific Conductivity (kSm2/kg)** | **R300K/R10K** | **X-Grain R300K/R10K** |
|---|---|---|---|---|---|
| **Aspect Ratio** | 0 | 0.073431 | 0.044602 | 0.001762 | 0.029762 |
| **Specific Conductivity (kSm2/kg)** | 0.073431 | 0 | 0.291164 | 0.064867 | 0.068813 |
| **Cross Grain Specific Conductivity (kSm2/kg)** | 0.044602 | 0.291164 | 0 | 0.098253 | 0.193952 |
| **R300K/R10K** | 0.001762 | 0.064867 | 0.098253 | 0 | 0.018776 |
| **Cross Grain- R300K/R10K** | 0.029762 | 0.068813 | 0.193952 | 0.018776 | 0 |

Correlations-- Restricted to **De-doped CNT Ribbons**, calculated over 5 samples with different aspect ratios

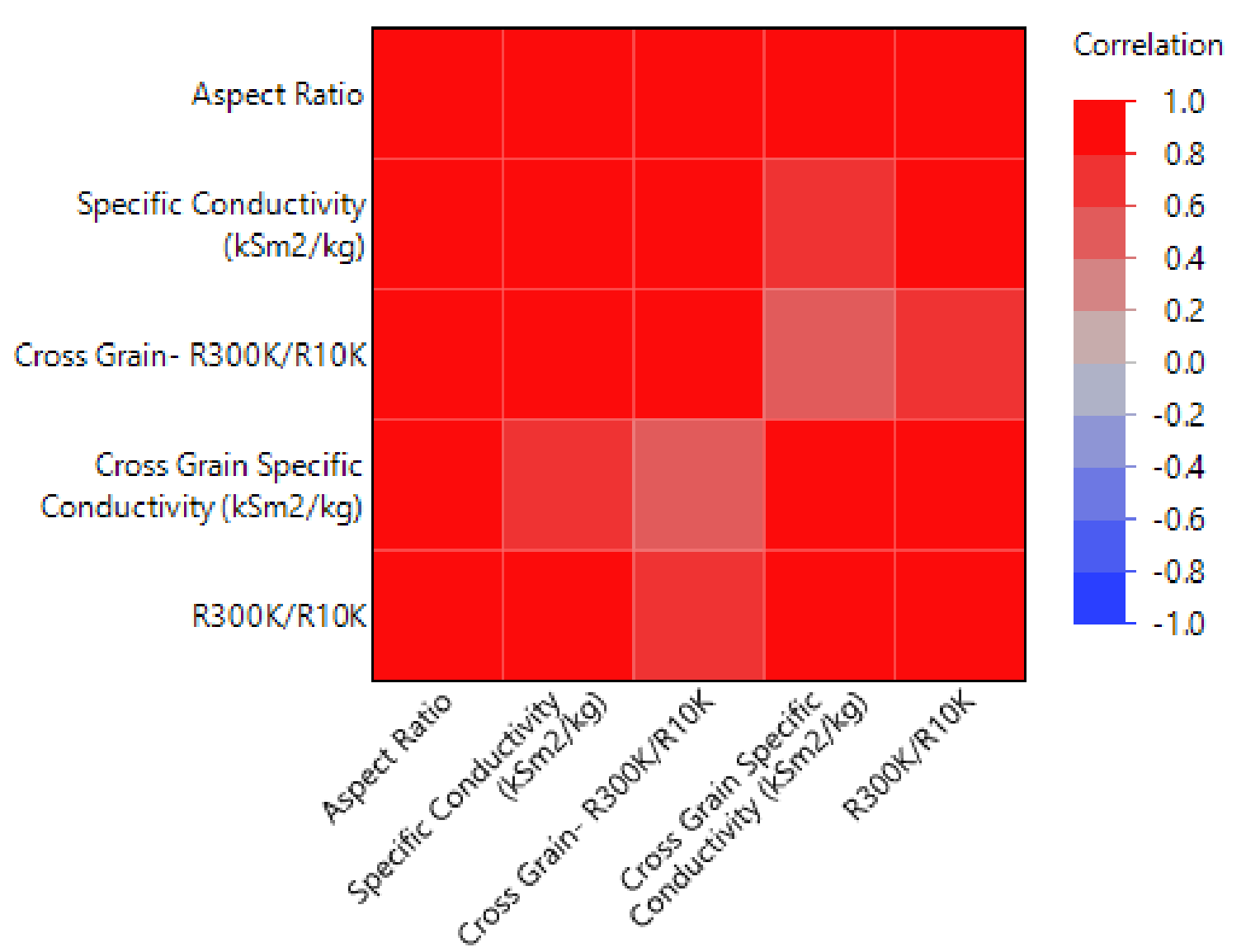

**Supplemental Table 2.2-3. Correlation table for de-doped CNT ribbons, for 5 CNT ribbons.**

| | **Aspect Ratio** | **Specific Conductivity (kSm2/kg)** | **X-Grain Specific Conductivity (kSm2/kg)** | **R300K/R10K** | **X-Grain R300K/R10K** |
|---|---|---|---|---|---|
| **Aspect Ratio** | 1 | **0.944496169** | 0.830389207 | **0.900463781** | 0.856722597 |
| **Specific Conductivity (kSm2/kg)** | **0.944496169** | 1 | 0.716097207 | 0.850846529 | 0.858969477 |
| **Cross Grain Specific Conductivity (kSm2/kg)** | 0.830389207 | 0.716097207 | 1 | **0.969135264** | 0.460597902 |
| **R300K/R10K** | **0.900463781** | 0.850846529 | **0.969135264** | 1 | 0.608310858 |
| **Cross Grain- R300K/R10K** | 0.856722597 | 0.858969477 | 0.460597902 | 0.608310858 | 1 |

**Supplemental Table 2.2-4. Associated p-values for correlation table for de-doped CNT ribbons.**

| | **Aspect Ratio** | **Specific Conductivity (kSm2/kg)** | **X-Grain Specific Conductivity (kSm2/kg)** | **R300K/R10K** | **X-Grain R300K/R10K** |
|---|---|---|---|---|---|
| **Aspect Ratio** | 0 | 0.015566 | 0.081685 | 0.037129 | 0.063685 |
| **Specific Conductivity (kSm2/kg)** | 0.015566 | 0 | 0.173648 | 0.06758 | 0.062215 |
| **Cross Grain Specific Conductivity (kSm2/kg)** | 0.081685 | 0.173648 | 0 | 0.006479 | 0.435 |
| **R300K/R10K** | 0.037129 | 0.06758 | 0.006479 | 0 | 0.276325 |
| **Cross Grain- R300K/R10K** | 0.063685 | 0.062215 | 0.435 | 0.276325 | 0 |

Correlations-- Restricted to **As-is/Doped CNT Ribbons & Fibers**, this is calculated over 11 samples with different aspect ratios

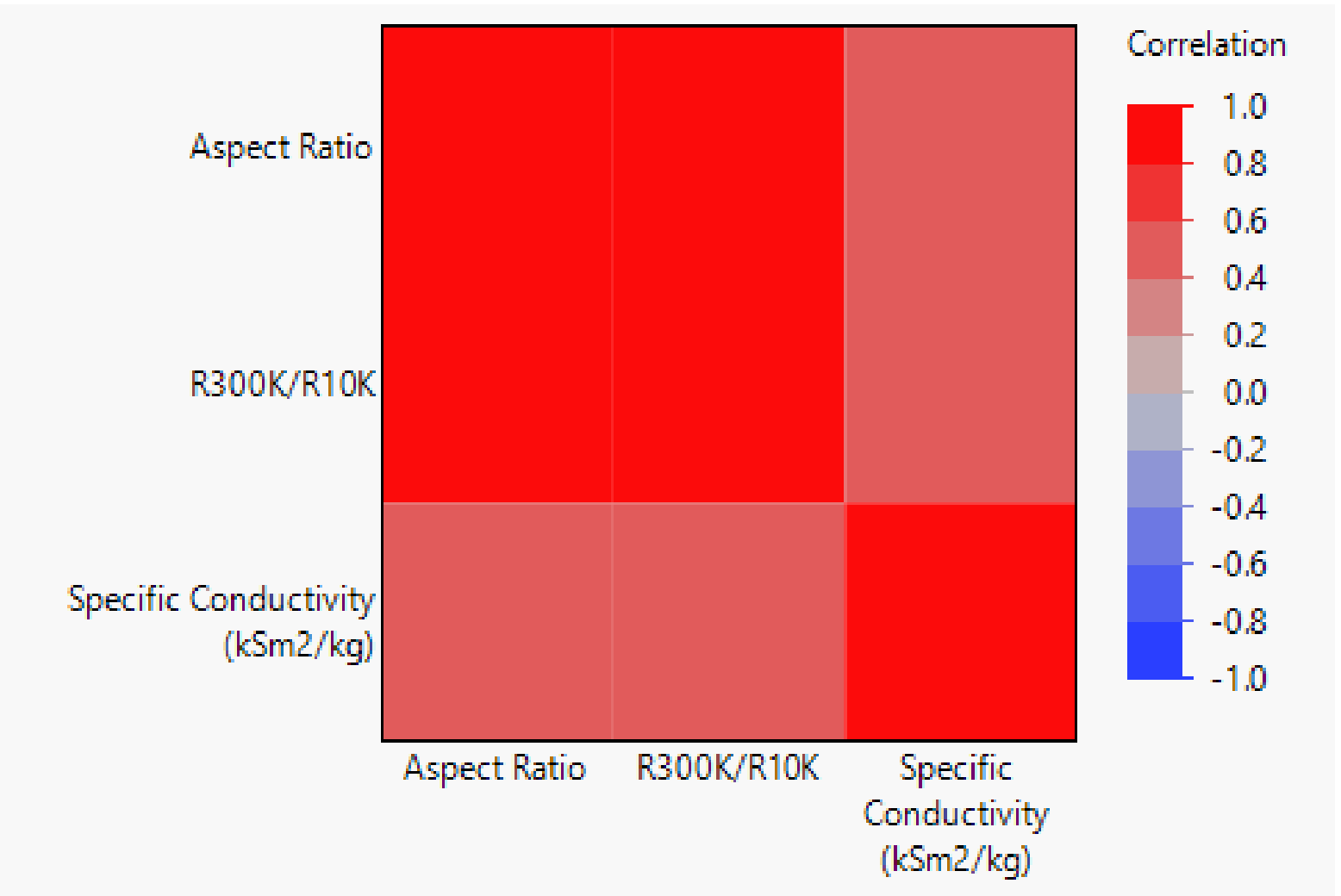

**Supplemental Table 2.2-5. Correlation table for As-is/ doped CNT fibers & ribbons, for 11 separate samples**

| | Aspect Ratio | Specific Conductivity (kSm2/kg) | R300K/R10K |
|---|---|---|---|
| **Aspect Ratio** | 1 | 0.55390085 | **0.938916538** |
| **Specific Conductivity (kSm2/kg)** | 0.55390085 | 1 | 0.522062775 |
| **R300K/R10K** | **0.938916538** | 0.522062775 | 1 |

**Supplemental Table 2.2-6. Associated p-values for correlation table for As-is/ doped CNT fibers & ribbons**

| | Aspect Ratio | Specific Conductivity (kSm2/kg) | R300K/R10K |
|---|---|---|---|
| **Aspect Ratio** | 0 | 0.077068 | 1.84E-05 |
| **Specific Conductivity (kSm2/kg)** | 0.077067788 | 0 | 0.081671 |
| **R300K/R10K** | 1.84352E-05 | 0.081671 | 0 |

Correlations-- Restricted to **De-doped CNT Ribbons & Fibers**, this is calculated over 6 samples with different aspect ratios

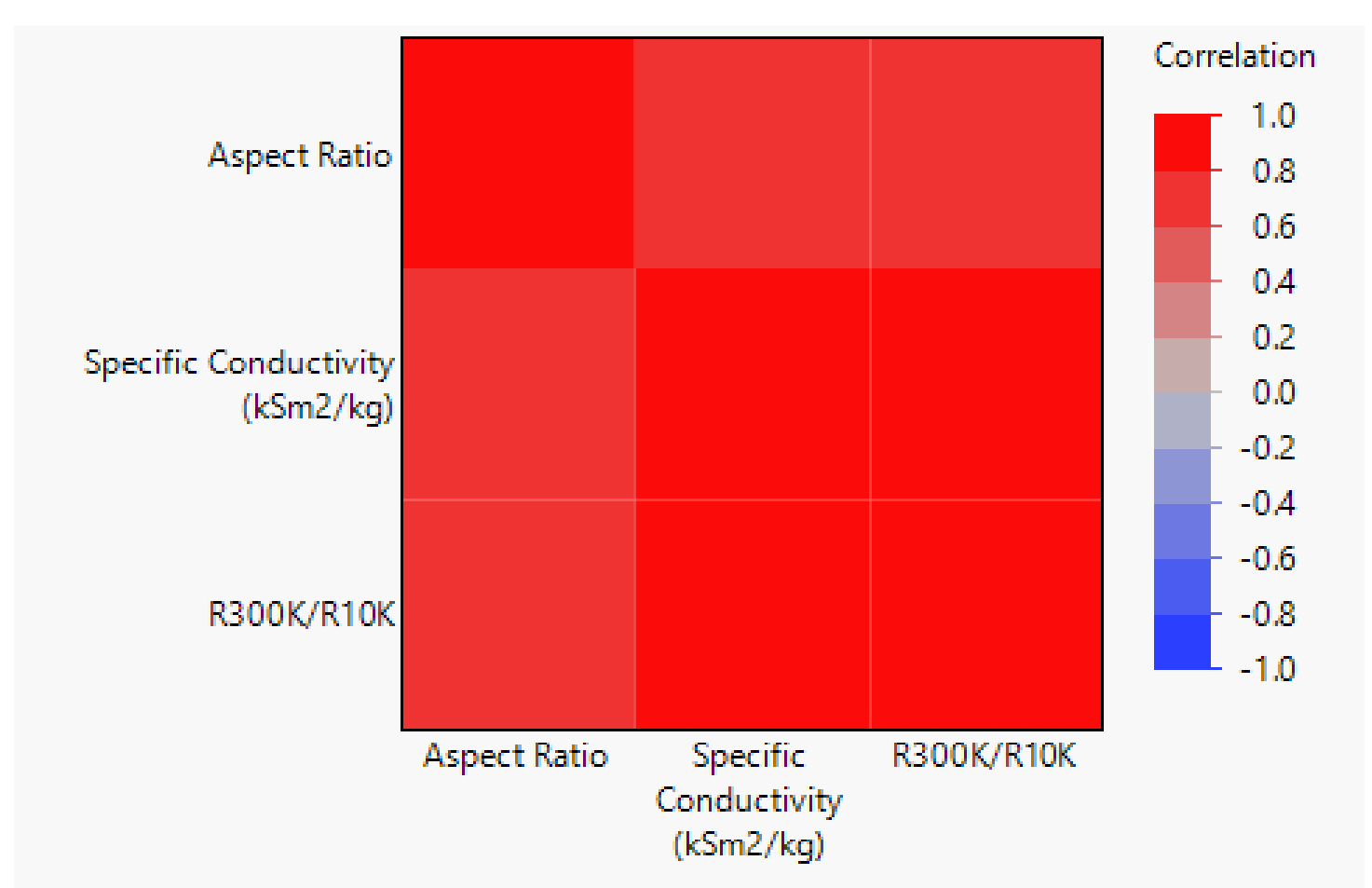

**Supplemental Table 2.2-7. Correlation table for de-doped CNT fibers & ribbons, for 6 separate samples**

| | **Aspect Ratio** | **Specific Conductivity (kSm2/kg)** | **R300K/R10K** |
|---|---|---|---|
| **Aspect Ratio** | 1 | 0.604687179 | 0.613278778 |
| **Specific Conductivity (kSm2/kg)** | 0.604687179 | 1 | **0.834153702** |
| **R300K/R10K** | 0.613278778 | **0.834153702** | 1 |

**Supplemental Table 2.2-8. Associated p-values for correlation table for de-doped CNT fibers & ribbons**

| | **Aspect Ratio** | **Specific Conductivity (kSm2/kg)** | **R300K/R10K** |
|---|---|---|---|
| **Aspect Ratio** | 0 | 0.20352 | 0.143052 |
| **Specific Conductivity (kSm2/kg)** | 0.203520133 | 0 | 0.038977 |
| **R300K/R10K** | 0.143051563 | 0.038977 | 0 |

Now we provide the fitted power laws for specific correlations.

**Supplemental Table 2.2-9. Power Law Analysis--- Log(Aspect Ratio) vs Log(R300K/R10K)**

| Category | Correlation on linear values | p-Value on Correlation | <u>Slope of Log/Log fit (power law exponent)</u> | error | Adj.RSq of Log/Log fit | Number of Points |
|---|---|---|---|---|---|---|
| As-is/Doped Ribbons & Fibers (measured \|\| CNT alignment) | 0.9322 | <0.0001 | 0.46182 | 0.07535 | 0.78523 | 11 |
| As-is/Doped Ribbons (measured XGrain alignment) | 0.9142 | 0.0298 | 0.25778 | 0.04419 | 0.89195 | 5 |
| De-doped Ribbons & Fibers (measured \|\| CNT alignment) | 0.6611 | 0.1059 | 0.35713 | 0.17251 | 0.35383 | 7 |
| De-doped Ribbons (measured XGrain alignment) | 0.8567 | 0.0673 | 0.39184 | 0.0808 | 0.84917 | 5 |

**Supplemental Table 2.2-10. Power Law Analysis--- Log(Specific Conductivity) vs Log(R300K/R10K)**

| | **Correlation on linear values** | **p-Value on Correlation** | **<u>Slope of Log/Log fit (power law exponent)</u>** | **error** | **Adj.RSq of Log/Log fit** | **Number of Points** |
|---|---|---|---|---|---|---|
| As-is/Doped & De-doped Ribbons & Fibers (measured \|\| CNT alignment) | 0.8272 | <0.0001 | 1.24011 | 0.12418 | 0.85309 | 18 |
| As-is/Doped & De-doped Ribbons (measured XGrain alignment) | 0.8908 | 0.0005 | 1.07751 | 0.17983 | 0.795 | 10 |

**Supplemental Table 2.2-11. Power Law Analysis--- Log(Specific Conductivity) vs Log(Aspect Ratio)**

| Category | Correlation on linear values | p-Value on Correlation | <u>Slope of Log/Log fit (power law exponent)</u> | error | Adj.RSq of Log/Log fit | Number of Points |
|---|---|---|---|---|---|---|
| As-is/Doped Ribbons & Fibers (measured \|\| CNT alignment) | 0.4686 | 0.146 | 0.62682 | 0.2319 | 0.38673 | 11 |
| As-is/Doped Ribbons (measured XGrain alignment) | 0.8447 | 0.0717 | 0.51941 | 0.22166 | 0.52892 | 5 |
| De-doped Ribbons & Fibers (measured \|\| CNT alignment) | 0.6417 | 0.1696 | 0.45904 | 0.16558 | 0.57213 | 6 |
| De-doped Ribbons (measured XGrain alignment) | 0.8305 | 0.0816 | 0.39262 | 0.24676 | 0.27688 | 5 |

Now we provide box-whisker plots for the various CNT categories we consider.

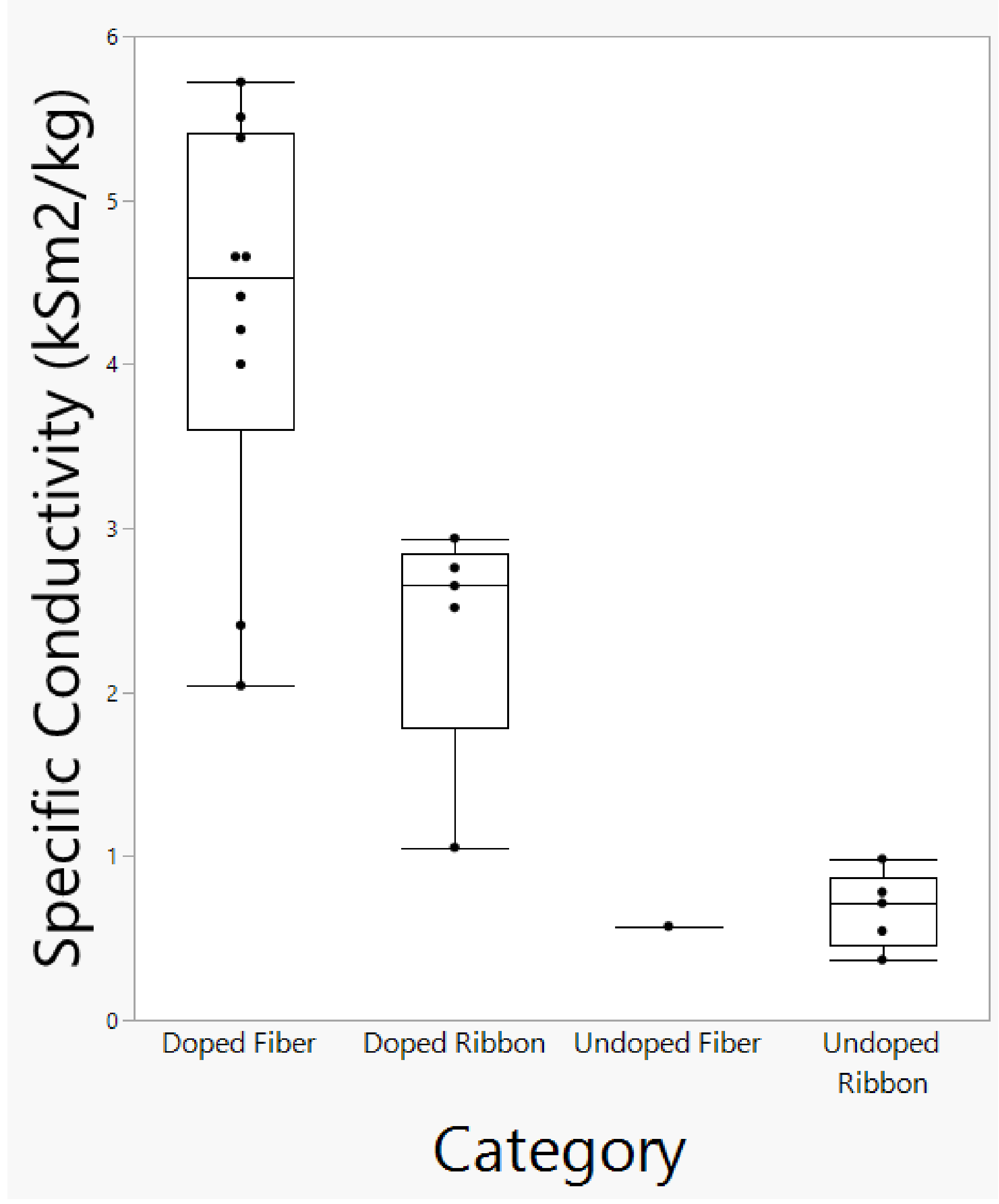

**Supplemental figure 2.2-1. Specific conductivity versus CNT fiber/ ribbons, as-is (doped)/ un-doped. The specific conductivity (kSm$^2$/kg) average and standard deviation are, respectively: 4.30 +/- 1.23; 2.38 +/- 0.76; 0.57; 0.67 +/- 0.23**

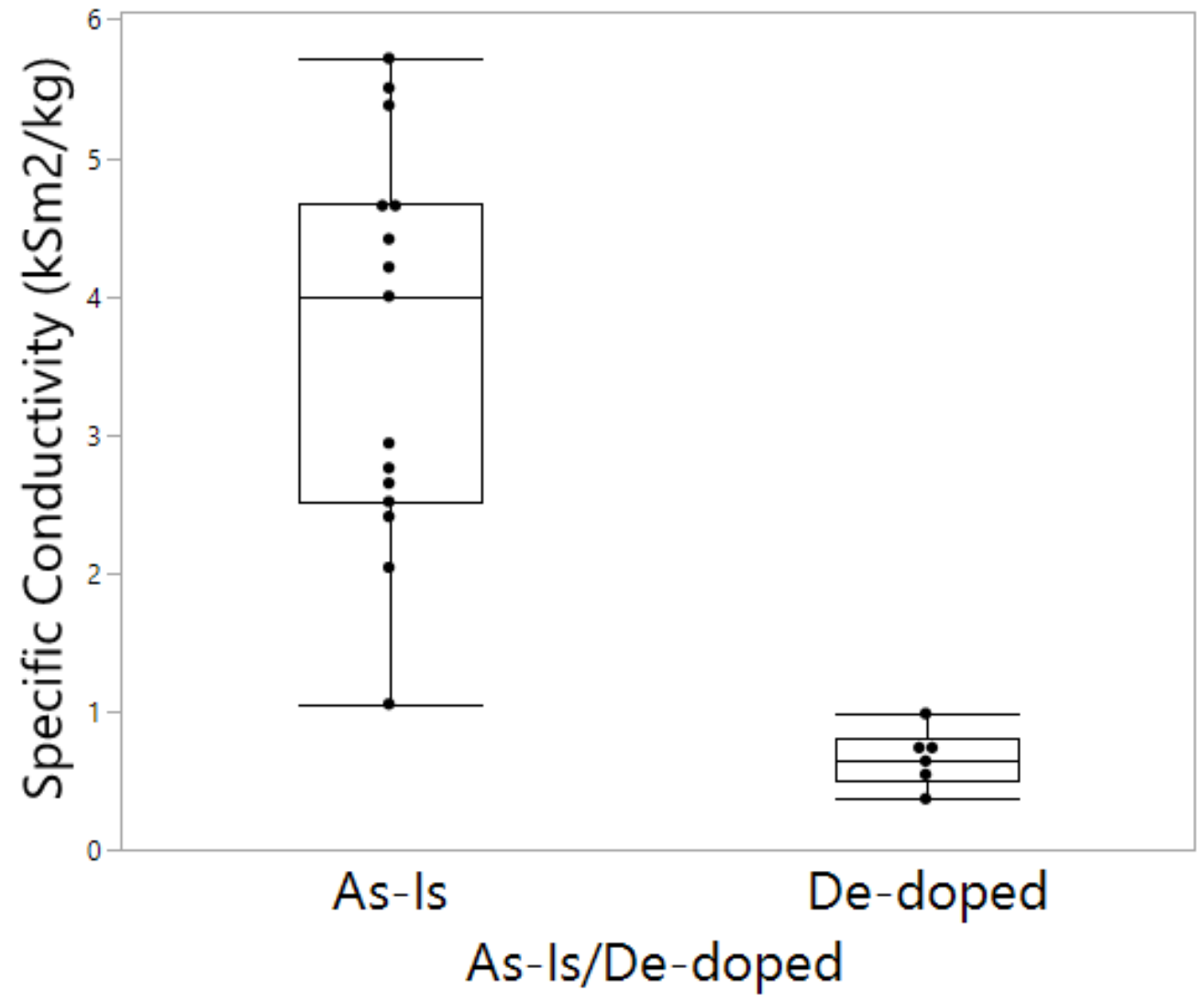

Supplemental figure 2.2-2. Specific conductivity versus as-is and de-doped CNT materials, without partitioning for morphology. Perpendicular microstructure alignment samples are not present. The specific conductivity (kSm$^2$/kg) average and standard deviation are, respectively: 3.66 +/- 1.42 and 0.65 +/- 0.21.

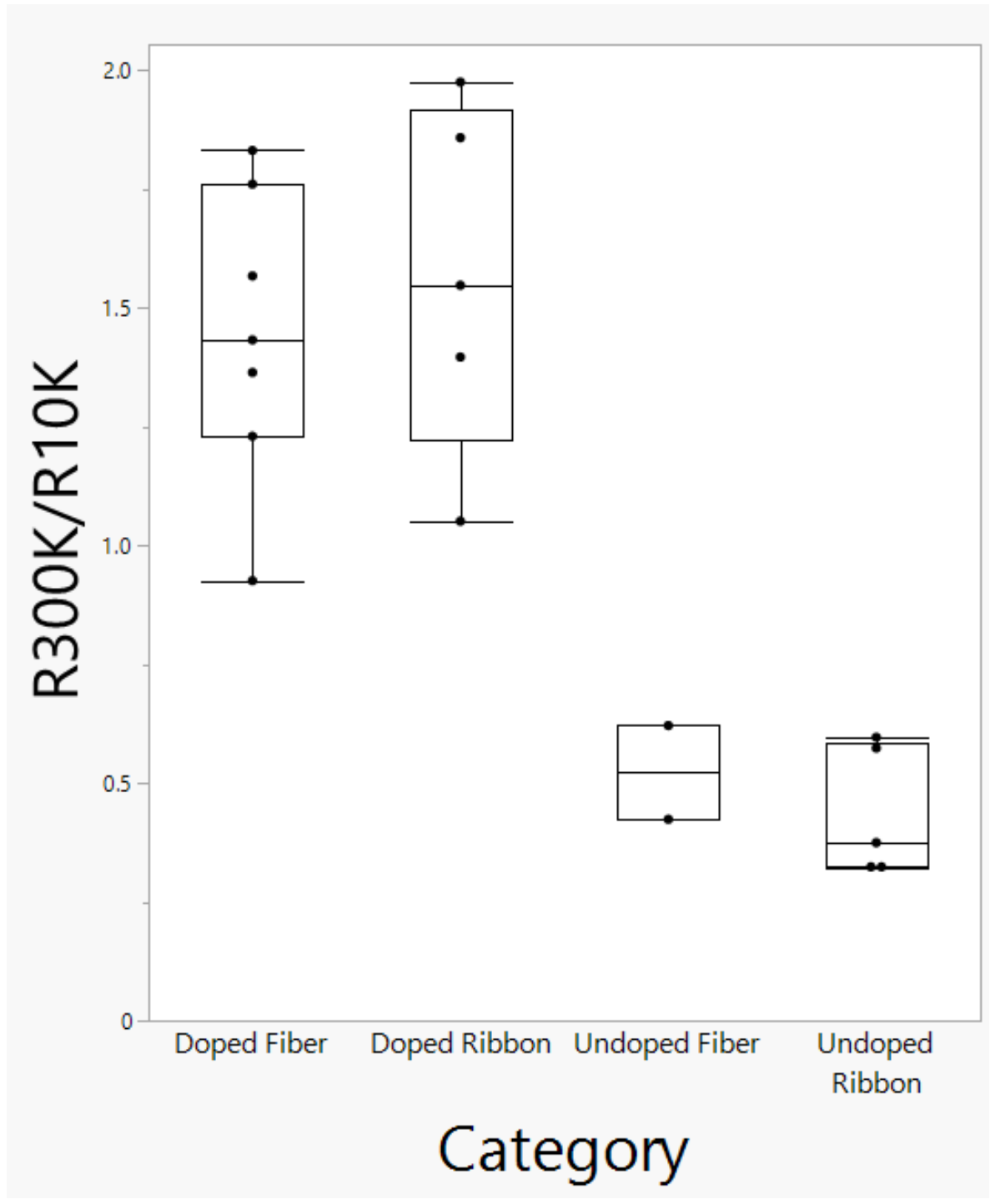

Supplemental figure 2.2-3. Cryogenic resistance ratio versus CNT fiber/ ribbons, as-is (doped)/ un-doped. Perpendicular microstructure alignment samples are not present. The average and standard deviation are, respectively: 1.44 +/- 0.31; 1.57 +/- 0.37; 0.52+/- 0.14; 0.44 +/- 0.14.

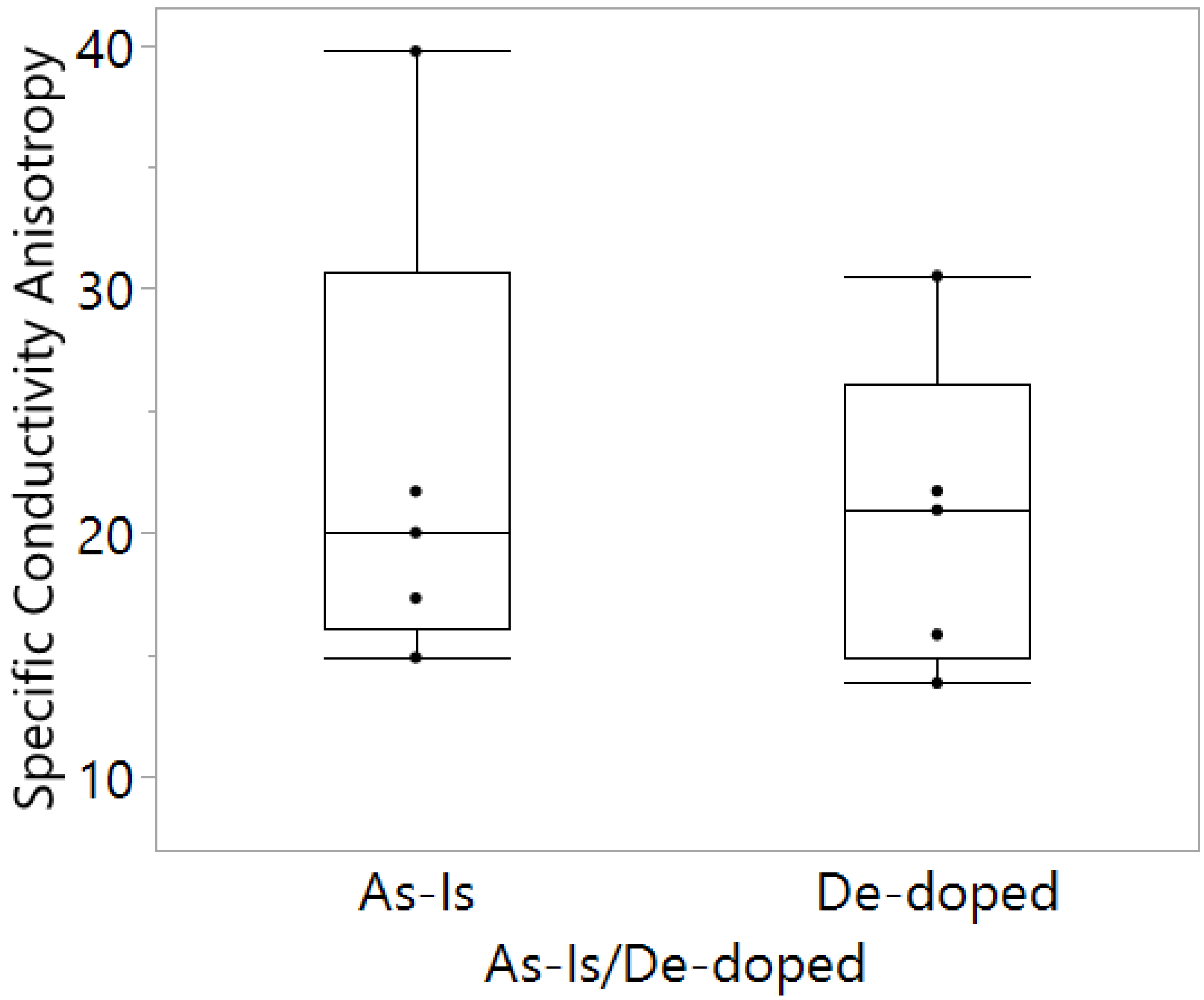

**Supplemental figure 2.2-4. The anisotropy in specific conductivity of the CNT ribbons, partitioned according to as-is or de-doped. The average and standard deviation are, respectively: 22.74 +/- 9.85; 20.57 +/- 6.48.**

**Supplemental section 2.3. Material correlations for fluctuation induced tunneling fit parameters**. Here, it is shown that aspect ratio is negatively correlated with both semi-conducting pre-factor $R_c$ and metallic exponent x; and positively correlated with prefactor $R_{metallic}$. There were no consistent relationships with junction transmission parameters ($T_1$ and $T_2$); further, parallel or perpendicular microstructure alignment did not affect fluctuation induced tunneling parameter values fits, although did scale the overall conductivity.

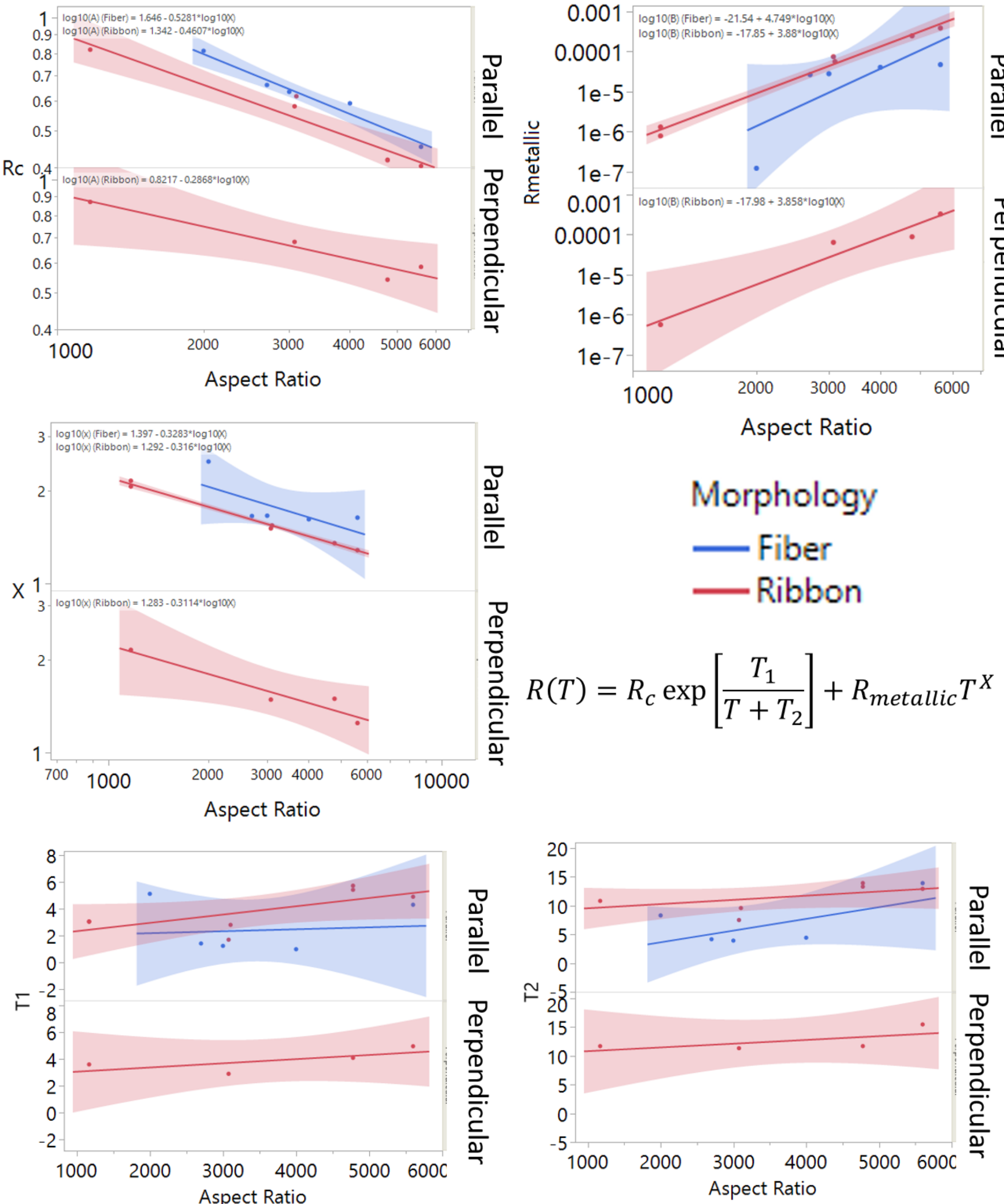

**Supplemental figure 2.3-1. Fits of the heterogenous conduction model with metallic term as a power law (see figure for equation), as a function of aspect ratio and partitioned according to morphology and microstructure alignment.**

**Supplemental section 2.4. Correlations with Metallic-like fraction.** The following are selected correlations and power laws associated with the metallic fraction present in the CNT materials, calculated by subtracting from the resistance versus temperature trace the fitted Arrhenius thermal activation.

**Supplemental Table 2.4-1. Correlation and power-law exponent for the metallic fraction (%) at 300 K vs aspect ratio.**

| | Correlation Strength | p-Value on Correlation | Power Law Exponent | Power Law Error |
|---|---|---|---|---|
| CNT fibers & Ribbons (as-is, \|\| CNT alignment) | 0.94 | <.0001 | 0.85 | 0.11 |
| CNT fibers & Ribbons (as-is, ⊥ CNT alignment) | 0.89 | 0.1062 | 0.84 | 0.15 |
| CNT ribbons (de-doped, \|\| CNT alignment) | 0.9304 | 0.0696 | 0.43 | 0.13 |
| CNT ribbons (de-doped, ⊥ CNT alignment) | 0.8 | 0.2 | 0.51 | 0.14 |

**Supplemental Table 2.4-2. Correlation and power-law exponent for the metallic fraction of specific conductivity (at 300 K) vs aspect ratio, showing no correlation.**

| | Correlation Strength | p-Value on Correlation |
|---|---|---|
| CNT fibers & Ribbons (as-is, \|\| CNT alignment) | -0.38 | 0.24 |
| CNT fibers & Ribbons (as-is, ⊥ CNT alignment) | -0.4 | 0.59 |
| CNT ribbons (de-doped, \|\| CNT alignment) | 0.26 | 0.74 |
| CNT ribbons (de-doped, ⊥ CNT alignment) | -0.13 | 0.87 |

**Supplemental section 3.0-- Pulsed High Magnetic Field at the National High Magnetic Field Laboratory (NHMFL) at Los Alamos National Laboratory (LANL), Los Alamos, NM.**

**Supplemental Section 3.1 Measurement setup**. Below is a sketch of the AC measurement circuit/ lock-in amplifier technique that measured the four-wire resistance of the sample in the noisy pulsed magnetic field environment.

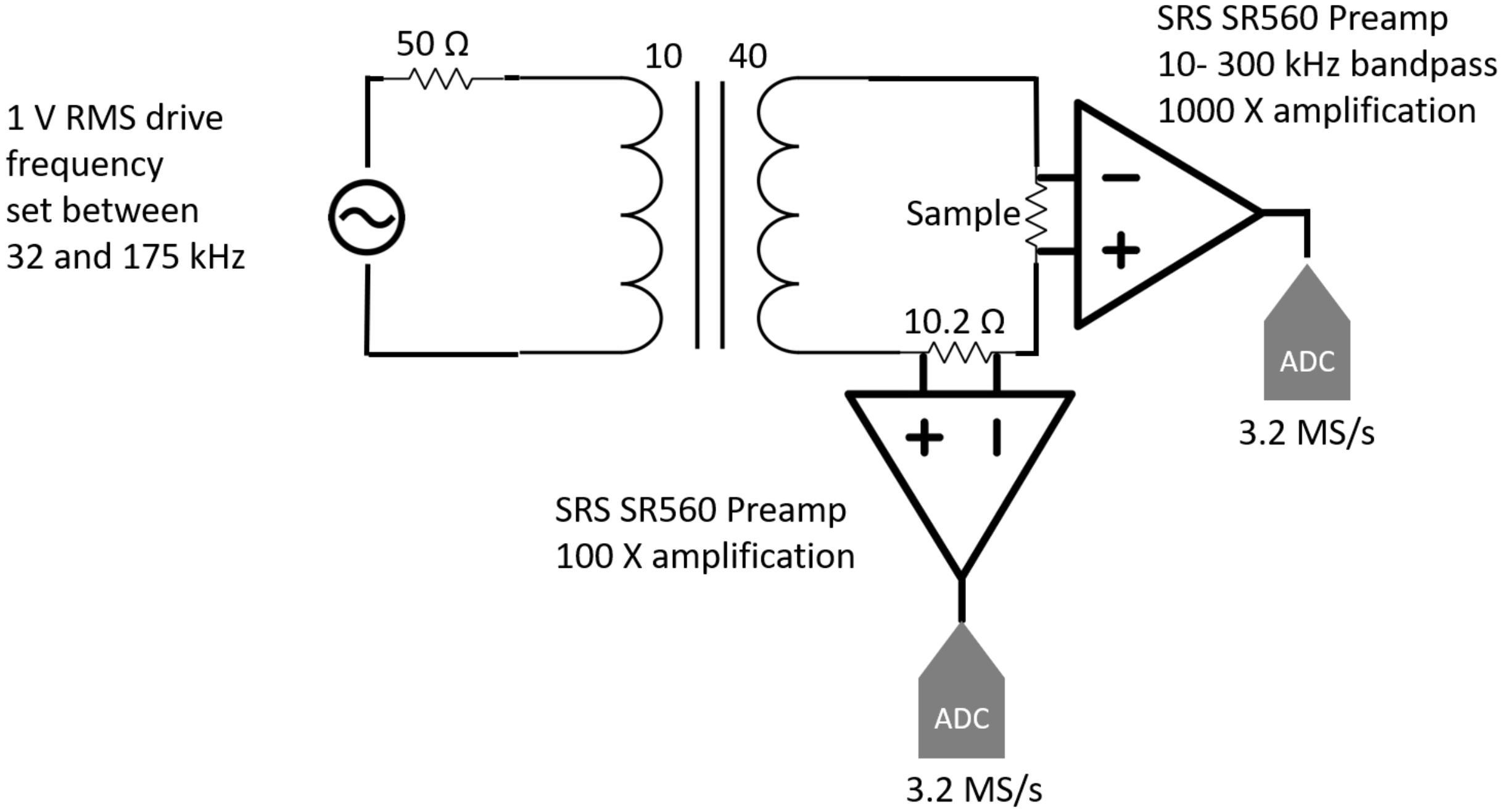

**Supplemental figure 3.1-1. AC four-wire measurement circuit showing an isolation transformer, independent current measurement across the 10.2 Ω resistor, and preamps with bandpass filters for both the independent current measurement and sample voltage drop.**

**Supplemental section 3.2 High Field Results**. Below are the complete set of pulsed magnetoresistance (MR) measurements for different CNT ribbons (as-is vs de-doped, AR: 4800 vs AR: 1200). In most cases, as indicated, the magnetic field was normal to the CNT ribbon surface and perpendicular to the probe current (transverse MR). As indicated in the last case, magnetic field was oriented parallel to both the probe current and CNT microstructure alignment (longitudinal MR). Two CNT ribbons from the same category were measured simultaneously for redundancy (with duplicated measurement circuits) and these are indicated by the red and black traces that overlap nicely.

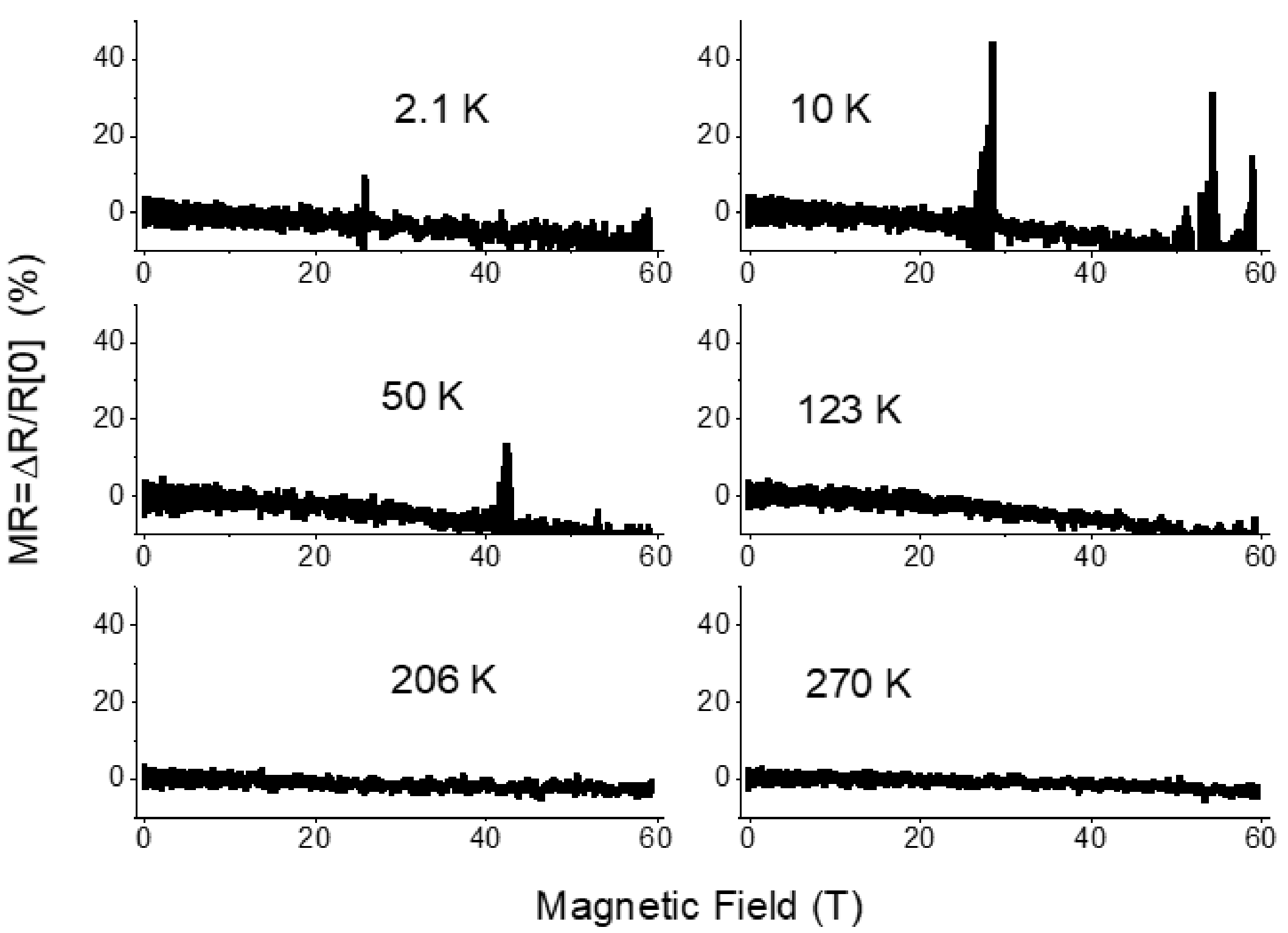

**Supplemental Figure 3.2-1. Pulsed transverse MR of as-is CNT ribbon with AR: 4800 at six temperature setpoints. Connection failure on the other sample prevented measurement and is not shown here.**

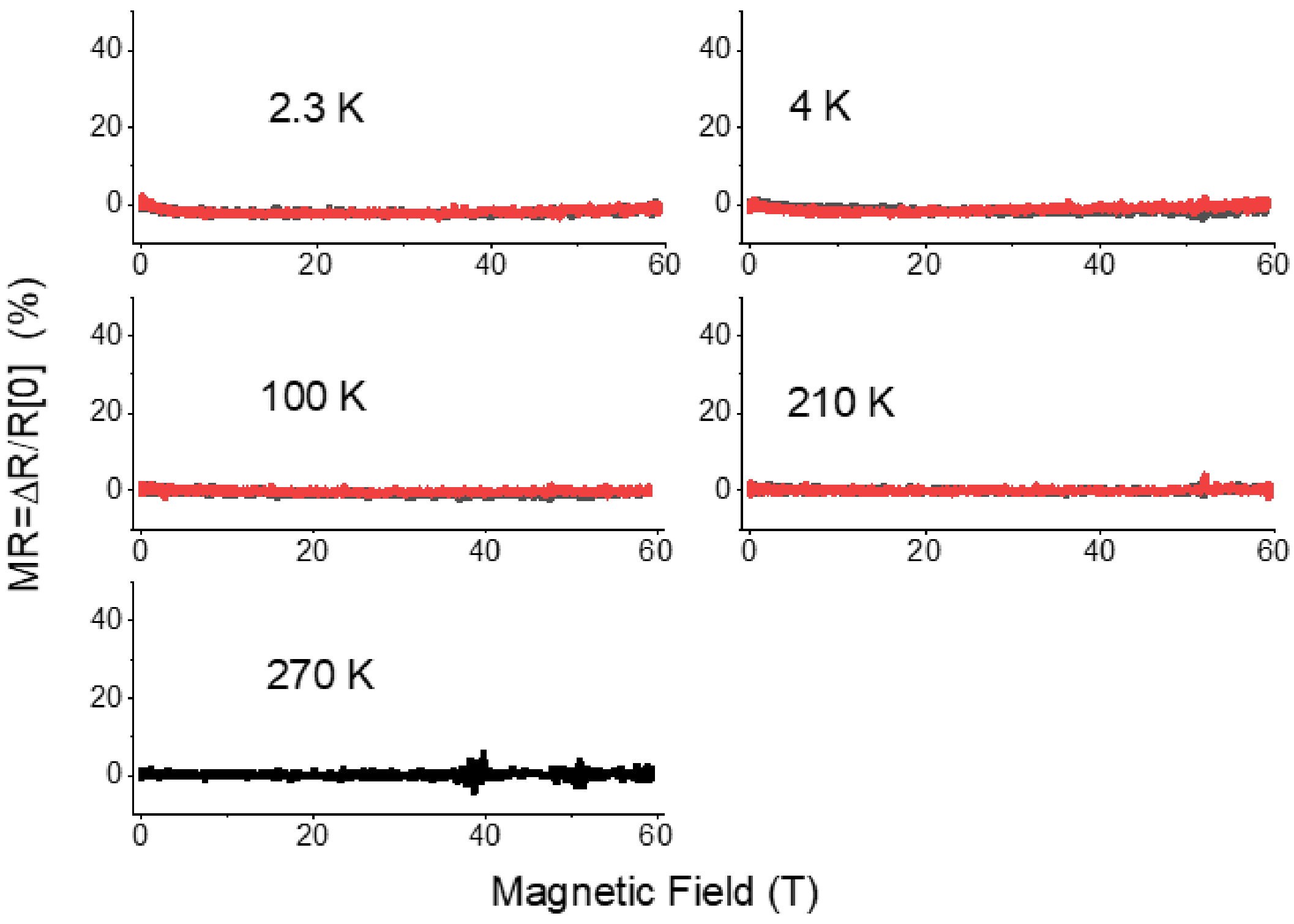

**Supplemental Figure 3.2-2. Pulsed transverse MR of as-is CNT ribbon with AR: 1200 at five temperature setpoints.**

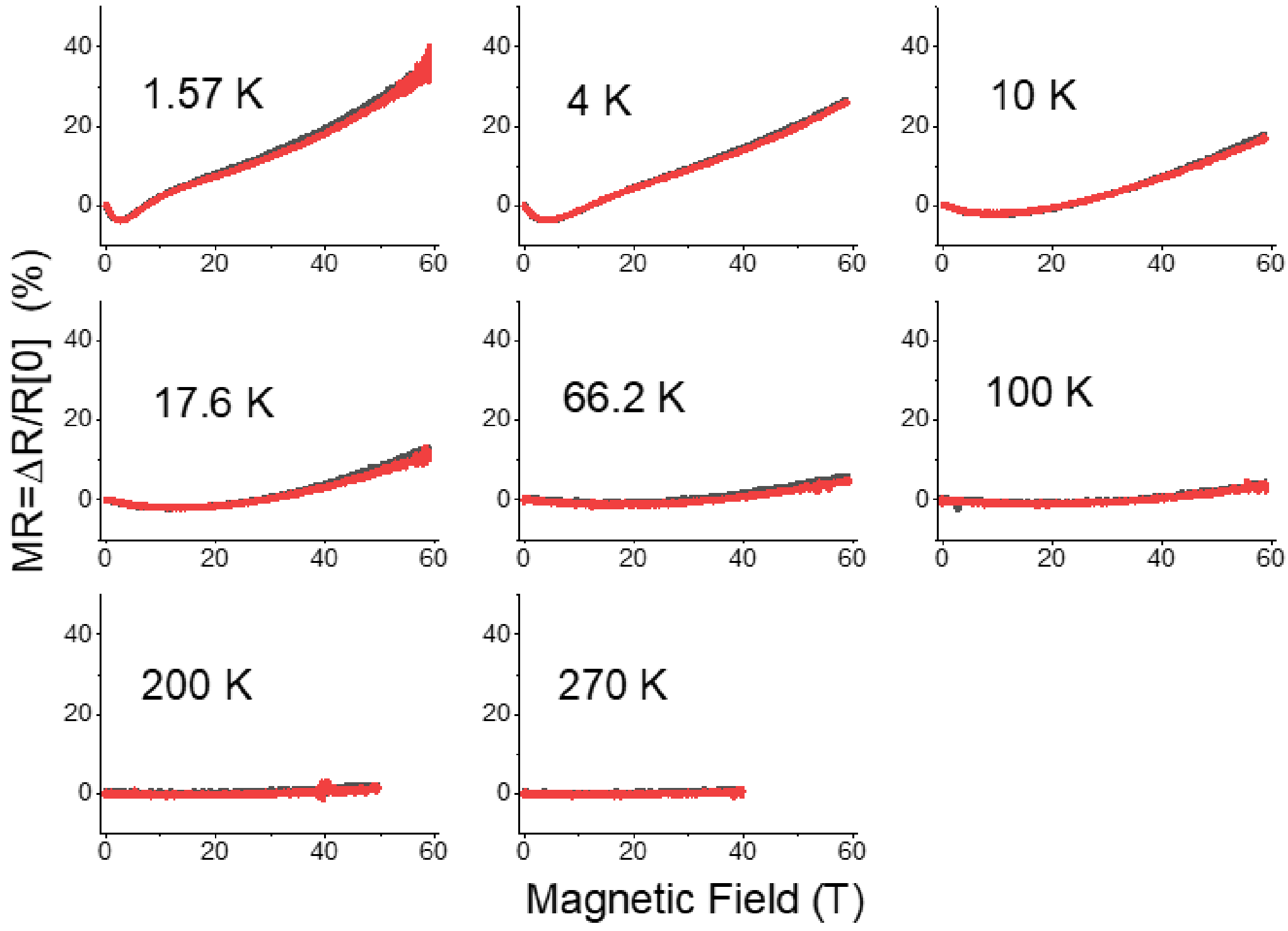

**Supplemental Figure 3.2-3. Pulsed transverse MR of de-doped CNT ribbon with AR: 4800 at eight temperature setpoints.**

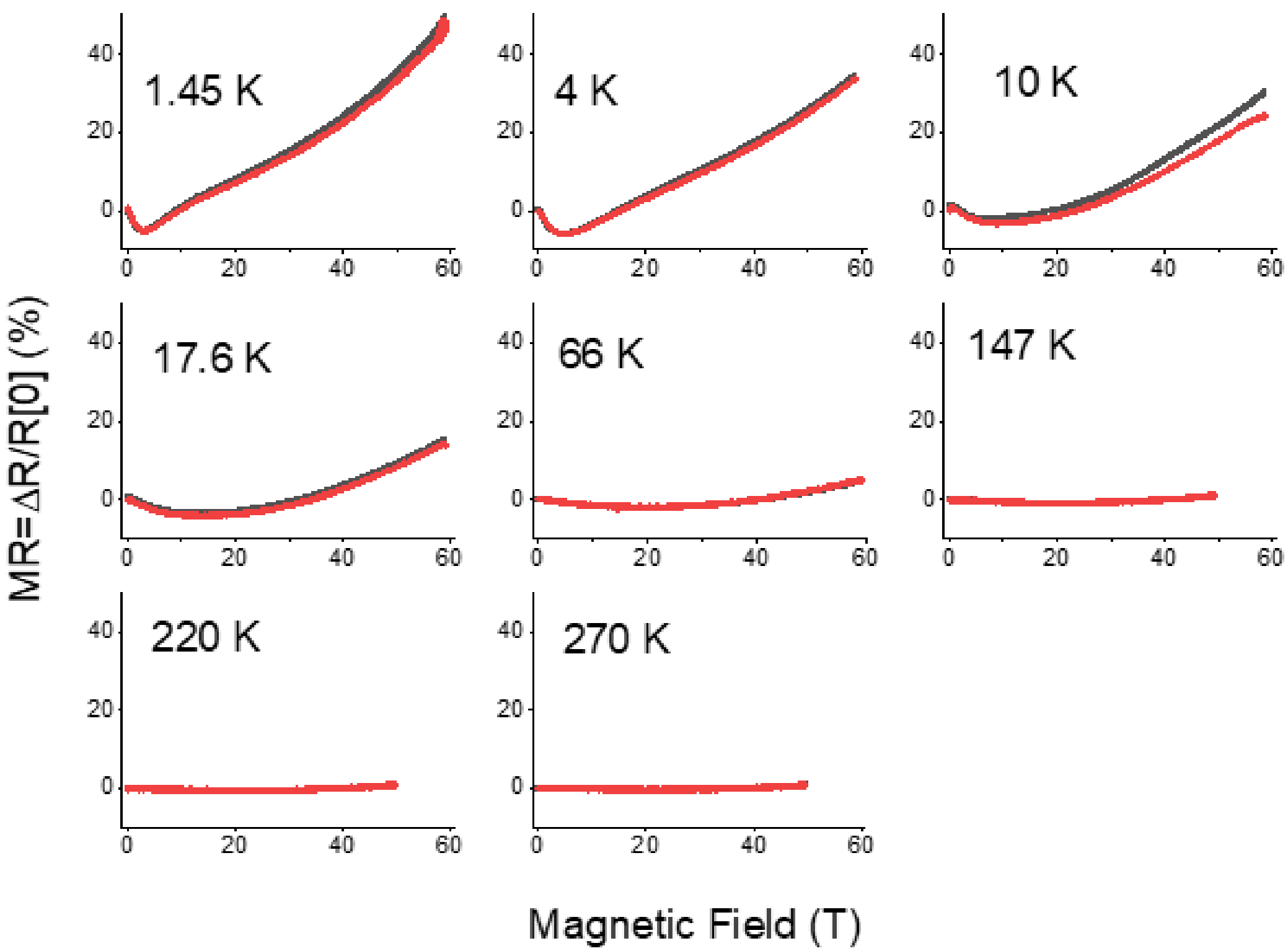

**Supplemental Figure 3.2-4. Pulsed transverse MR of de-doped CNT ribbon with AR: 1200 at eight different temperature setpoints.**

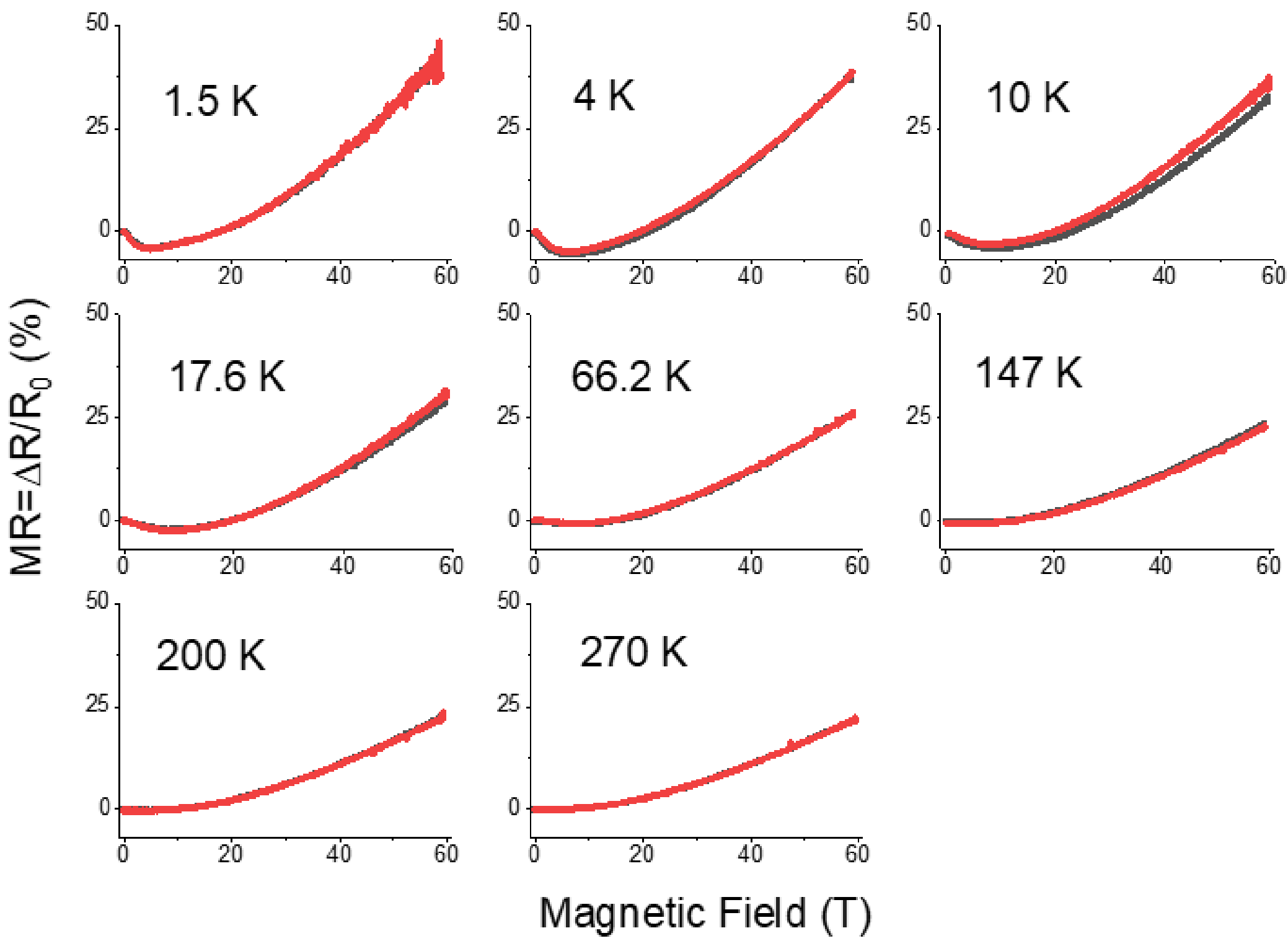

**Supplemental Figure 3.2-5. Pulsed longitudinal (magnetic field is parallel to probe current and CNT microstructure) MR of de-doped CNT ribbon with AR: 1200 at eight different temperature setpoints. Note that due to time constraints these $H_2$ annealed CNT ribbons were not vacuum baked in the measurement setup and therefore are not completely de-doped like the other materials.**

**Supplemental section 4.0-- Low-Magnetic Field MR**

**4.1 Transverse MR.** Magnetoresistance MR(%) was also measured in our Quantum Design Physical Properties Measurement System (PPMS) up to 9 T. First, we explore how transverse MR changes with applied magnetic field in a few specific examples, and then over the whole experimental space using contour maps. Next, we fit the low-magnetic field data to weak localization models (3D, 2D, and 1D) and show their resulting phase coherence length versus temperature profiles. Finally, we measure MR in CNT fiber samples that rotate in a steady DC magnetic field, for a variety of set temperatures and DC magnetic fields.

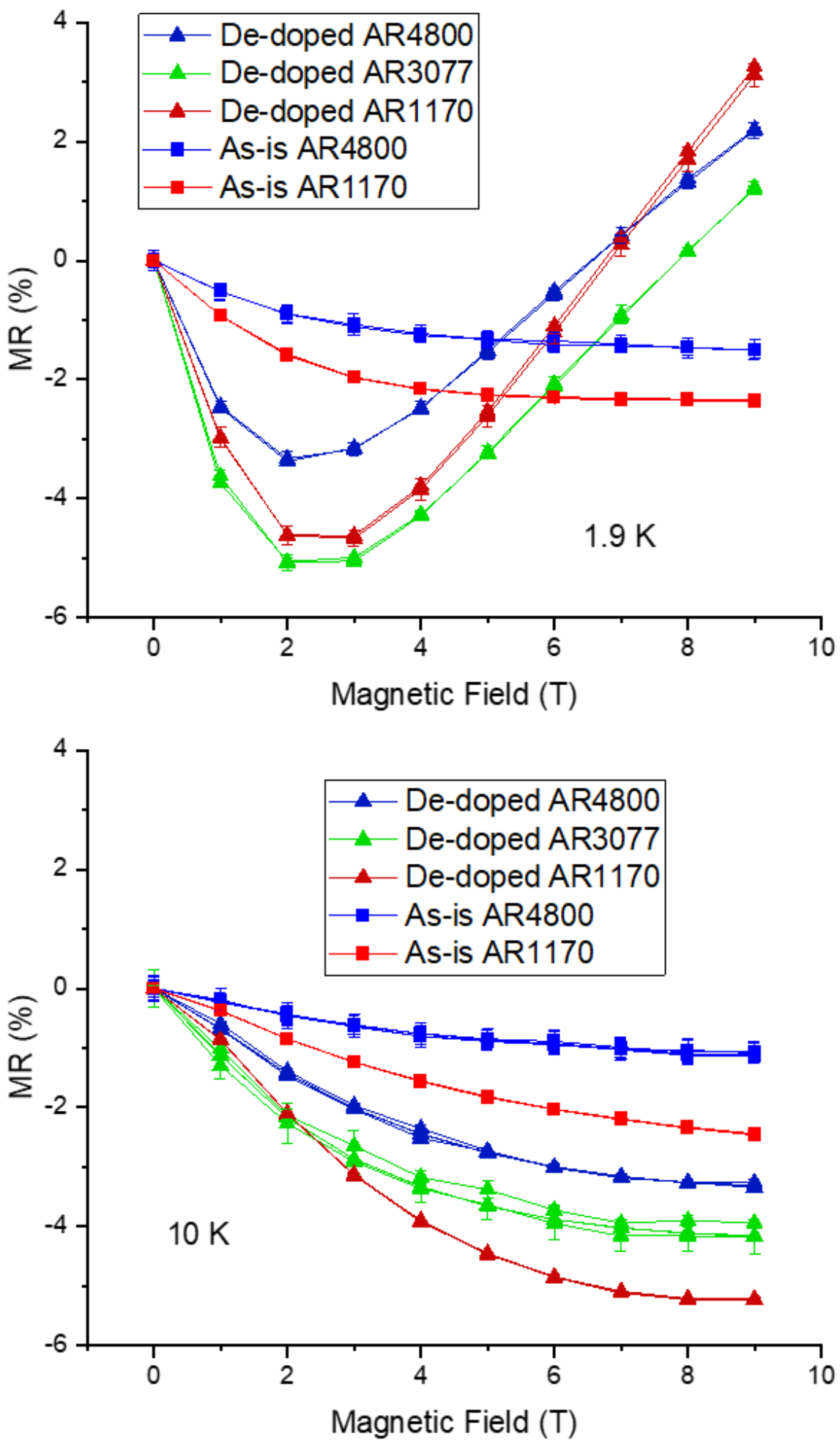

**Supplemental figure 4.1-1. Selected examples of how (*MR*(%) = 100% (*R*(*H*)-*R*(0))/*R*(0)) changes with applied magnetic field *H* at both 1.9 K (Top) and 10 K (Bottom), for various aspect ratios (AR) and doping status. This is the case for field perpendicular to probe current. We see mostly negative MR except a +MR upturn for de-doped CNTs at 1.9 K.**

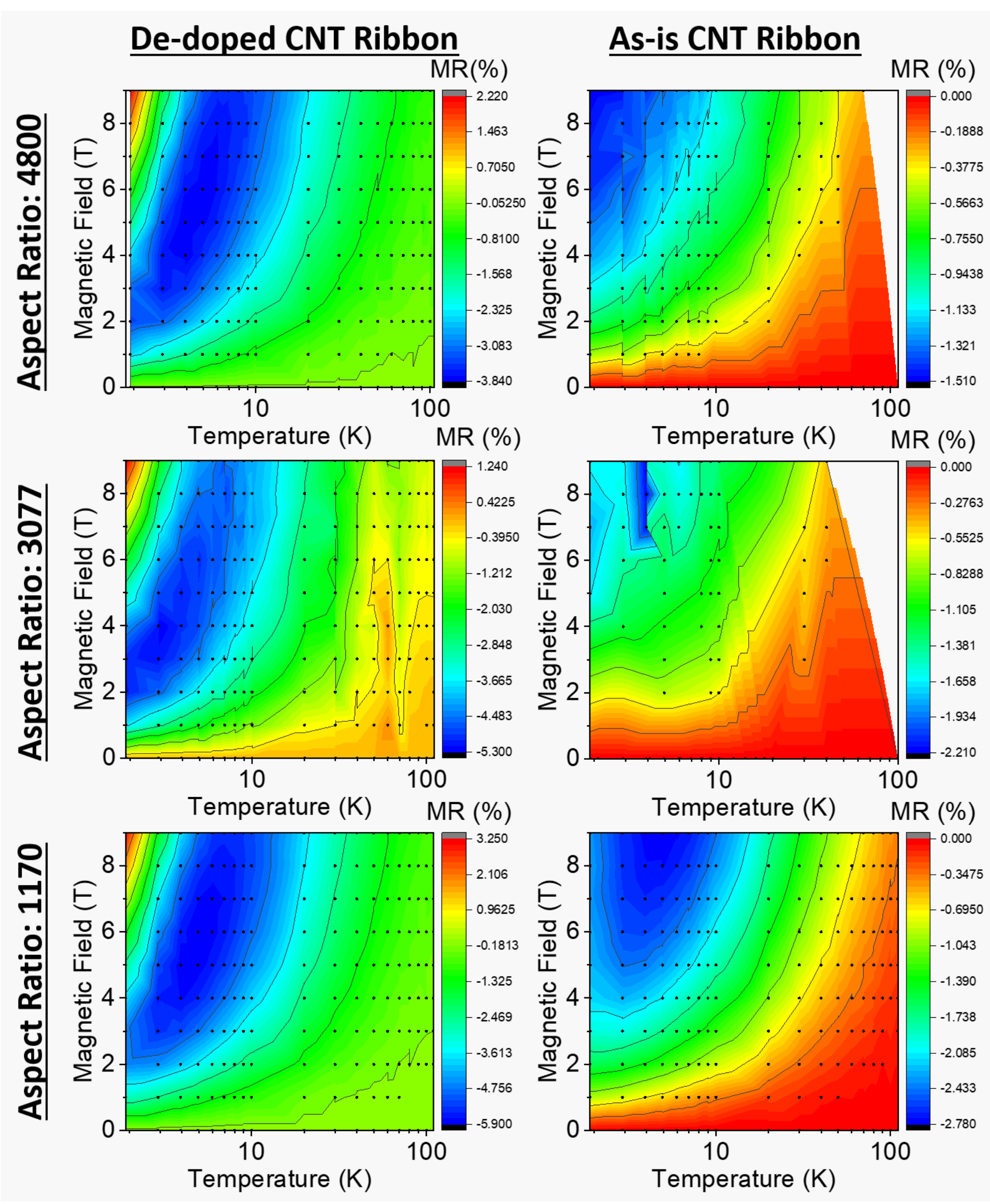

**Supplemental figure 4.1-2. Contour maps showing transverse MR(%) as a function of magnetic field (T) and temperature (K), for as-is vs de-doped, and for different aspect ratios (AR).**

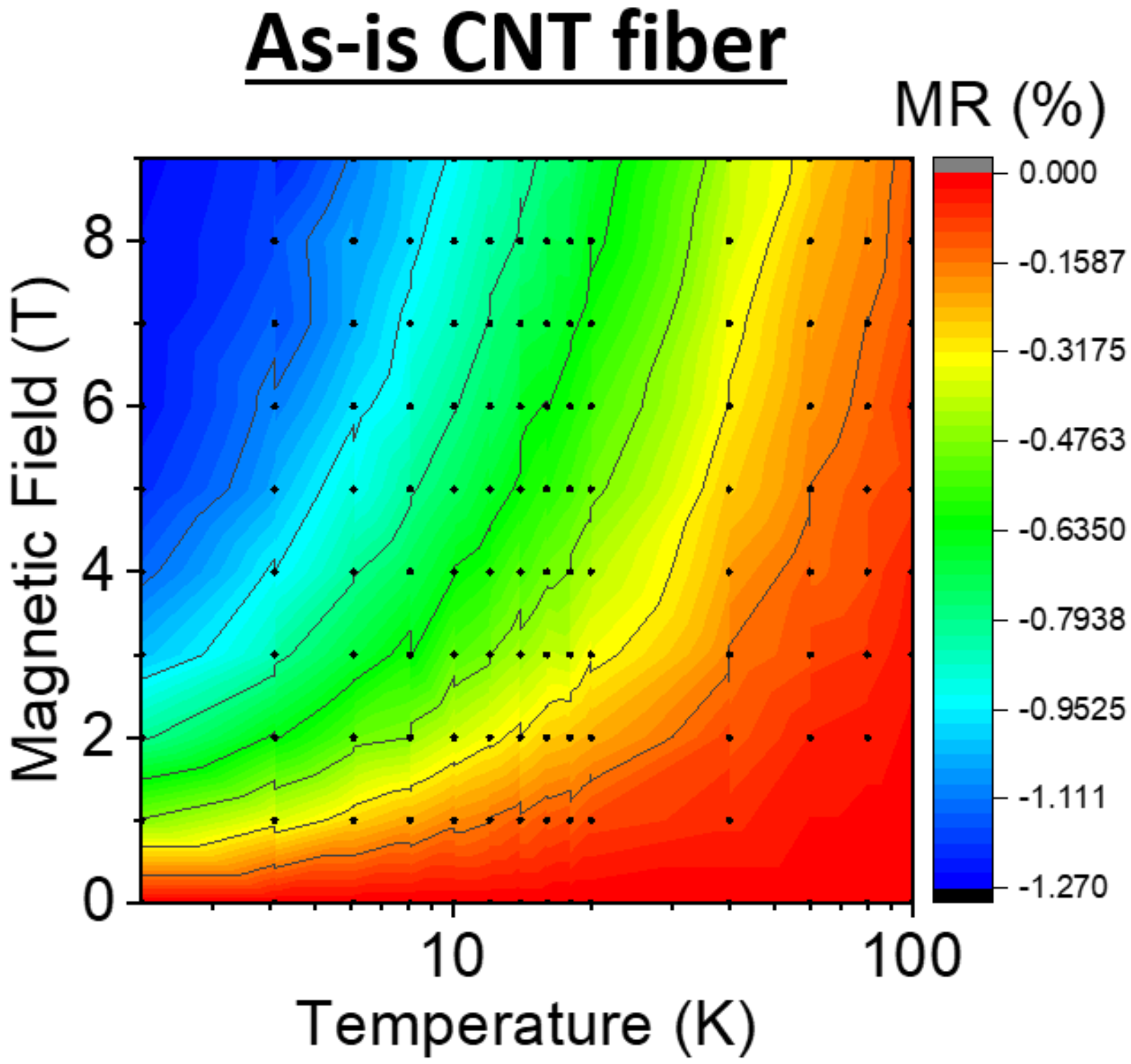

Supplemental figure 4.1-3. Transverse MR contour map for an as-is CNT fiber.

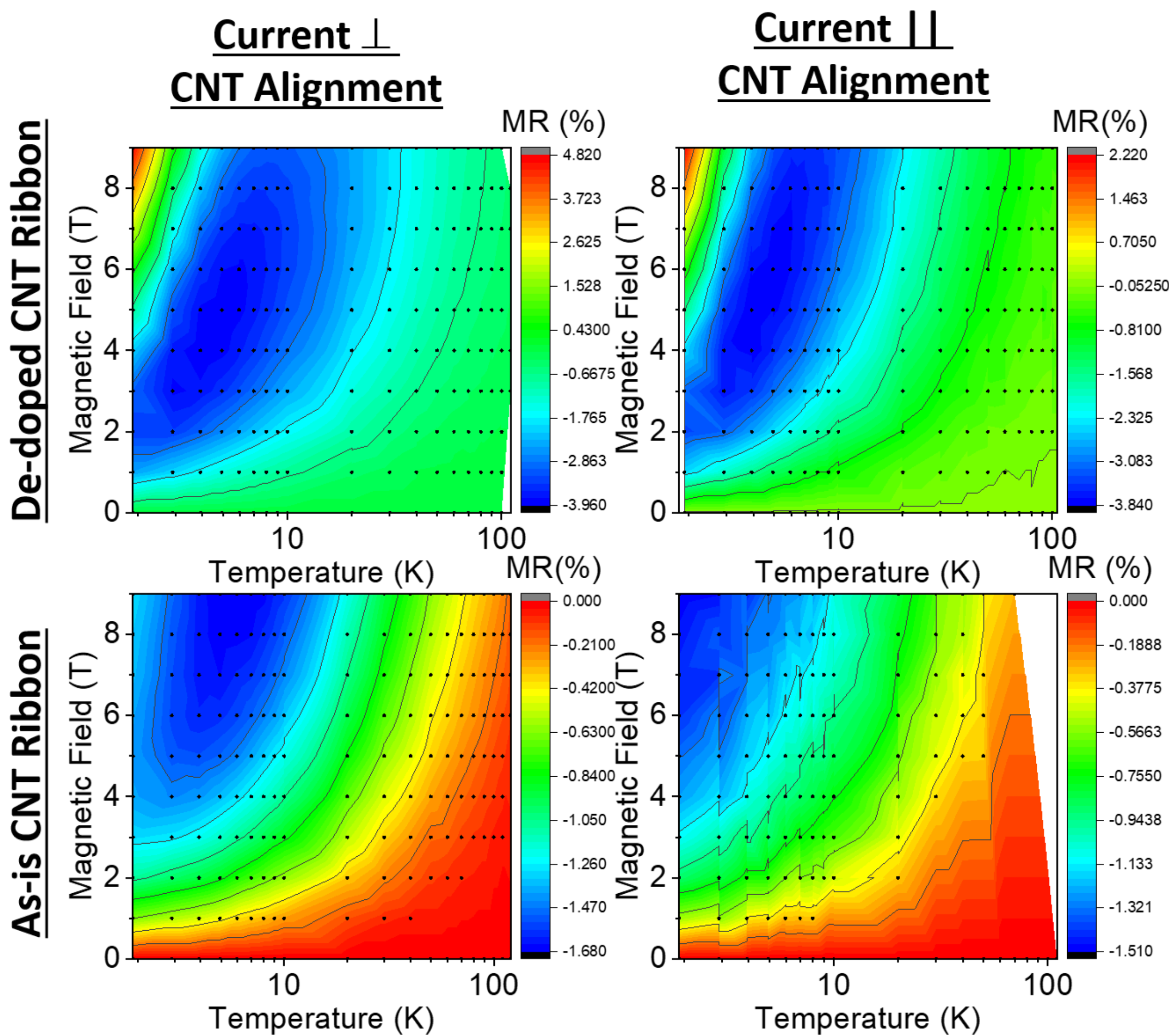

**Supplemental figure 4.1-4. Transverse MR contour map for an as-is and de-doped CNT ribbon, where probe current is perpendicular to probe current compared to the typical case of probe current parallel to microstructure alignment. This is for aspect ratio 4800.**

**Supplemental section 4.2 Weak Localization Fitting Low-magnetic Field MR**. Below are the results of the fits of negative MR to 1D, 2D, and 3D weak localization models where magnetic field is applied perpendicular to the probe current/ microstructure alignment. From the high magnetic field experimentation on as-is CNT conductors, we know that negative MR is the only contributing signal. For de-doped, we know there is a positive MR that appears at the coldest temperatures that will confound the weak localization analysis; for this reason MR below 10 K was excluded from weak localization analysis for the de-doped. When only one de-phasing mechanism is present, we should see power law behavior between the phase coherence length and temperature. In 3D and 2D weak localization, we will see that the phase coherence length saturates below 10 K; similar deviations from the power law is observed my multiple authors[2,3,4]. Nonlinear fitting was accomplished in OriginPro with tabular fitting results and analysis details provided in the supplemental database. Note that rather than MR, weak localization models use the change in conductance $\Delta C = C(H)-C(0) = R(H)^{-1}-R(0)^{-1}$ for resistance $R$ and magnetic field $H$.

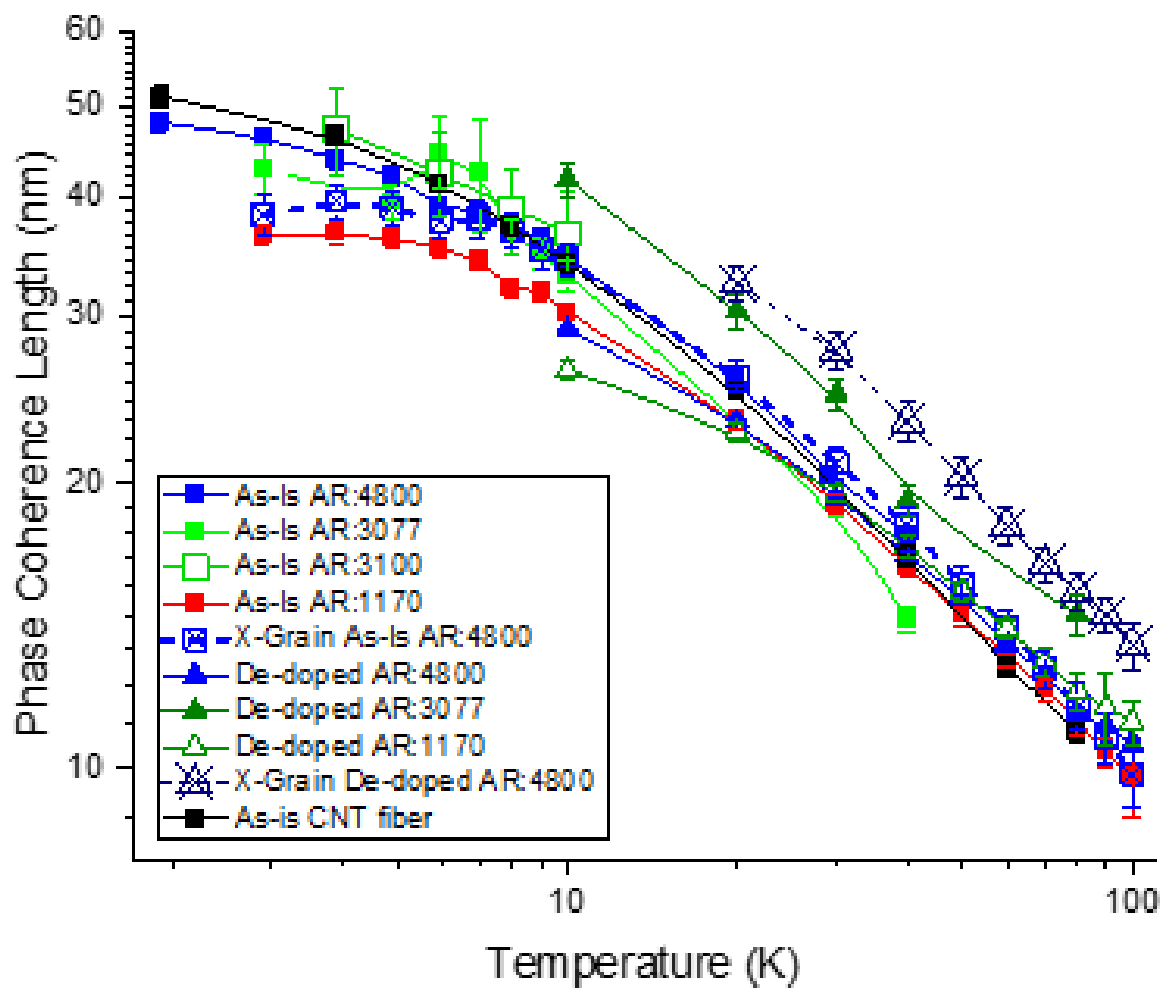

**Supplemental figure 4.2-1. Phase coherence length versus temperature from fitting negative transverse magnetoresistance (MR) data to the 2D weak localization model[2,3,4,5,6,7,8,9]. The 2D weak localization model is: $\Delta C(H)= a*(e^2/(2\pi^2\hbar))*(\ln(H/H_\varphi)+\psi((H_\varphi/H)+\frac{1}{2}))$ where $H_\varphi$ is the temperature dependent dephasing magnetic field and, when only one dephasing mechanism is acting, obeys a power law according to $H_\varphi=B\,T^p$, where $p$ is an exponent related to the inelastic dephasing mechanism. $H_\varphi$ is related to the coherence length $L_\varphi$ of the backscattered charge carriers as $H_\varphi= \hbar/(4\,e\,L_\varphi^2)$. a and $B$ are fitting constants and $\psi$ is the Degamma function. In our analysis, the temperature independent prefactor a applied to the 2D weak localization model was globally fitted across all temperatures for a given sample, which more conservatively constrained the fitting (yielding one fitted prefactor for all temperatures). The other 2D weak localization fitting term, $B$, associated with the temperature dependent phase coherence length, was locally fit (yielding a phase coherence length for every temperature, as depicted). A few points had fitting $R^2$<80% and are excluded from analysis and are not shown above; otherwise, the fitting $R^2$ averaged 97% +/- 0.2% across all points depicted here. Above 10 K, the slope/ exponent $p$ is near 1 for all CNT conductor varieties and points to an electron electron dephasing mechanism. Note that a wide variety of CNT conductor categories (as-is vs de-doped, parallel vs X-grain, multiple aspect ratios) respond similarly.**

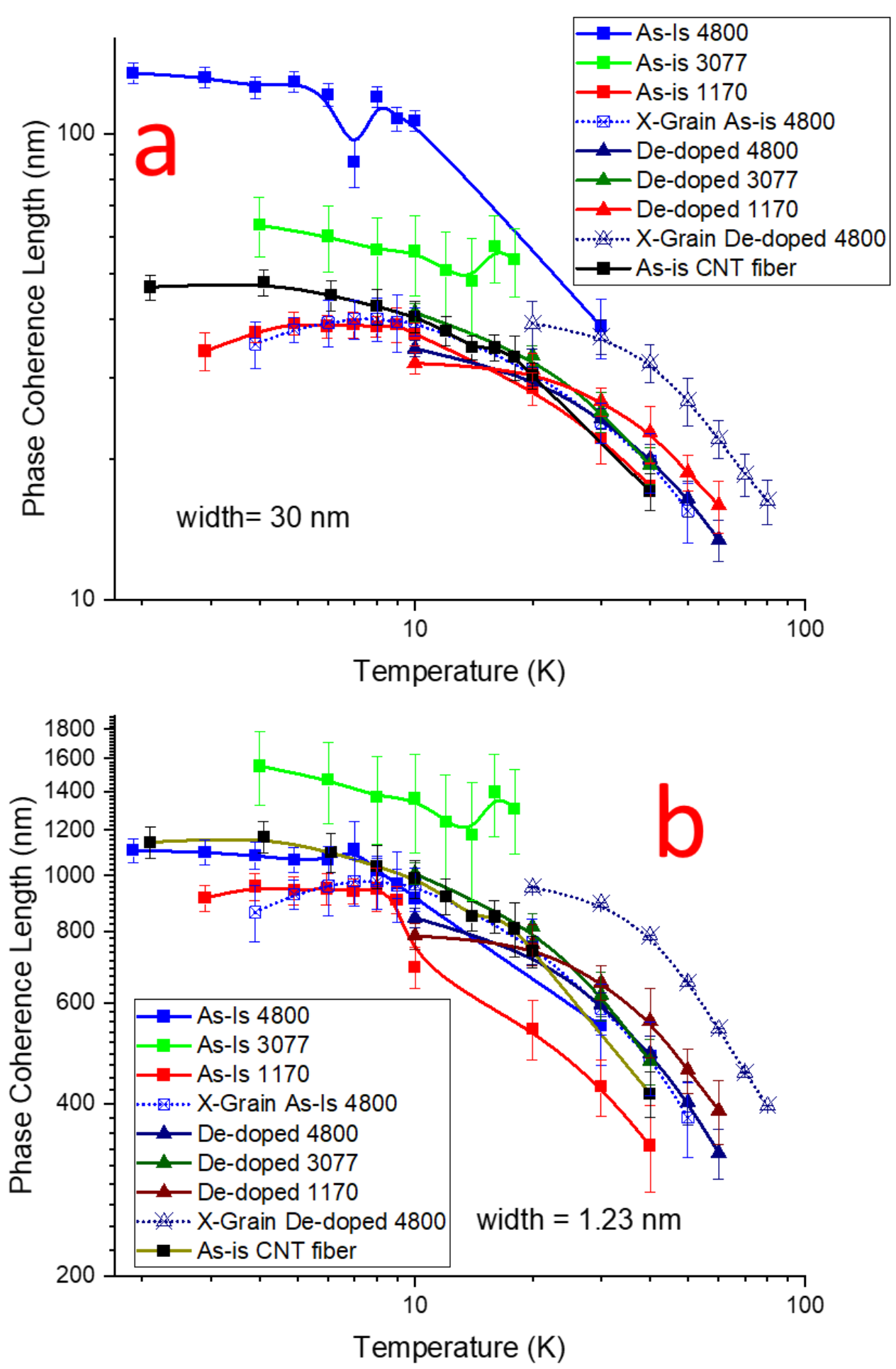

**Supplemental figure 4.2-2. Phase coherence length versus temperature from fitting transverse negative magnetoresistance (MR) data to the 1D weak localization model (used for InAs wires[10] [11] [12] [13] and considered for CNTs [4]), for a user selected conductor width of (a) 30 nm—approximate width of a bundle and (b) 1.23 nm—approximate width of a single CNT. The 1D model is: $C(H)= C_0 - (2e/(hL))*((1/L_\varphi^2)+(e^2H^2w^2/(3\hbar^2))^{-1/2}$ where $C_0$ is the conductivity before the weak localization correction, $L$ is the length of the 1D conductor, and $w$ is the width. The temperature independent prefactor applied to the 1D weak localization model was globally fitted across all temperatures for a given sample, which more conservatively constrained the fitting (yielding one fitted prefactor for all temperatures). The temperature dependent phase coherence length $L_\varphi$ was locally fit (yielding a phase coherence length for every temperature, as depicted). The 1D weak localization model also had an extra fitting term for the 1D conductor width, $w$, which we show both cases considered above and there was no significant difference in the goodness of fit. The 1D weak localization model also requires the conductance at a given temperature $C_0$ before the weak localization correction. This was estimated by taking the conductance value at 9 T where all weak localization effects should vanish. A few points had the fitted $R^2$<80% and are excluded from analysis and not depicted here; otherwise, the fitting $R^2$ averaged 91% +/- .03% across all points depicted here.**

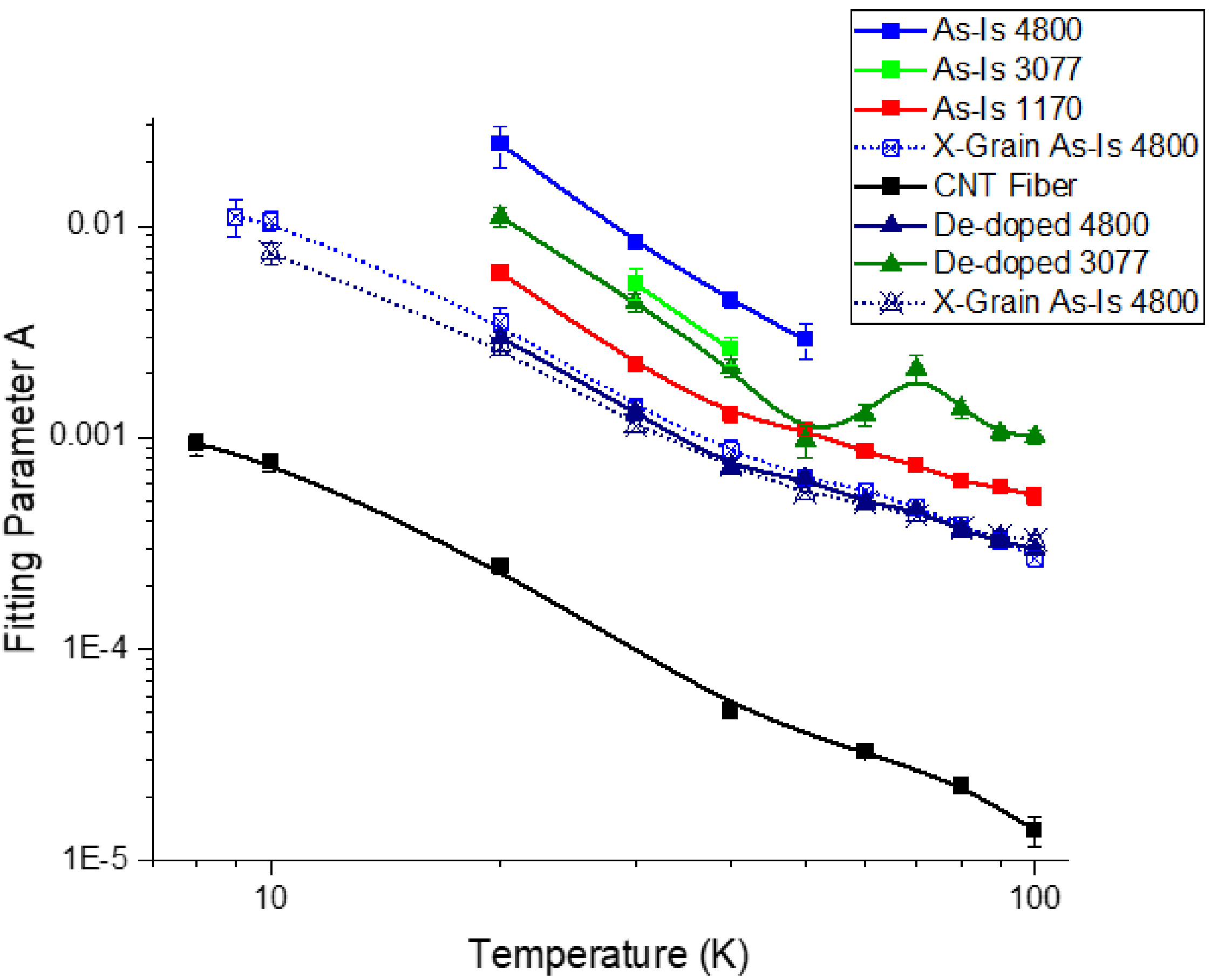

**Supplemental figure 4.2-3. For 3D weak localization[1,7,14,15], $\Delta C(H)=a*(1/12\mathrm{pi}^2)(e/\hbar)^2(G_0 L_\varphi^3)\, H^2 = A\, H^2$ where $a$ is a fitting constant, $G_0$ is the conductance quantum, and $L_\varphi$ is the temperature dependent phase coherence length of the backscatter. For a single de-phasing mechanism, the phase coherence length should follow the temperature power law $H_\varphi=B\, T^p$ and $H_\varphi= \hbar/(4\, e\, L_\varphi^2)$. For 3D weak localization, $p$=3 for inelastic electron-phonon scattering; p=2 (clean limit) and 3/2 (dirty limit) for inelastic electron-electron scattering; $p$=1 for near the metal-insulator transition. This magneto-conductance equation applies only when the magnetic field is sufficiently small ($g\, \mu_B\, H << k_B T$) where $g$ is the g-factor g≈2, $\mu_B$ is the Bohr magneton, and $k_B$ is Boltzmann constant. To balance satisfying this smallness criteria with keeping enough experimental data to fit, we only considered magnetic field and temperature combinations smaller than $H/T < 3\, k_B/(\, g\, \mu_B)$. Fits were notably worse than the 2D weak localization case and, rather than fitting globally as in the previous case, it was determined it was more meaningful to fit independently for every temperature.**

**Supplemental section 4.3 MR vs Magnetic Field Orientation.** Below we show the effect of resistance from an increasing magnetic field, depending if the magnetic field is perpendicular or parallel the CNT microstructure alignment. This is from the Quantum Design PPMS with the 9 T maximum field where the probe current is always in the direction of the CNT microstructure alignment. In the de-doped case, we see a larger positive MR component when the magnetic field is parallel to the CNTs.

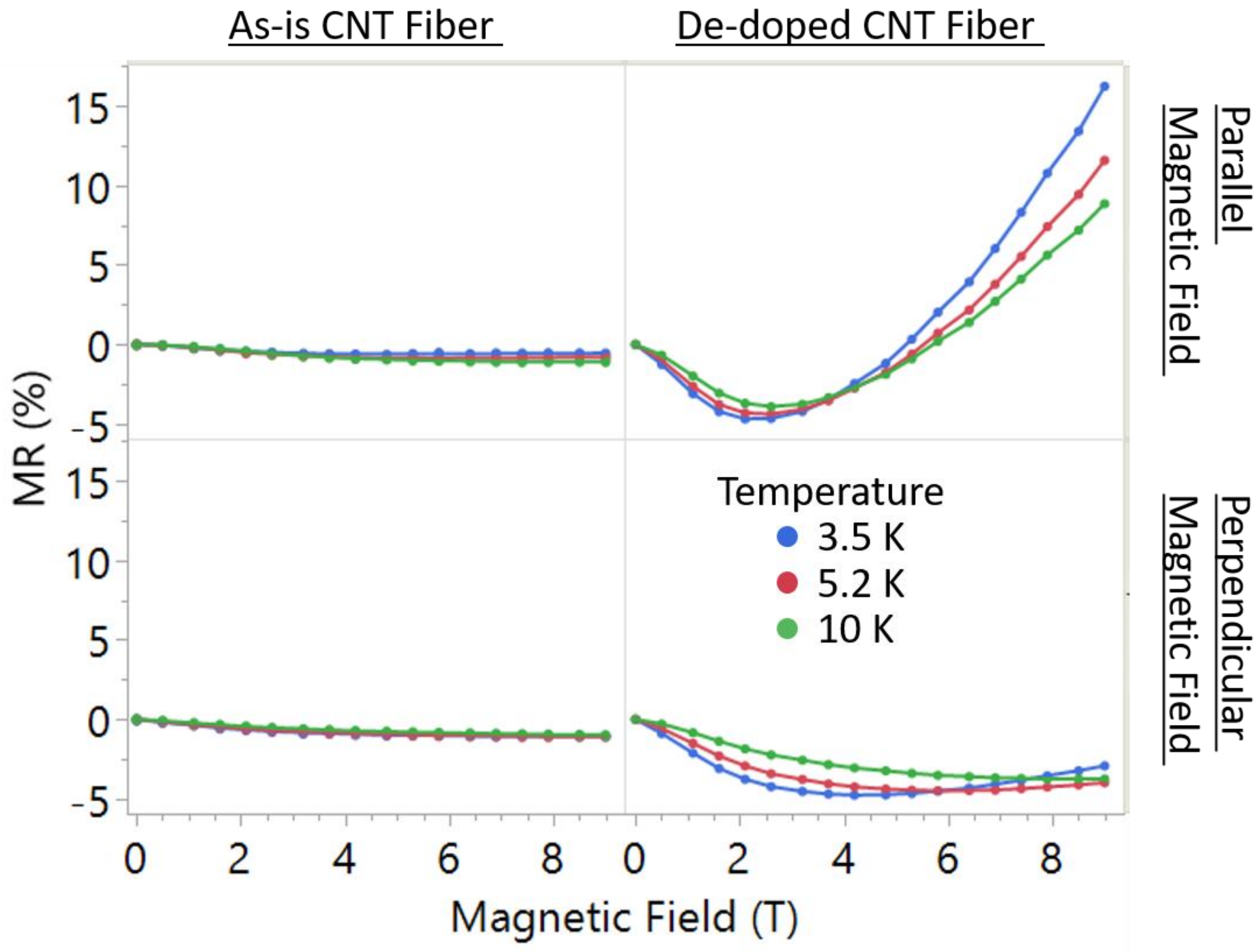

**Supplemental figure 4.3-1. MR (%) as a function of magnetic field for an as-is (left) and de-doped (right) CNT fiber when the magnetic field is parallel the CNT microstructure alignment (top) and perpendicular (bottom). In this configuration, the probe current is always in the direction of the CNT microstructure alignment.**

Now for the case for probe current perpendicular to the CNT microstructure alignment, MR versus magnetic field when the field is parallel or perpendicular the CNTs. This is for a de-doped CNT ribbon (AR: 1200) only.

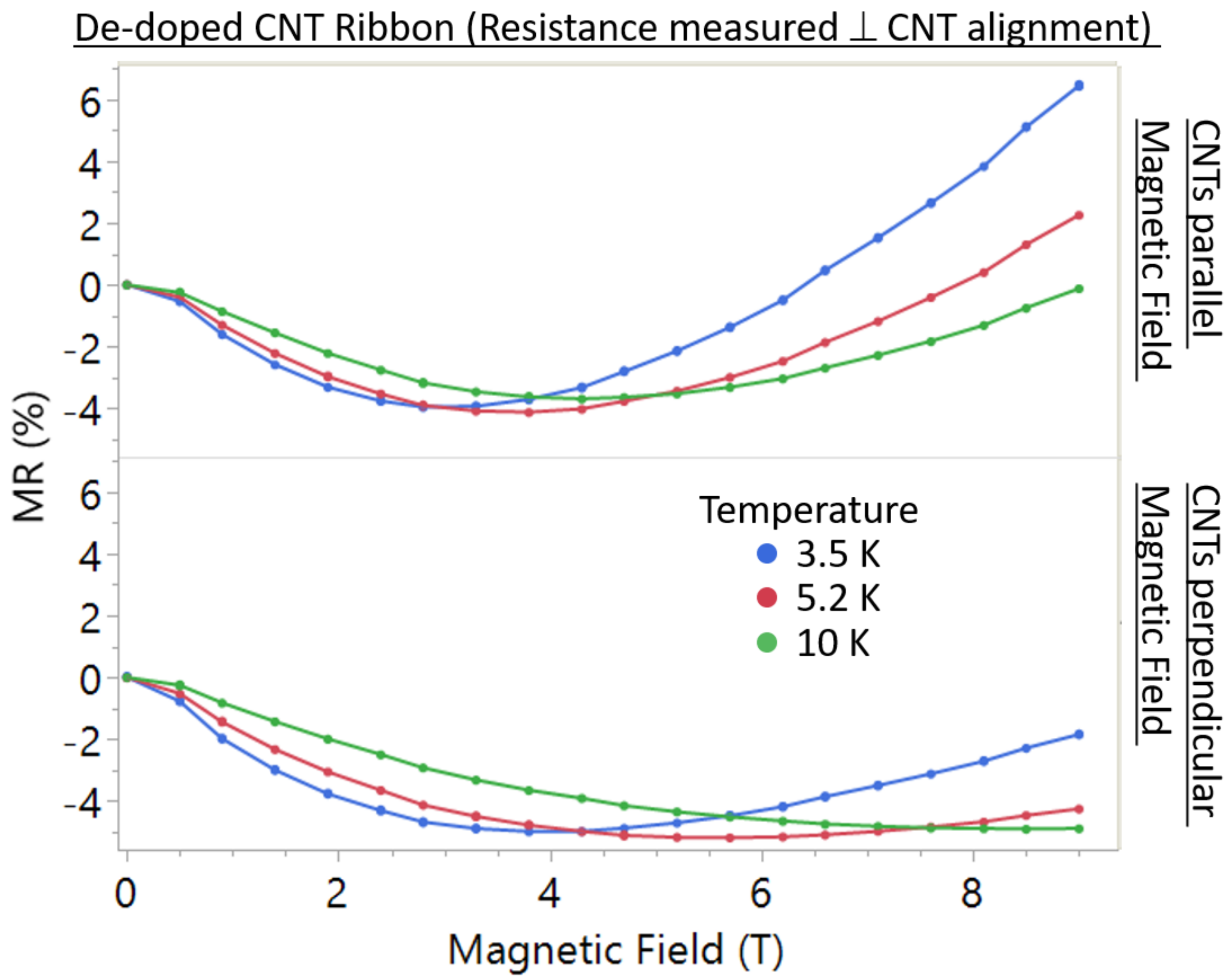

**Supplemental figure 4.3-2. Exploring MR (%) versus magnetic field for cross-grain—or when the probe current is perpendicular the CNT microstructure alignment, for a de-doped CNT ribbon (AR:1200). In this configuration, the probe current is always perpendicular the magnetic field as well. We consider two field orientations: field parallel the CNTs (top) and field perpendicular the CNTs (bottom).**

Now we consider MR (%) of CNT ribbons or fiber in a steady DC magnetic field (either at 1, 4.5 or 9 T) that rotates continuously in the magnetic field. Supplemental figure 4.3-3 shows the primary setup below where probe current is aligned with the CNT microstructure alignment and both are perpendicular to the axis of rotation. At 0°, the magnetic field is perpendicular to both the probe current and CNT microstructure alignment. At 90°, the magnetic field is parallel to both the probe current and CNT microstructure alignment. Multiple temperatures are attempted at each field. Note that, for every temperature, we also collected MR as a function of angle at zero magnetic field; this was to capture a baseline for the MR calculation and was generally not a significant factor. The complete dataset is in the supplemental database.

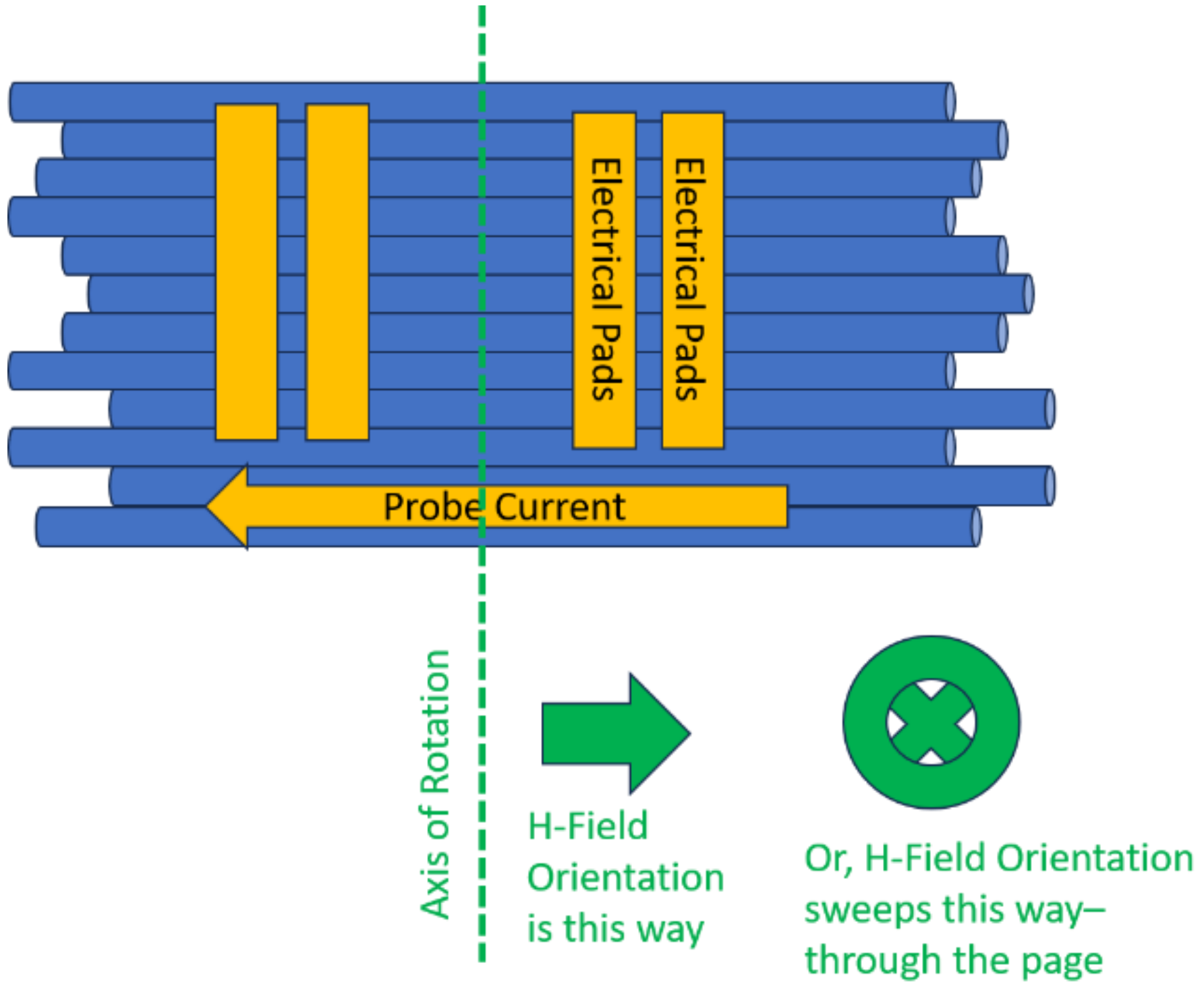

**Supplemental figure 4.3-3. Depiction of the MR orientation study where the probe current was parallel to the CNT microstructure alignment, and both are perpendicular to the axis of rotation. CNT microstructure is shown as horizontal blue tubes. As the sample rotated in the constant DC field, both the CNT and probe current changed its relative orientation with field.**

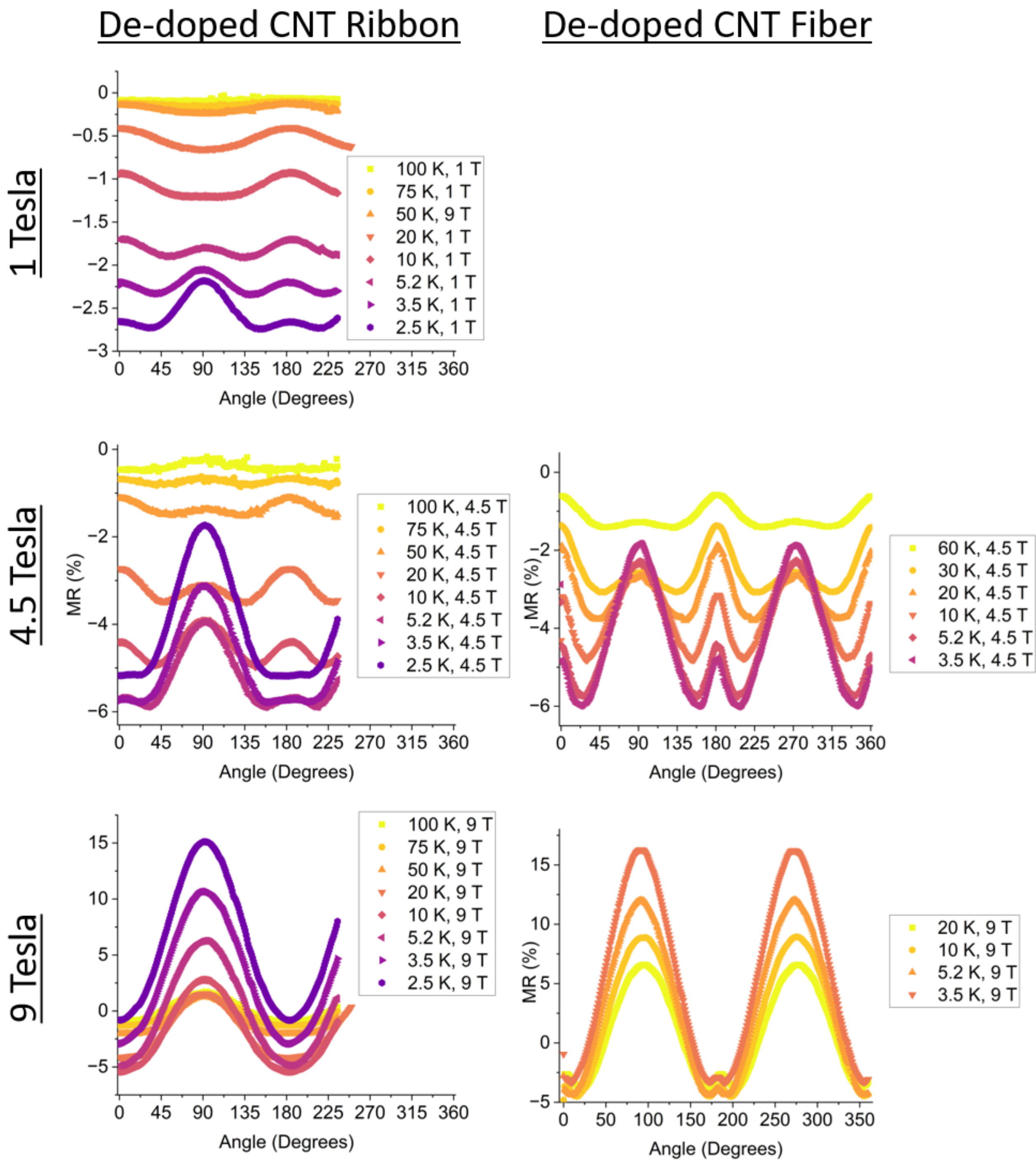

**Supplemental figure 4.3-4. For de-doped CNT ribbon (AR 1200) and CNT fiber, magneto-resistance MR (%) as a function of angle between the static magnetic field (at 1, 4.5 or 9 T) and probe current, where probe current is parallel to CNT microstructure alignment. 0° represents probe current perpendicular to field and 90° represents parallel. For the fiber (right column), the sample rotated a full 360°, although for the ribbon (left column), the sample only rotated 230° because of logistical constraints. 1 T data was not collected for the fiber.**

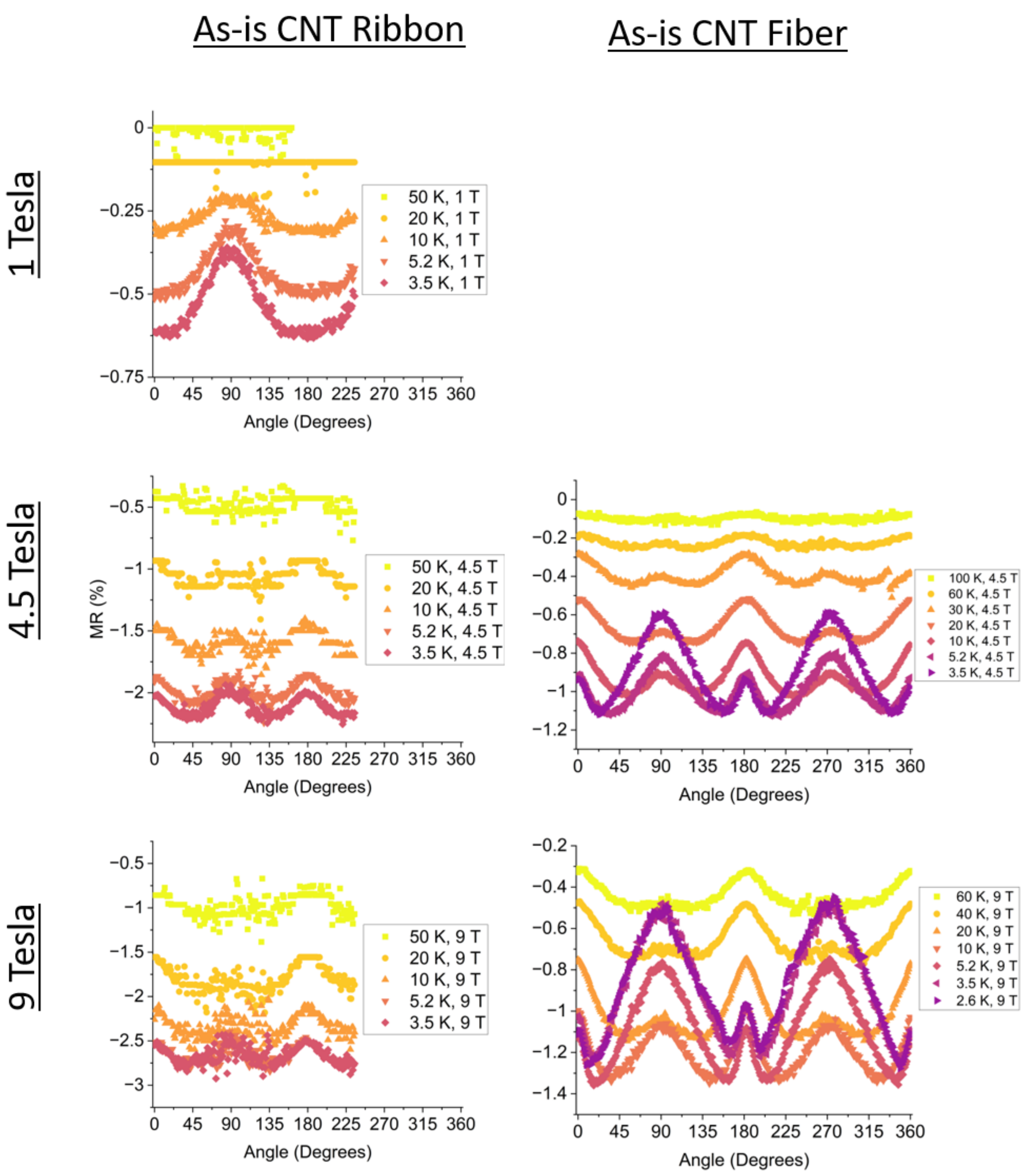

**Supplemental figure 4.3-5. Same as above, except now it is for as-is CNT ribbons and fibers that are still doped from their acid-based manufacturing process. The MR is lower and the samples are more conductive, which hinders signal to noise.**

The rotation of the sample in field leads to periodic modulation of the MR as a function of angle and therefore this periodic modulation can be analyzed in terms of its Fourier components. Care was taken so that components could be compared across samples; for example, interpolation functions kept the sampling rate the same across all samples (approximately 1 degree between datapoints). Further, samples had different angular measurement ranges, which were truncated to a standardized 0 to 180° for all samples before the Fourier transform. The non-zero real part for 180° rotation (two-fold symmetry) and the 90° rotation (four-fold symmetry) is shown below (supplemental figure 4.3-6) as a function of temperature and further partitioned by magnetic field strength and doping status. CNT fibers and ribbon were attempted and displayed similar responses. The imaginary components are near zero because of the even symmetry of materials and are not shown.

As shown in supplemental figure 4.3-6, the real two-fold coefficient is negative for all data sets and becomes more negative as temperature decreases or the magnetic field increases. A negative coefficient is equivalent to the two-fold term rotated by 90°, so the maxima of the two-fold modulation occurs at angle = 90° and 270° (when the field is parallel to the CNT alignment). The de-doped sample exhibits a substantially more negative coefficient than the as-is sample, indicating a stronger anisotropic MR in the parallel-field configuration. Also shown in supplemental figure 4.3-6, the real four-fold coefficient is positive for all data sets and there is the trend of more positive values for higher magnetic field; de-doped has greater magnitude than as-is. The trend for temperature is not as straightforward, although generally colder temperatures imply greater four-fold components. The fact the four-fold coefficients are positive mean that this four-fold MR component is maximized when the field is either parallel or perpendicular the CNTs.

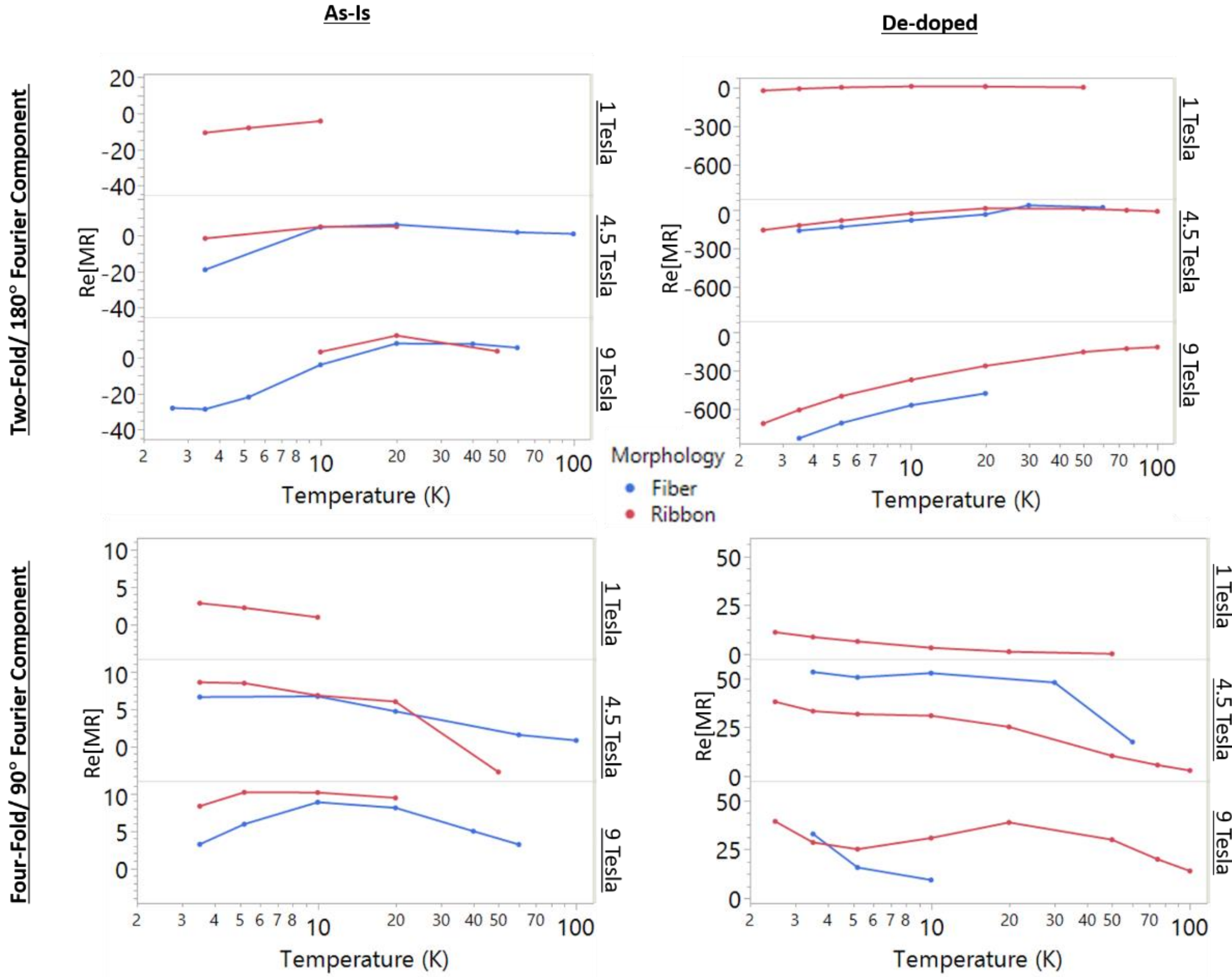

**Supplemental figure 4.3-6. Rotating the sample in field leads to a periodic waveform between MR (%) and angle. Here, we show the real part of the Fourier transform components as a function of temperature, for both CNT fibers (blue) and ribbon (red) where the probe current is in the direction of microstructure alignment. Four plots are shown representing quadrants where columns indicate as-is (left) or de-doped (right) status. The top row indicates the Fourier component associated with a 180° rotation and the bottom row is 90° rotation. Each quadrant plot is split between the applied static DC field: 1 T (top), 4.5 T (middle), and 9 T (bottom).**

We also hooked up a CNT ribbon in an alternative configuration. Here, the probe current is perpendicular to the microstructure alignment, while the probe current is parallel to the axis of rotation (supplemental figure 4.3-7). At 0° the magnetic field is perpendicular to the CNT microstructure alignment and at 90° the magnetic field is parallel to the CNT alignment. The intent of this experiment was that any MR modulation here had to be only from the CNT orientation and there could be no effect from weak localization, because the probe current orientation with respect to magnetic field was fixed. In the previous case, CNT microstructure alignment and probe current were pinned together. This picture is only true if indeed the probe current kept its fixed perpendicular orientation to probe current. Alternatively, it is quite possible the probe current actually takes a tortured path going across the grain of the microstructure alignment and develops significant lateral components that experience the field modulation from rotation. Figure supplemental figure 4.3-7 below shows that this “cross grain” MR are qualitatively similar to the “with grain” MR shown above, through the MR magnitude is generally lower. This is shown more explicitly with the Fourier components shown in figure supplemental figure 4.3-8.

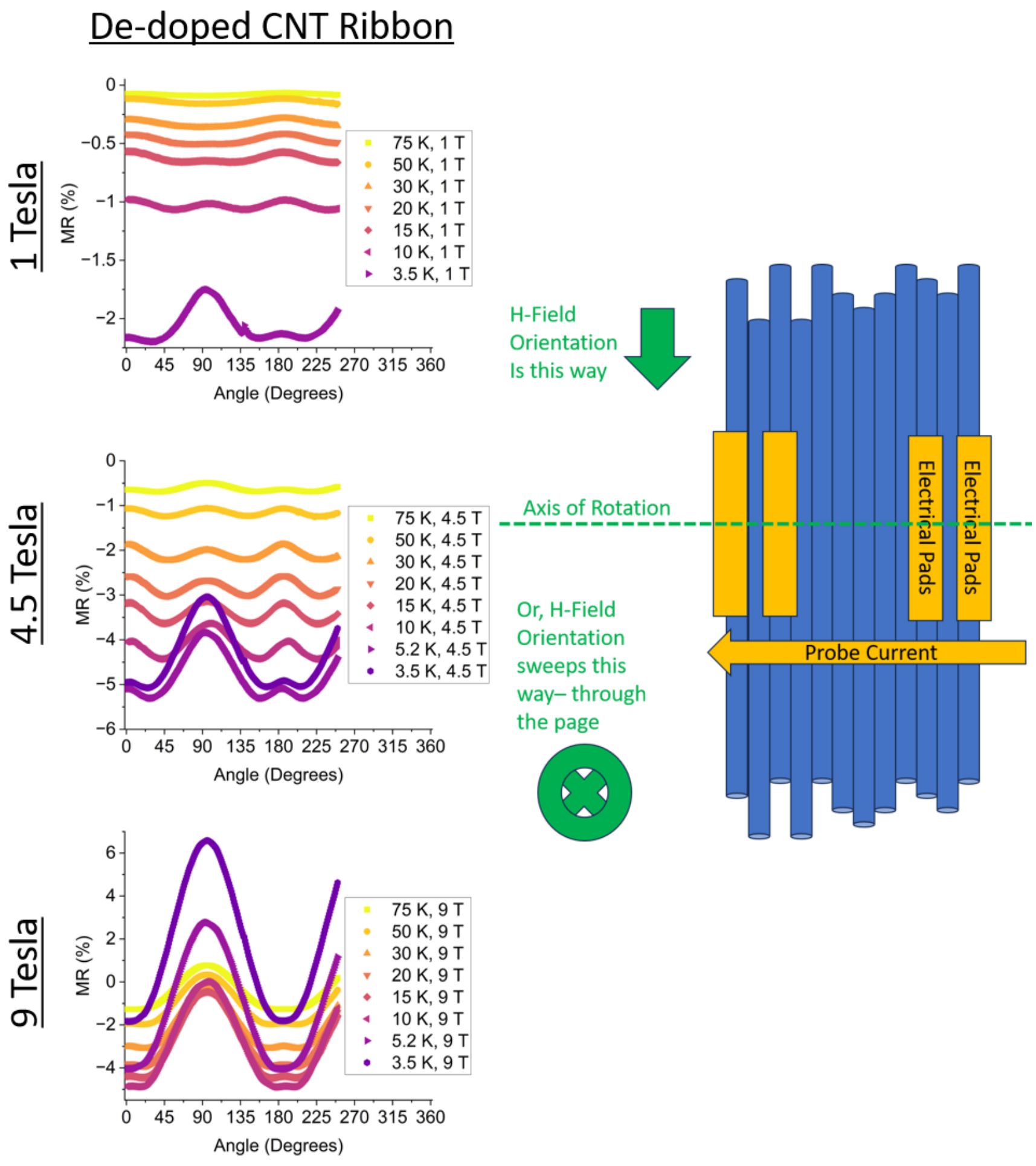

**Supplemental figure 4.3-7. For de-doped CNT ribbon (AR 1200) measured across grain, magneto-resistance MR (%) as a function of angle now between the static magnetic field (at 1, 4.5 or 9 T) and CNT microstructure alignment. 0° represents CNT alignment perpendicular to field and 90° represents parallel. In this setup, probe current is always perpendicular CNT microstructure alignment.**

**De-doped CNT Ribbon**

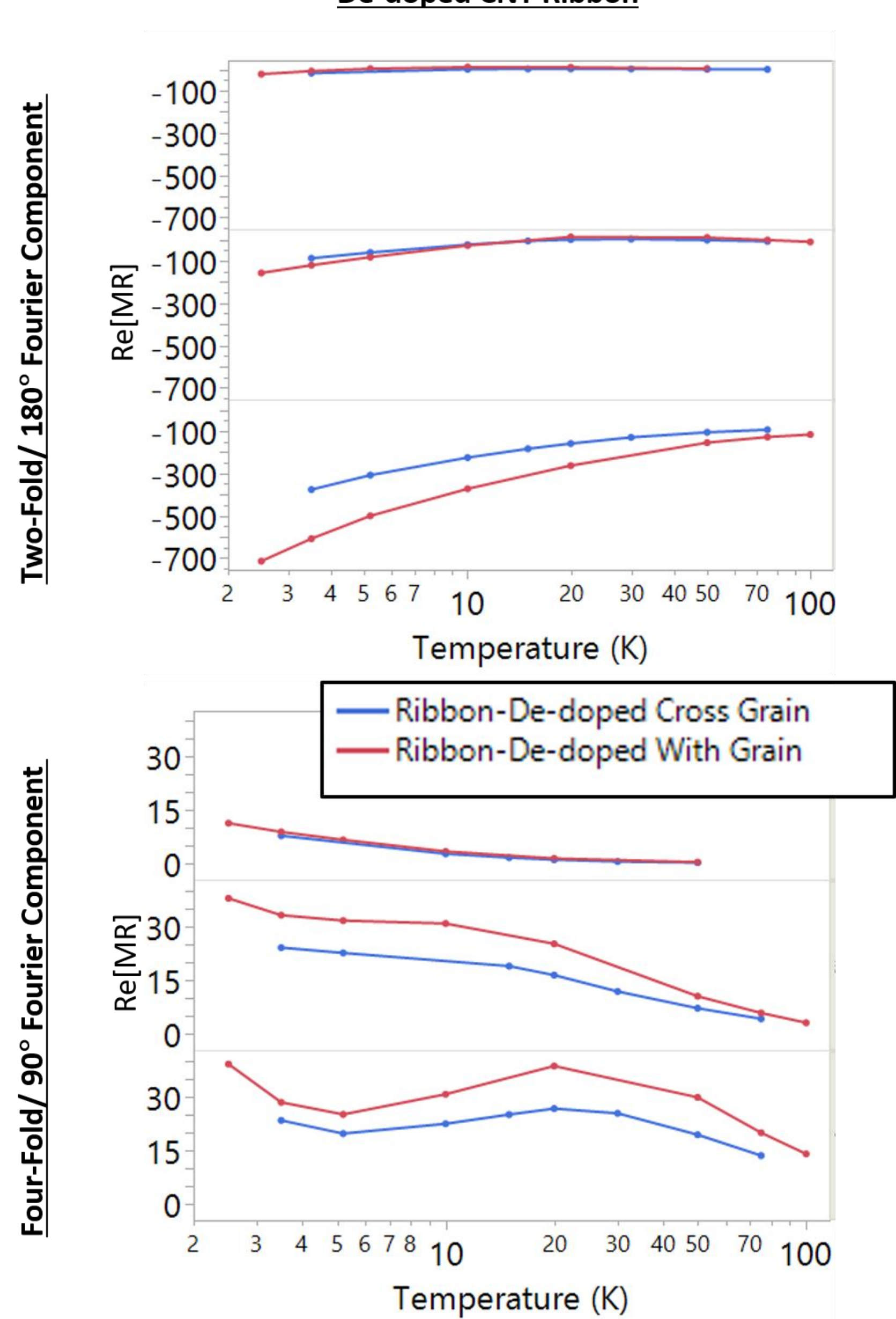

Supplemental figure 4.3-8. Comparing the Fourier components (180°-- top, 90°–bottom) where resistance is measured perpendicular to the CNT microstructure alignment (cross grain, blue) versus parallel the microstructure alignment (with grain, red), for de-doped CNT ribbon (AR: 1200) Traces are qualitatively similar.

As another attempted configuration, a de-doped CNT ribbon was hooked up in four wire resistance measurement such that probe current, CNT microstructure alignment, and axis of rotation were all parallel, with the field perpendicular to them all. This means that as the sample rotated, the CNT alignment and probe current did not change their orientation with the perpendicular field. The CNT ribbon has a planar morphology so at 0° the field is normal the ribbon's surface and at 90° it is parallel.

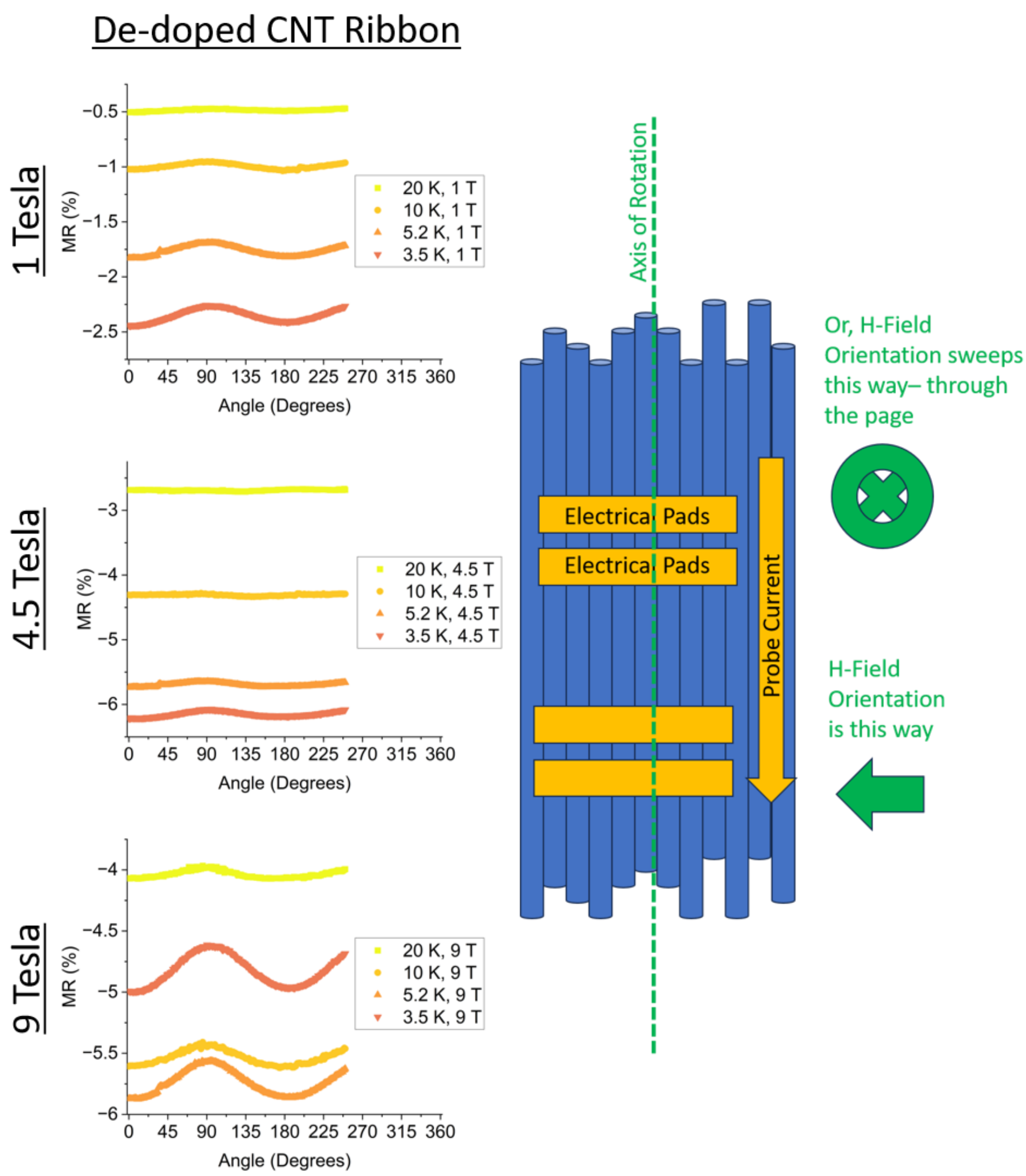

**Supplemental figure 4.3-9. For de-doped CNT ribbon (AR 1200), MR (%) as a function on angle between the magnetic field and the planar surface of the ribbon where at 0° the field is normal the ribbon's surface and at 90° it is parallel. The probe current, CNT microstructure alignment, and axis of rotation are all parallel with each other and all are perpendicular to the field.**

For field orientation measurements, it is customary to show the MR modulation with angle at the selected DC magnetic field and again at zero magnetic field (where the MR modulation should nominally be zero, to isolate any experimental artifacts). Below we show a representative case (as-is and de-doped CNT fiber, at 3.5 and 20 K) of the raw resistance as a function of angle, for 0 T, 4.5 T, and 9 T. This shows that the 0 T case has a constant resistance when sweeping angle. Other examples can be found in the database.

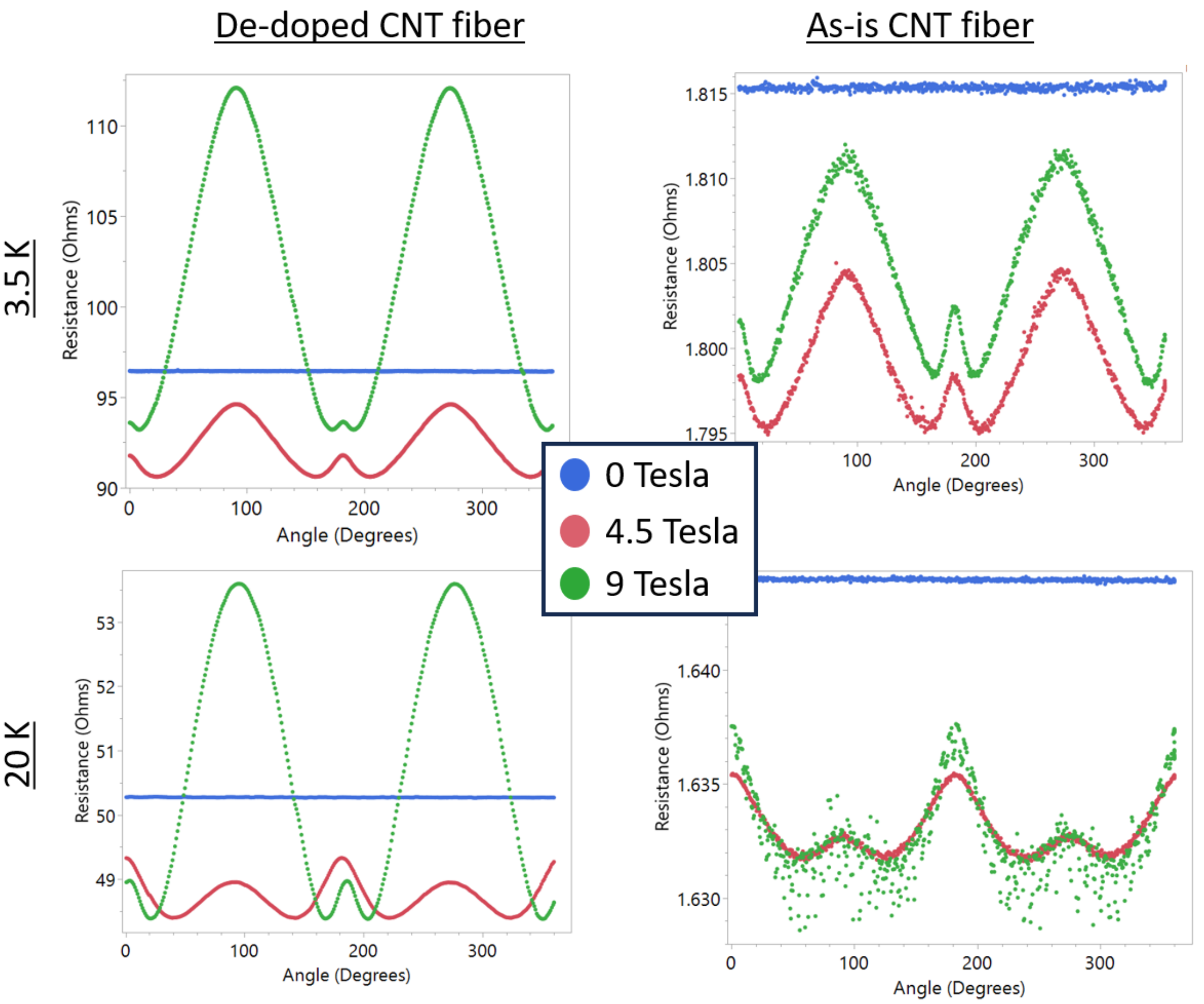

**Supplemental figure 4.3-10. Representative case (de-doped and as-is CNT fiber, at 3.5 and 20 K) showing the raw resistance as a function of angle for three fields: 0 T, 4.5 T, and 9 T. As shown the sample resistance does not change when rotating the sample when the field is zero.**

**Supplemental section 4.4 Low-magnetic field MR—Hall measurements.** Hall measurements were accomplished with our in-house Quantum Design PPMS. CNT ribbons were cut in a Hall bar geometry [16,17] with Hall voltage probe tabs in the middle and probe current was injected at the ribbon ends. We also implemented the standard voltage probes for standard four-wire resistance measurement, which would follow immediately after each Hall measurement. In both cases of the Hall voltage measurement and the standard four-wire resistance measurement, the probe current was flipped to isolate thermo-electric effects. The DC magnetic field *H*, normal to the CNT sample surface and perpendicular to the probe current, would ramp from positive to negative values to generate a corrected Hall voltage *V* according to $(V(H)-V(-H))/2$ that isolated out other confounding magneto-resistance effects. This corrected Hall voltage is what we see in the plots below in terms of a Hall resistance and applied magnetic field. The Hall data is contained in the database.

First, we consider molybdenum shim for validation of the Hall measurement process. Note that molybdenum is a hole majority conductor[18,19].

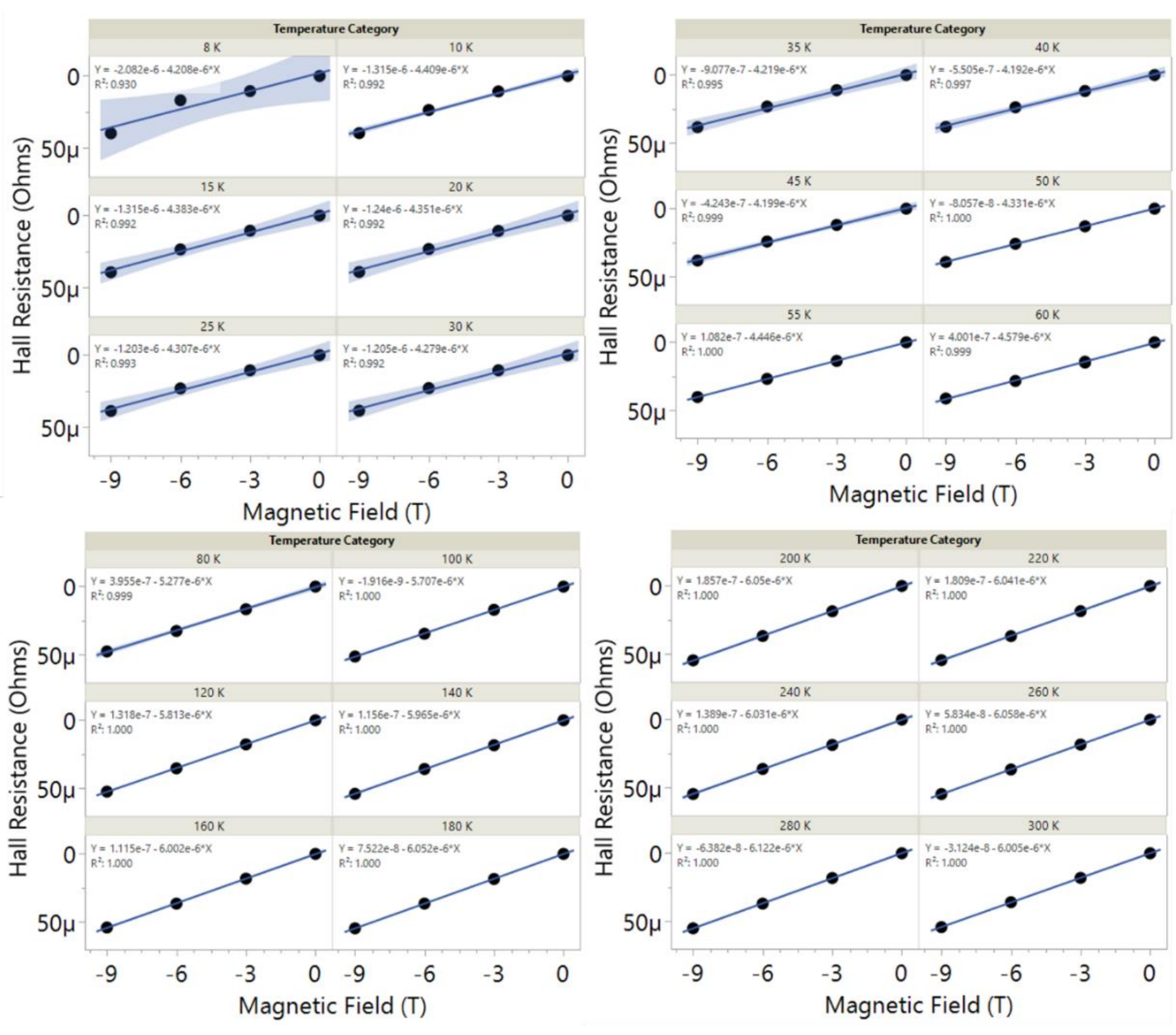

**Supplemental figure 4.4-1. Hall resistance versus magnetic field for room temperature down to 8 K for a strip of molybdenum film, for process validation.**

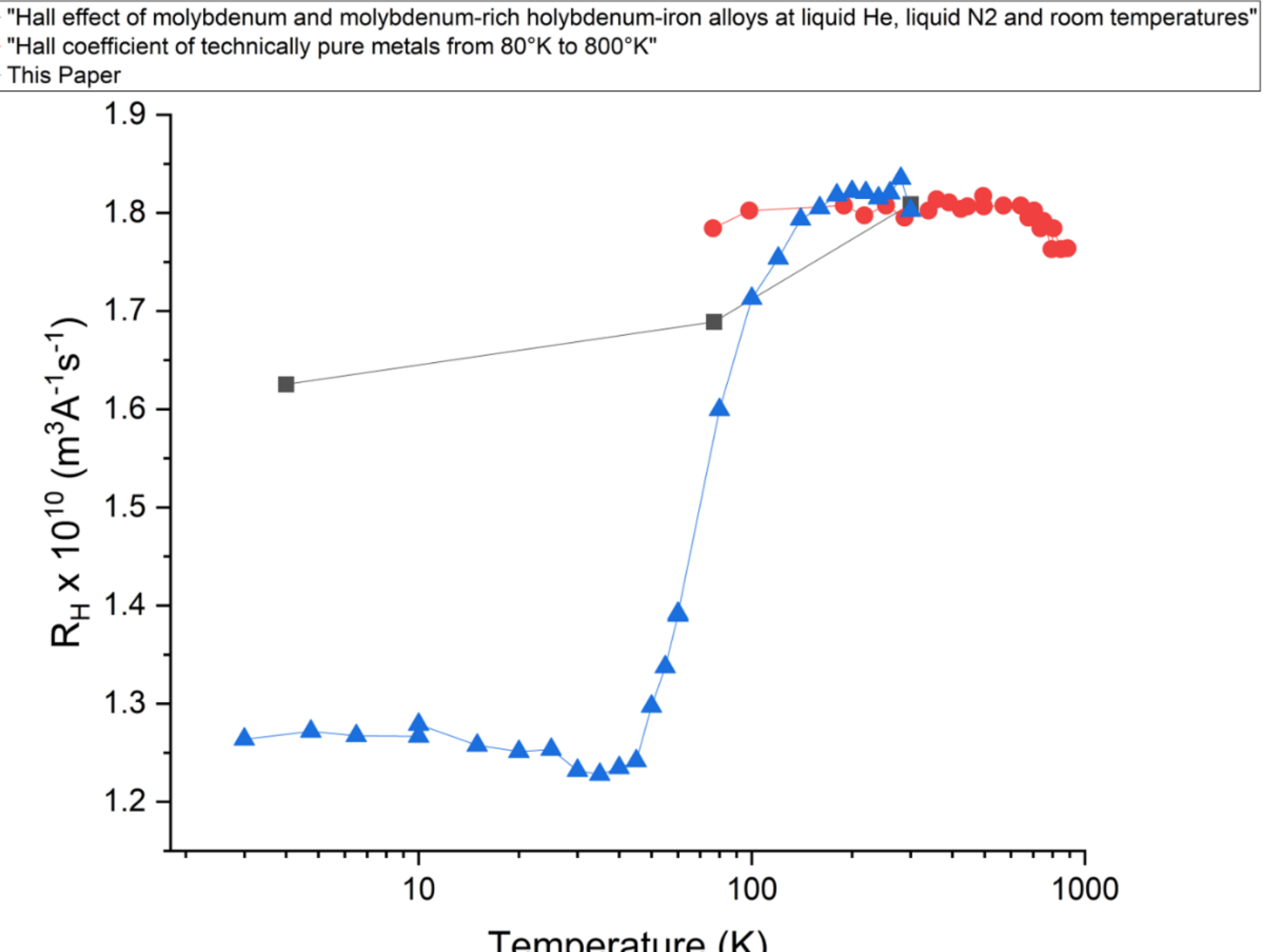

**Supplemental figure 4.4-2. The Hall coefficient $R_H$ of molybdenum calculated from the above magnetic field sweeps, as a function of temperature. Superimposed are $R_H$ values from literature. The molybdenum sheet is 30 µm thick. Note that these are hole majority conductors.**

To determine carrier density and electronic mobility, Hall measurements were conducted on as-is and de-doped CNT ribbon. As-is CNT materials has relatively temperature independent carrier density with holes as the majority carrier. The de-doped CNT materials had mostly hole conduction, although their traces were noisier and the was a dramatic change to thermally activated electron conduction above 300 K. The carrier density was unphysically high ($10^{28}$ to $10^{30}$ $m^{-3}$); this similar Hall anomaly was found in unaligned CNTs Buckypapers[20] where their Hall voltage was reduced by the heterogeneous nature of the CNT agglomerations.

Hall response to field for an as-is CNT ribbon with aspect ratio 4800

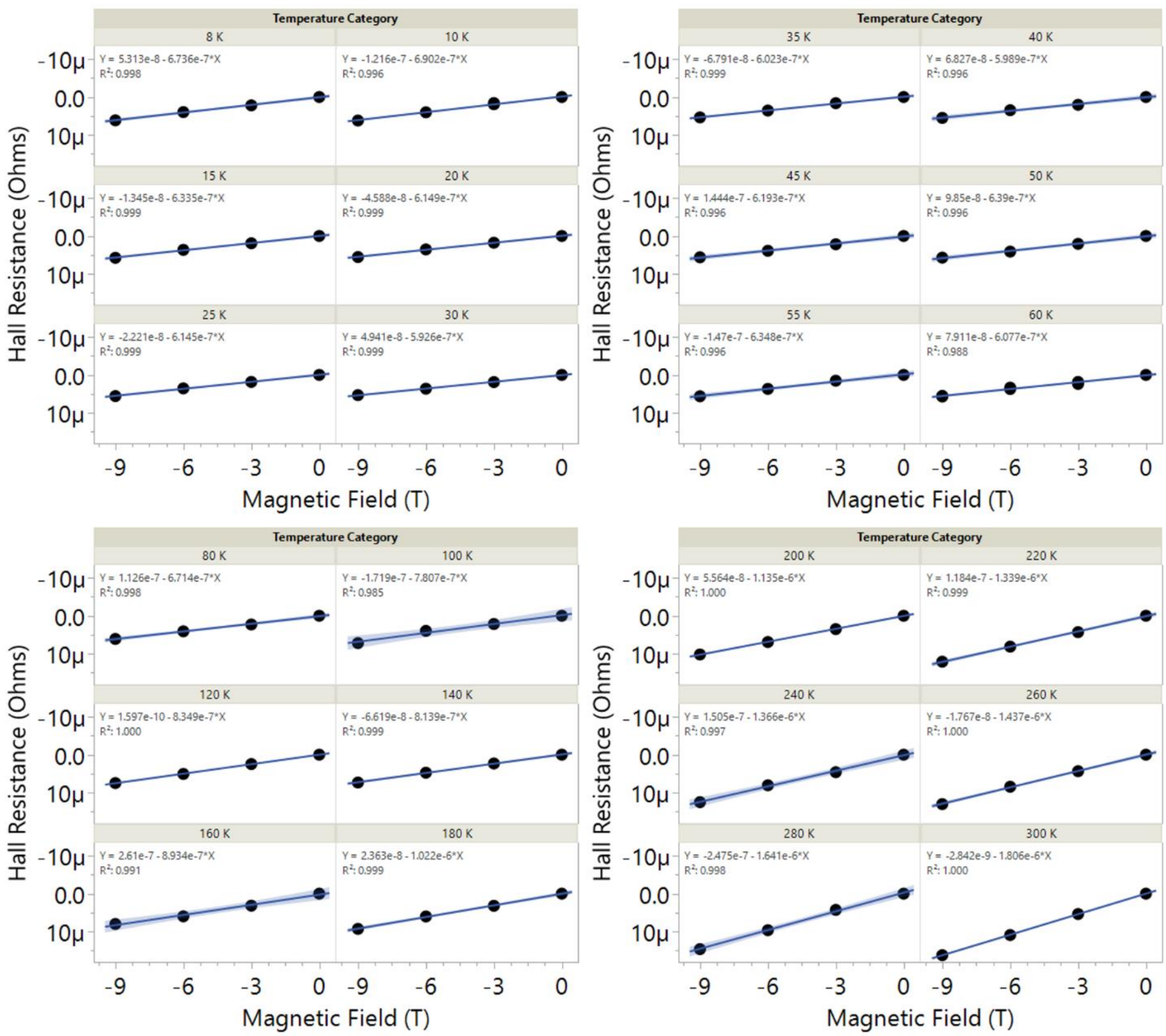

**Supplemental figure 4.4-3. Hall resistance as a function of magnetic field for as-is CNT ribbon with aspect ratio 4800, for set temperatures 300 to 8 K.**

Now, we consider de-doped CNT ribbon with aspect ratio 4800. Note that the goodness of fit is more varied, and the slope changes sign at approximately 315 K.

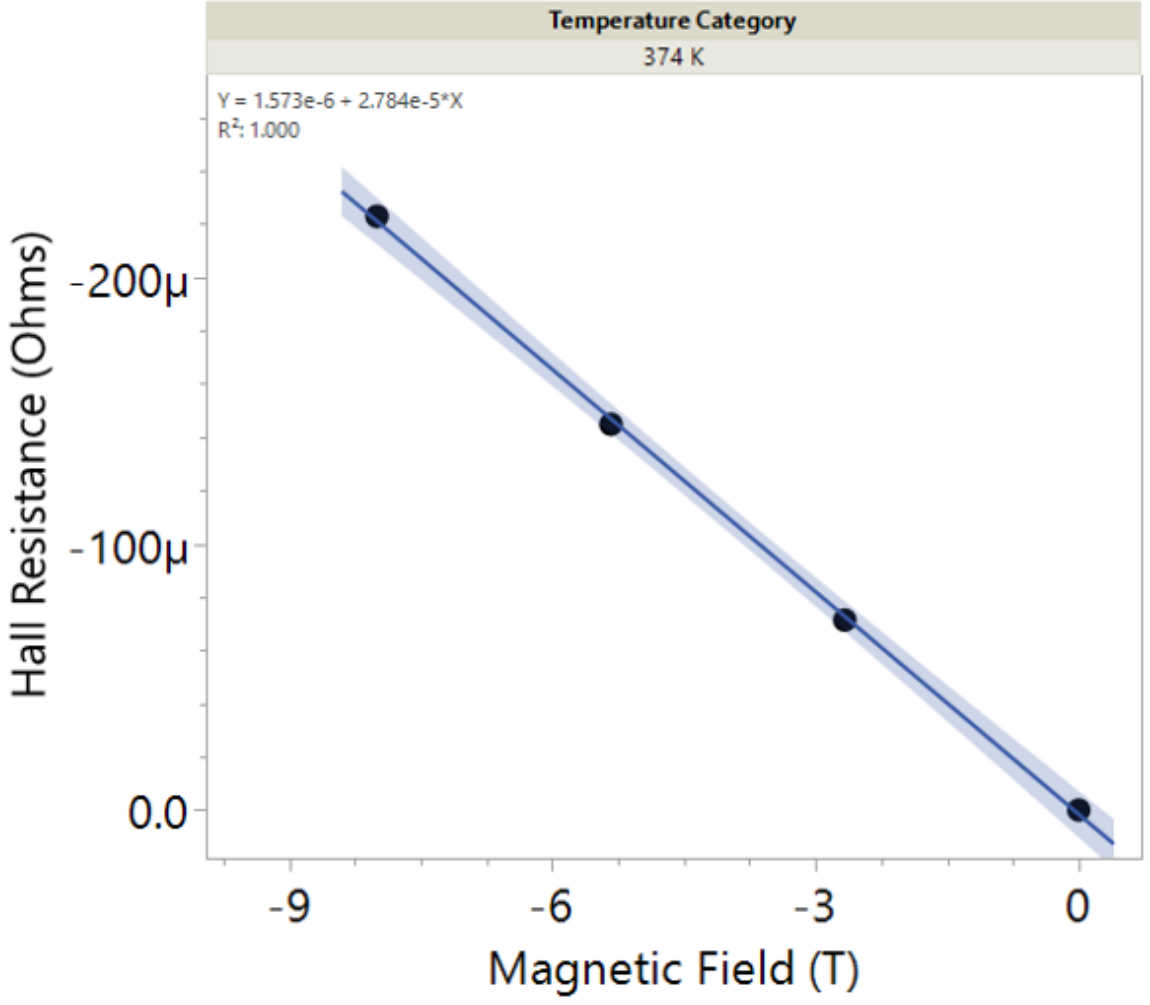

**Supplemental figure 4.4-4. Hall resistance as a function of magnetic field: de-doped CNT ribbon with aspect ratio 4800, for 374 K.**

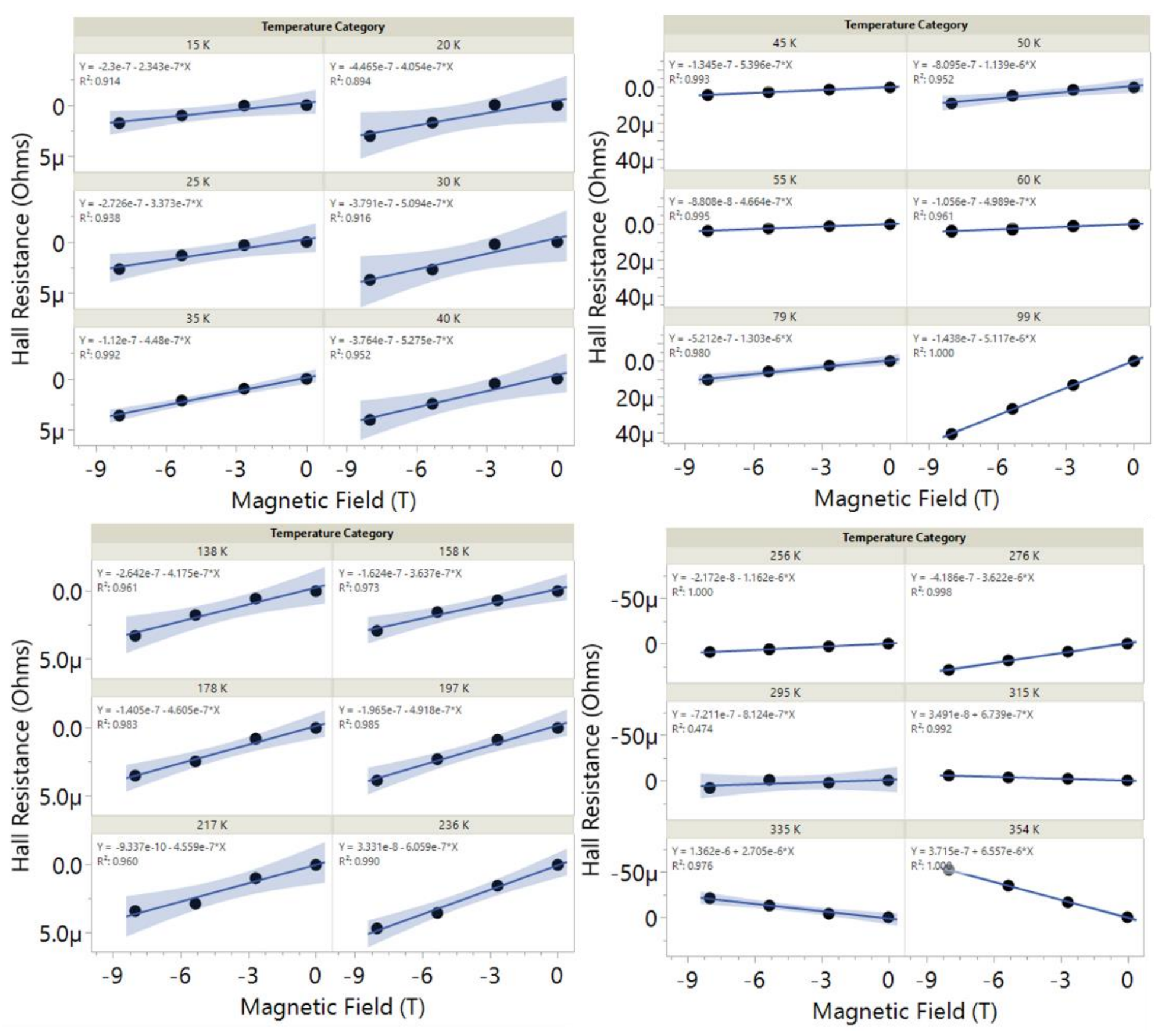

**Supplemental figure 4.4-5. Hall resistance as a function of magnetic field: de-doped CNT ribbon with aspect ratio 4800, for 354 to 15 K.**

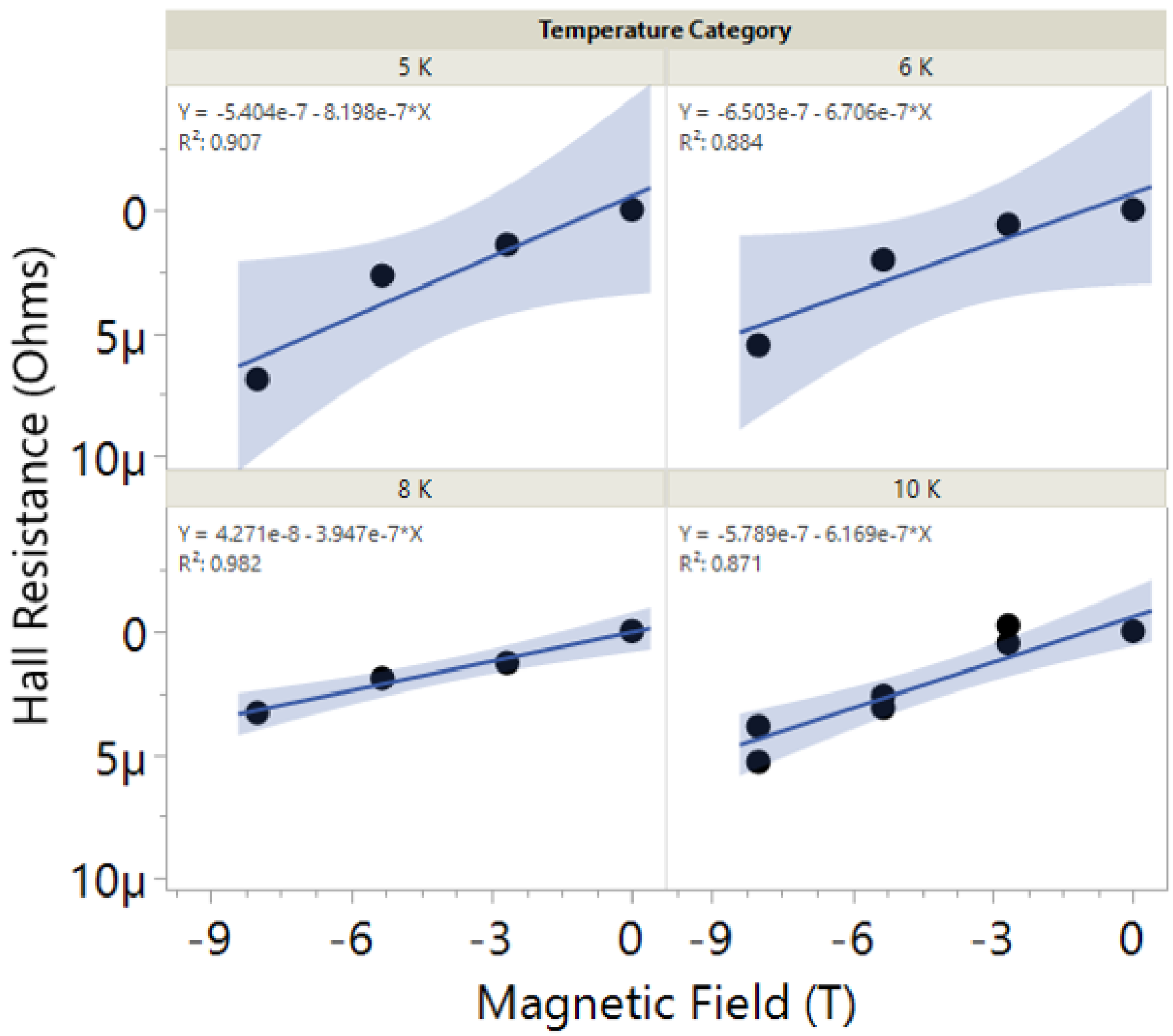

**Supplemental figure 4.4-6. Hall resistance as a function of magnetic field: de-doped CNT ribbon with aspect ratio 4800, for 10 to 5 K.**

Below is the magnitude of the Hall coefficient verses temperature, calculated from the slope of the Hall resistance verses magnetic field, for the as-is and de-doped CNT ribbons with aspect ratios 4800, 3100, and 1200. Below that plot is the associated carrier density, assuming one charge carrier. These plots show the absolute value, where the carrier density sign (indicating either holes or electrons) is depicted by color. Most, but not all, magnetic field sweeps led to a linear change of the Hall resistance. Linear fits with $R^2$ less than 90% are excluded below.

As shown, the as-is traces are more regular than the de-doped CNT materials. Above 300 K the de-doped CNT materials consistently had a change in slope in Hall resistance versus magnetic field and indicated a change from majority hole to majority electron conduction. The magnitude of this reversed slope grows in magnitude with increased temperature showing thermal activation of the electrons.

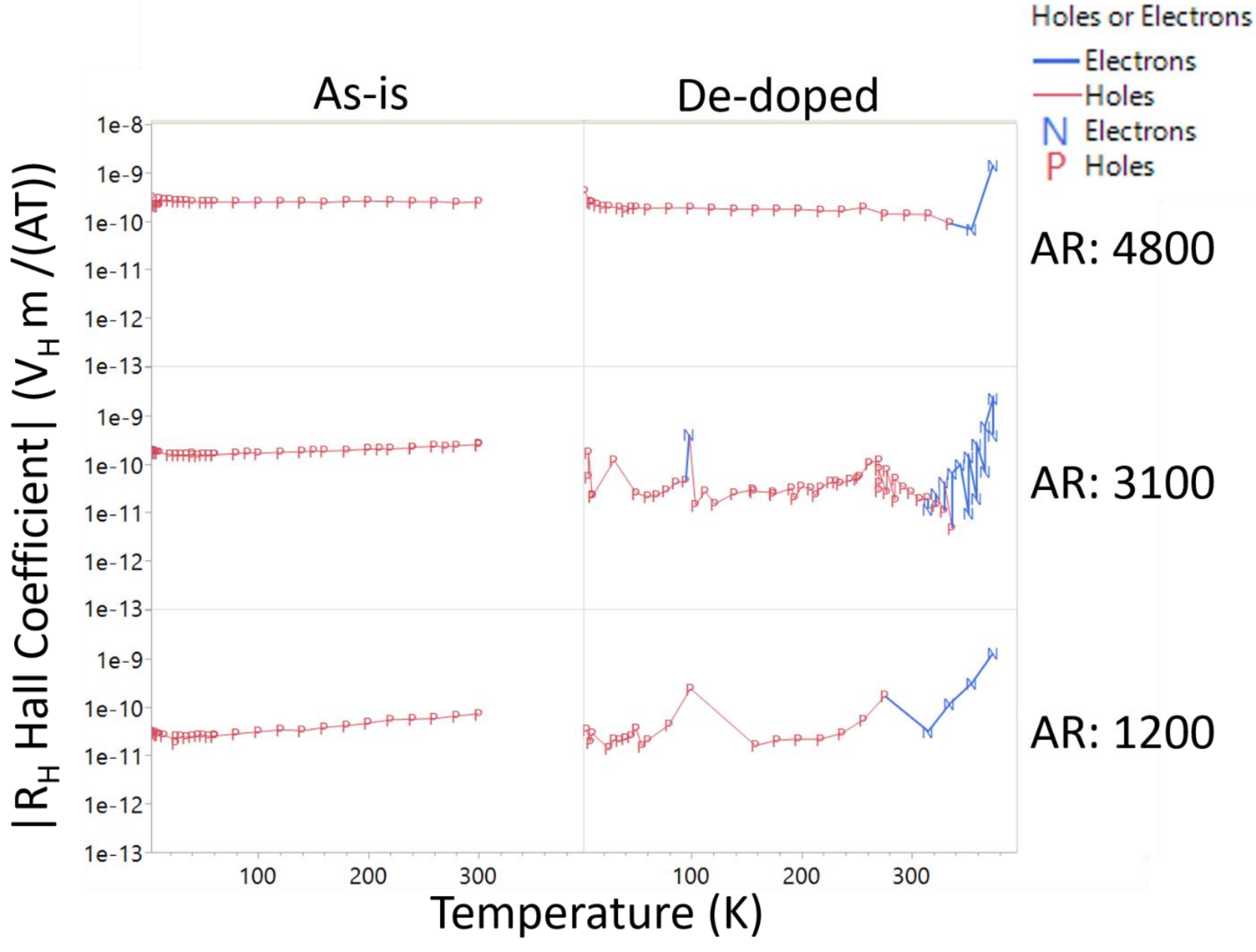

**Supplemental figure 4.4-7. Magnitude of the Hall coefficient $R_H$ as a function of temperature for as-is and de-doped with three different aspect ratios. Red P's are holes and blue N's are electrons. When the linear fit between the Hall voltage and magnetic Field had an RSq less than 90%, the point was excluded.**

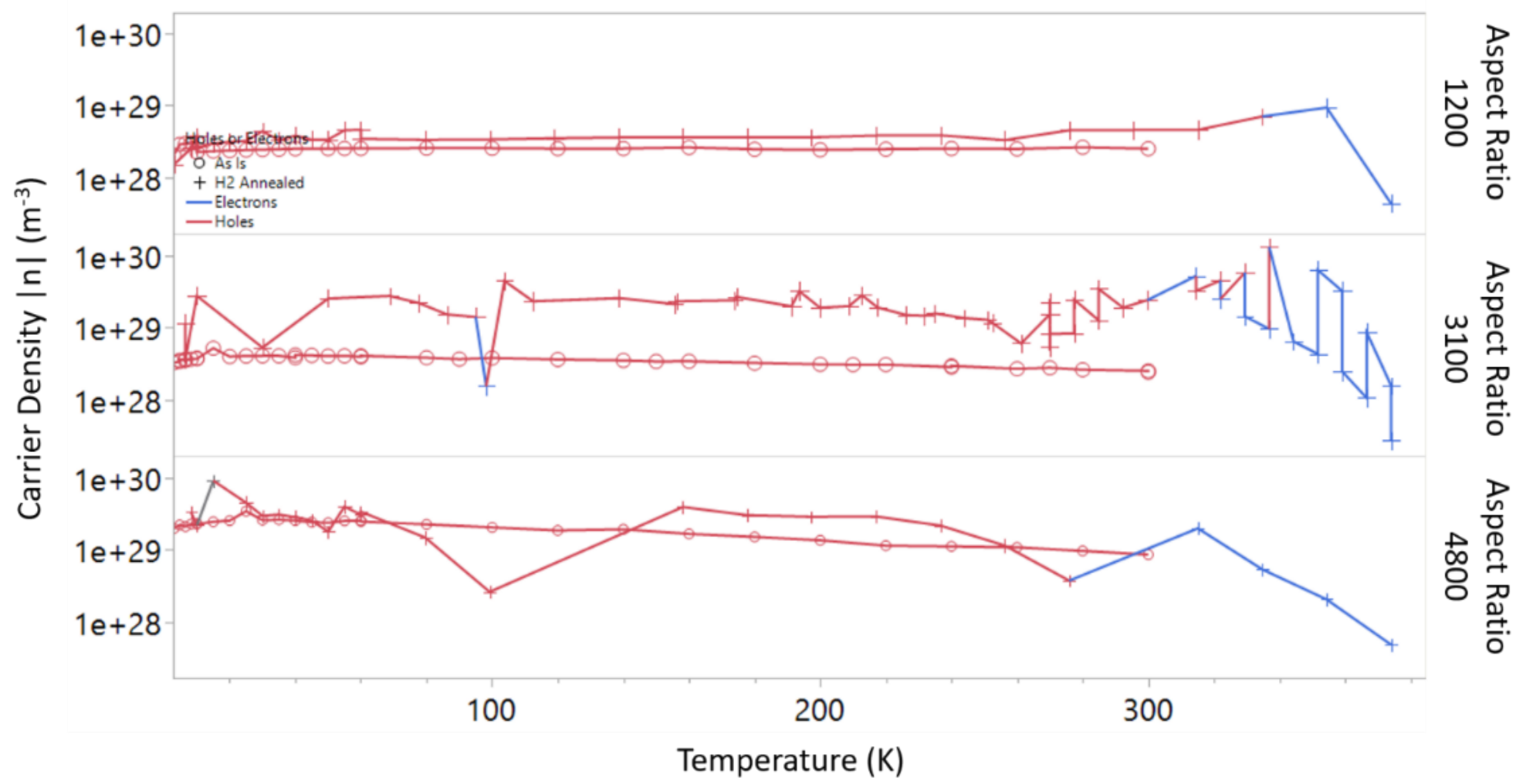

**Supplemental figure 4.4-8. Absolute value of the carrier density n as a function of temperature for as-is ( ◦) and de-doped (+) with three different aspect ratios. Red is hole conduction and blue is electron conduction. When the linear fit between the Hall voltage and magnetic Field had an RSq less than 90%, the point was excluded.**

Below we plot temperature versus the RSq from the linear fit between Hall voltage and the applied magnetic field, for different aspect ratios, in the as-is and de-doped state. The as-is materials have a more consistent high RSq value, where-as the de-doped have a sporadic RSq response.

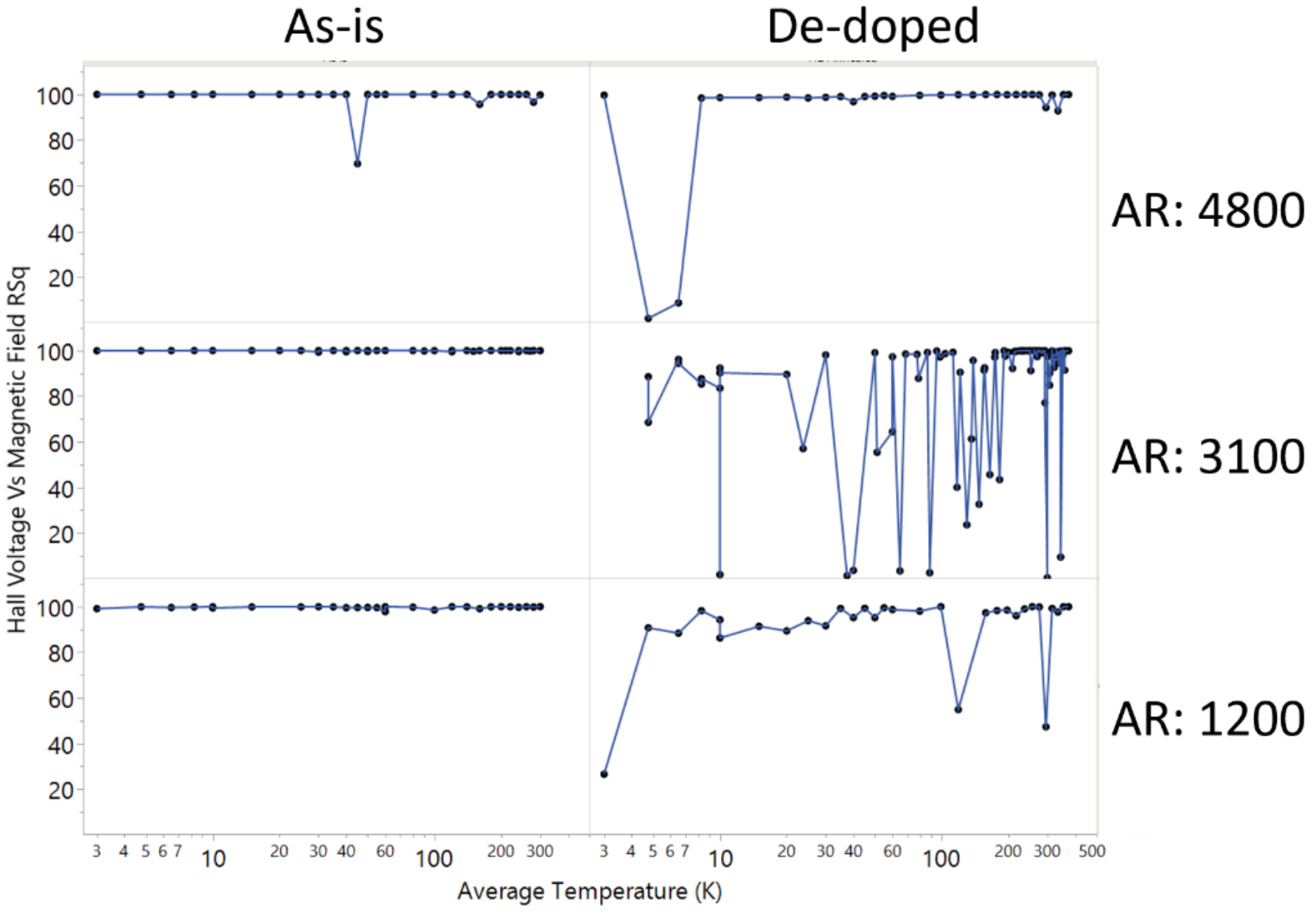

**Supplemental figure 4.4-9. As a function of temperature, the RSq of the linear fit between the Hall voltage and applied magnetic field, for the as-is and de-doped, for three different aspect ratio materials.**

**4.5. High Voltage Field Effect Transistor Setup.** As an alternative to Hall measurements, we probed CNT-ribbon carrier density and mobility with a high-field FET platform (capable of up to 50 kV / 100 A) inside a high vacuum chamber. The gate was a 1/8" x 6" diameter 260 Brass alloy disk overlaid with a flawless Dupont 300HN Polyimide/Kapton film (thickness: 0.003", width: 3.75"). A de-doped CNT ribbon was pressed flat onto the Kapton. Four silver-paint contacts provided a four-wire connection; all protruding wire filaments were secured to reduce electric stresses. The measurement circuit (composed of power supply, two multimeters (OWON XDM1041 Portable Digital Multimeter) for measurement of voltage drop across the sample and voltage drop across a in-series resistor of known value) was fully isolated from the high-voltage supply. A Teflon cylinder and additional Kapton tape plus American Sealants Inc. 388 Electrical-Grade pure silicone insulated exposed high-voltage and high electric stress regions throughout the sample chamber. The series capacitance across the Kapton film to the sample was measured to be 132 pF at 1000 Hz with a Matrix 5200 LCR Meter. An electrically isolated turbo-pump evacuated the chamber to 2E-5 Torr as measured from an ionization gauge, and a Hipotronics 880HL-PD high voltage DC power supply / hipot tester delivered the gate voltage.

Validation with a 1 Ω resistor showed no change as the gate swept from 0 to 10 kV. The fully de-doped CNT ribbon—annealed at 1000 °C, baked at 100 °C under vacuum, and transferred without air exposure—was then tested. Sweeping the gate from 0 to 10 kV (electric field = 0 to 260 kV/mm) left its resistance (1.102 ± .002 Ω) unchanged outside of the noise limit. No field-effect response was therefore observed in de-doped CNT ribbons.

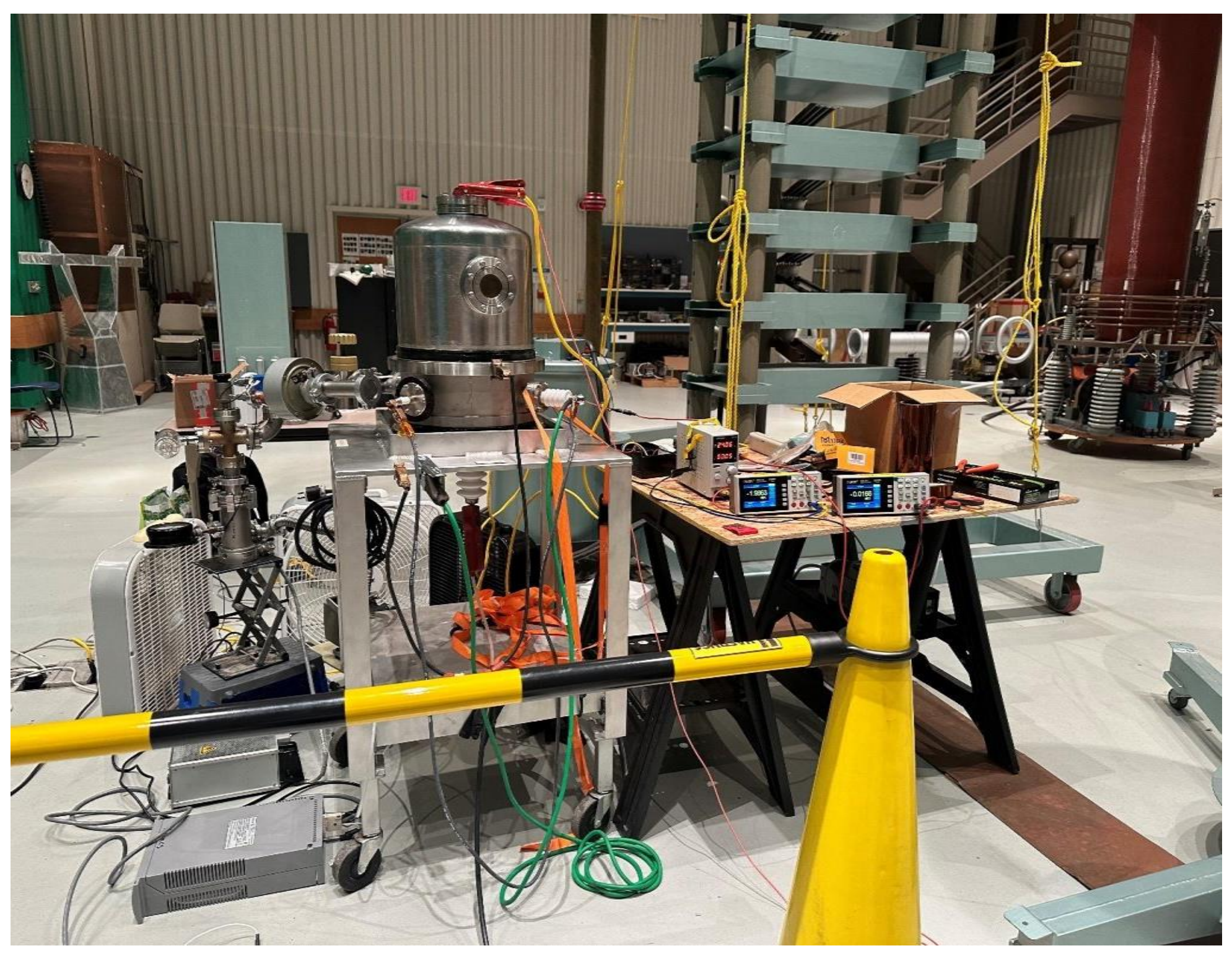

**Supplemental figure 4.5-1. High voltage vacuum chamber and, on right table, an isolated four-wire measurement circuit.**

## Supplemental Section 5.0—Tight-Binding Simulations

We had two parallel-effort tight-binding simulation campaigns, the first was to investigate the angular dependent MR and the second was to understand bundle transport. The first campaign primarily considered one individual chiral metallic CNT that was rotated in a static magnetic field. Non-armchair metallic CNTs have a small, few meV, bandgap caused by the curvature of a tube induced by rolling of the graphene plane, which can be modulated by <60 T magnetic fields applied parallel through the tube[21]. The second campaign primarily explored electronic transport through CNT bundles and measured the impact from (i) different bundle sizes, (ii) changes in metallicity and doping degree, and (iii) magnetic field, with the intent to select or refute justification for 2D and 3D weak localization models.

**Supplemental Section 5.1 Individual, non-armchair metallic CNT.** Calculation of the angle-dependent linear conductance in an individual CNT. The magnetic-field $\boldsymbol{B}$ modifies the bands crossing the charge-neutrality point in a CNT (Dirac bands) through two mechanisms, the Aharonov-Bohm (orbital) and the Zeeman (spin) effects. The first couples to the valley (K/K') degree of freedom through its associated transverse momentum with coupling strength $\mu_{orb}$, and the second couples to the spin projection on the magnetic-field direction, with the strength given by the Bohr magneton $\mu_B$ and the $g$ factor which in CNTs is 2. The Hamiltonian matrix describing the edges of the Dirac bands, i.e. the states with $k_z$=0, is

$$\mathrm{H}_{\pm}(B) = \pm \frac{E_{gap}}{2} \tau_0 \sigma_0 \pm \mu_{orb} B_z \tau_z \sigma_0 + \mu_B \tau_0 (B \cdot \sigma), \tag{1}$$

where +/- applies to the conduction/valence band, and $\tau$, $\sigma$ are the Pauli matrices acting in the valley and spin space, respectively. The lowest four eigenvalues $E_{-i}(\boldsymbol{B})$ of this Hamiltonian yield the shifted energies of the top of the valence bands, and the highest four $E_{+i}(\boldsymbol{B})$ those of the bottom of the conduction band. They mark the activation or deactivation of the transmission channel associated with each band within our window of energy, corresponding to changes in the doping-level, or in other words the Fermi level of the CNT (see the Supplemental Fig. 5.1-1 for illustration).

The linear conductance is then calculated within the Landauer-Büttiker approach, where

$$G(E_F, T, B) = \frac{e^2}{h} \int_{-\infty}^{\infty} d\,\epsilon \left( -\frac{\partial f_{FD}(\epsilon, E_F, k_B T)}{\partial \epsilon} \right) \Big|_{V_b = 0} T(\epsilon, B), \tag{2}$$

$$f_{FD}(\epsilon, E_F, k_B T) = \frac{1}{1 + exp\left( \frac{(\epsilon - E_F - V_b)}{k_B T} \right)}, \tag{3}$$

$V_b$ is the bias voltage, $T$ the temperature, $E_F$ the Fermi level, $e$ is electron charge, $h$ is Plank's constant, and $k_B$ is Boltzmann's constant. Our transmission $T(\varepsilon)$ for an infinitely long CNT is equal to the number of 1D conducting channels active at energy $\varepsilon$, which we split below into valence ($E_{-i}$) and conduction ($E_{+i}$) band contributions,

$$T(\epsilon, B) = \sum_{i=1}^{4} \theta\left( \epsilon - E_{+i}(B) \right) + \theta(E_{-i}(B) - \epsilon). \tag{4}$$

where Θ is the Heaviside step function. After performing the integral, we find that

$$G(E_F,T,B) = \frac{e^2/h}{2}\sum_{i=1}^{4}\left\{\left(1 - tanh\left(-\frac{E_F - E_{+i}(B)}{2k_BT}\right)\right) + \left(1 - tanh\left(\frac{E_F - E_{-i}(B)}{2k_BT}\right)\right)\right\}, \quad (5)$$

which is the quantity used for the calculation of the angle-dependent MR.

The figures below show how the curvature induced bandgap of a chiral metallic single wall CNT changes with parallel field (left most panel of 5.1-1). Next, it shows how the bandgap changes with orientation angle between the CNT and static field, due to its parallel component (right three panels of 5.1-1). Multiple temperatures are considered (5.1-2) as well as conductance (5.1-3). Chiral metallic CNTs dominate the transport through a de-doped bundle at cryogenic temperatures; gapless armchair CNTs are not numerous enough and semiconducting CNTs, while present, are frozen out.

The features due to the individual MR responses of the CNT are (5.1-4): a fourfold periodicity at charge neutrality point and magnetic fields slightly larger than needed to close the gap, and a twofold periodicity in the angle-dependent MR at other doping levels and higher field. The response of the whole fiber or ribbon would require averaging over the individual components of the network, and taking into account additional network-wide effects such as weak localization. The averaging alone would bring down the response of the system to a more reasonable order of magnitude. It would also on the one hand suppress the four-fold periodic component, which relies on the specific parameters of the CNT, while preserving the twofold periodic component which is due solely to its cylindrical topology and the Aharonov-Bohm effect. On the other hand, in a network, the weak localization effects can enhance the negative MR for perpendicular fields, strengthening the four-fold periodic component of the MR. Estimating from the magnitude of the experimentally measured weak localization, we expect this enhancement to be significantly weaker than the individual orbital response of the CNTs.

In our chiral metallic CNT (5.1-4), the four-fold periodicity is strongest at $H$ = 4.5 T and decreases at $H$ = 9 T, as in the experimental results shown in Supplementary Figures 4.3-4 and 4.3-6 for the de-doped fiber. The strong fourfold periodic component indicates that the distribution of diameters and chiralities in the fiber is narrow and the doping level is consistently near the charge neutrality point, so that the fourfold component survives the averaging.

The weakness of the four-fold periodic component in the ribbon can be explained e.g. by their wider distribution of diameters and chiralities, or less uniform doping levels.

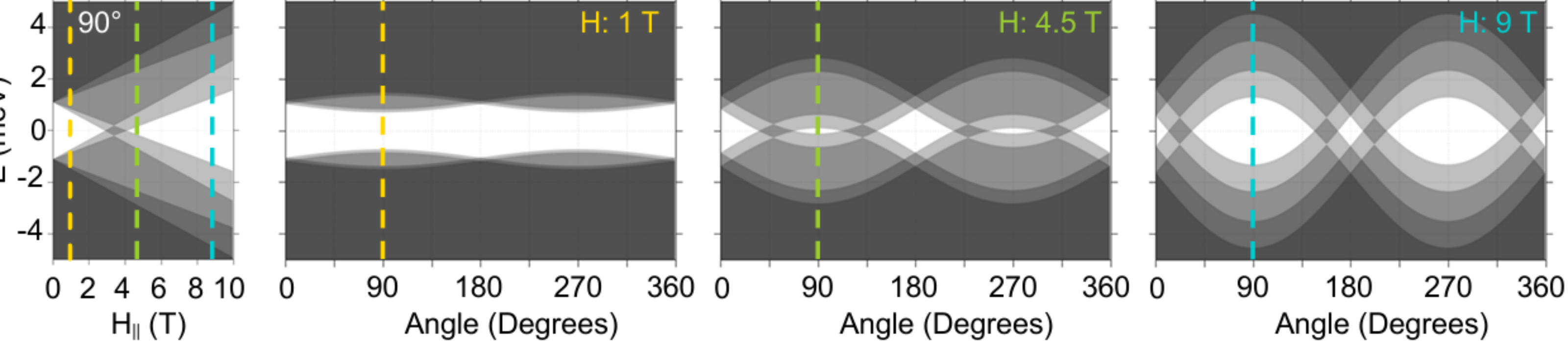

**Supplemental figure 5.1-1. Left—Small, curvature induced bandgap of an individual non-armchair metallic single-wall CNT versus applied parallel magnetic field, with temperature at 0 K. As depicted, darker shades indicate the simulated linear conductance in units of $G=e^2/h$. Yellow (1 T), green (4.5 T), and blue (9 T) slices are in the adjacent plots showing how changing the orientation angle of the CNT with respect to the constant field changes the bandgap. 0° is magnetic field perpendicular to the CNT alignment and probe current; 90° is parallel where maximum magnetic flux threads through the CNT.**

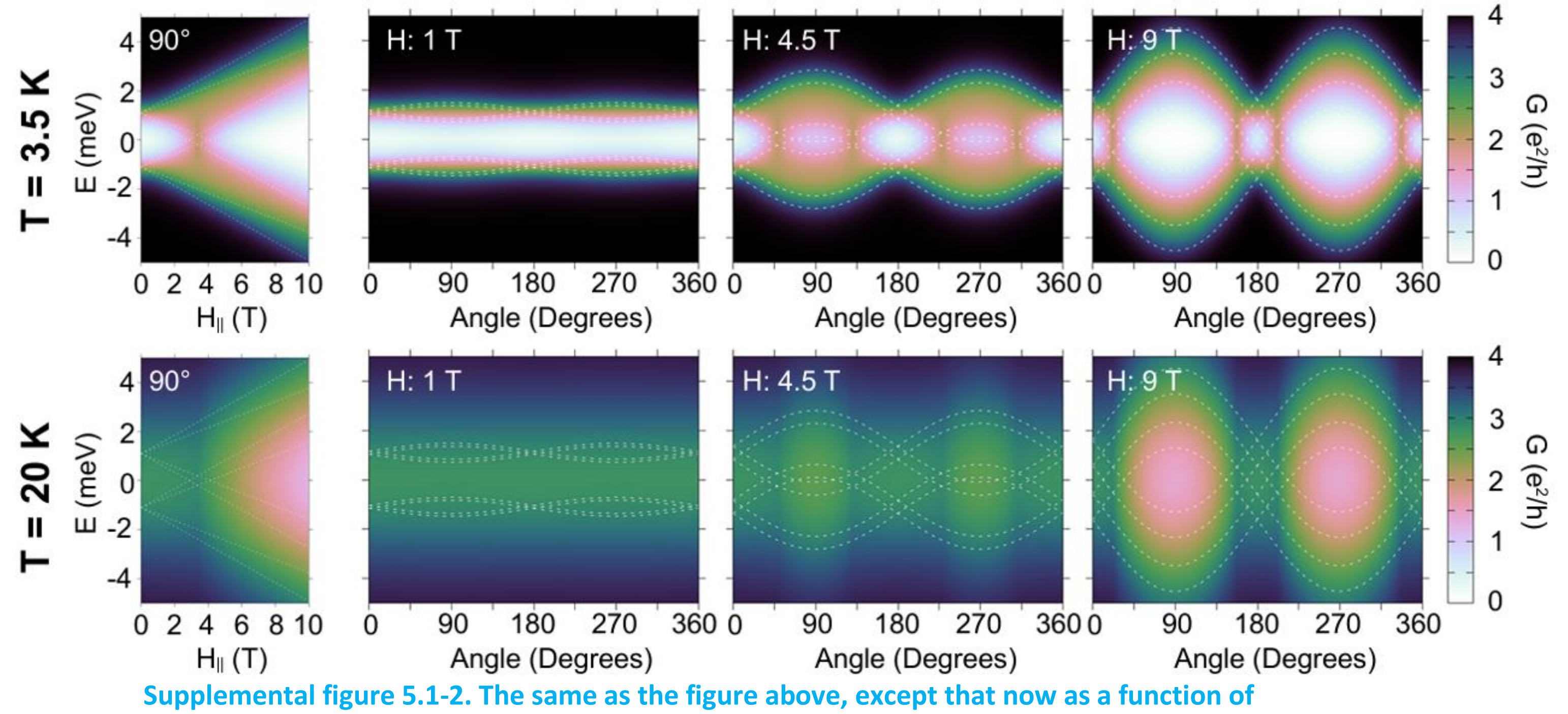

**Supplemental figure 5.1-2. The same as the figure above, except that now as a function of temperature. Top—3.5 K and bottom—20 K. Color indicates the simulated quantum conductance in units of $G=e^2/h$. White dashed line represents band edges.**

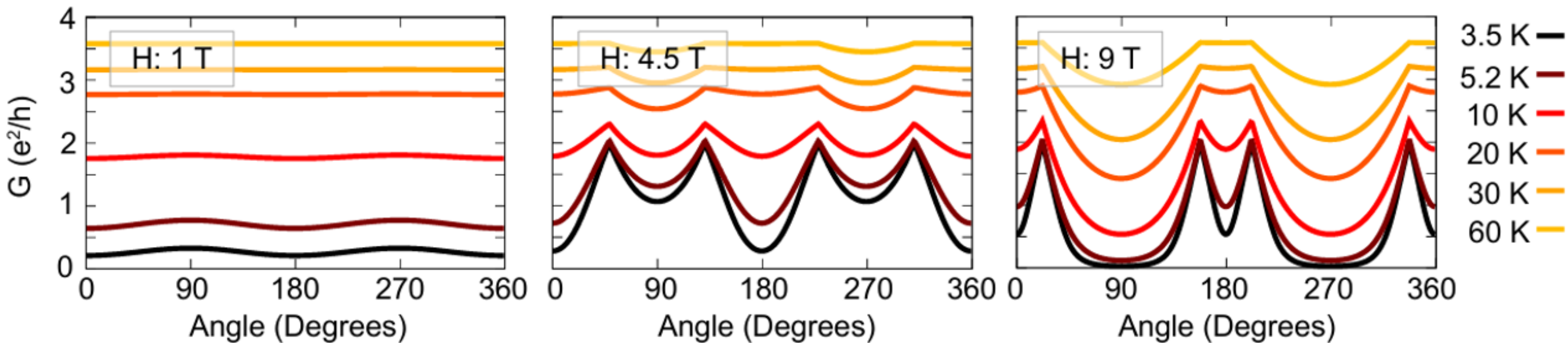

**Supplemental figure 5.1-3. Taking a slice at the Fermi-level $E_f$=0, simulated conductance versus orientation angle for three different magnetic fields and across multiple temperatures, as depicted.**

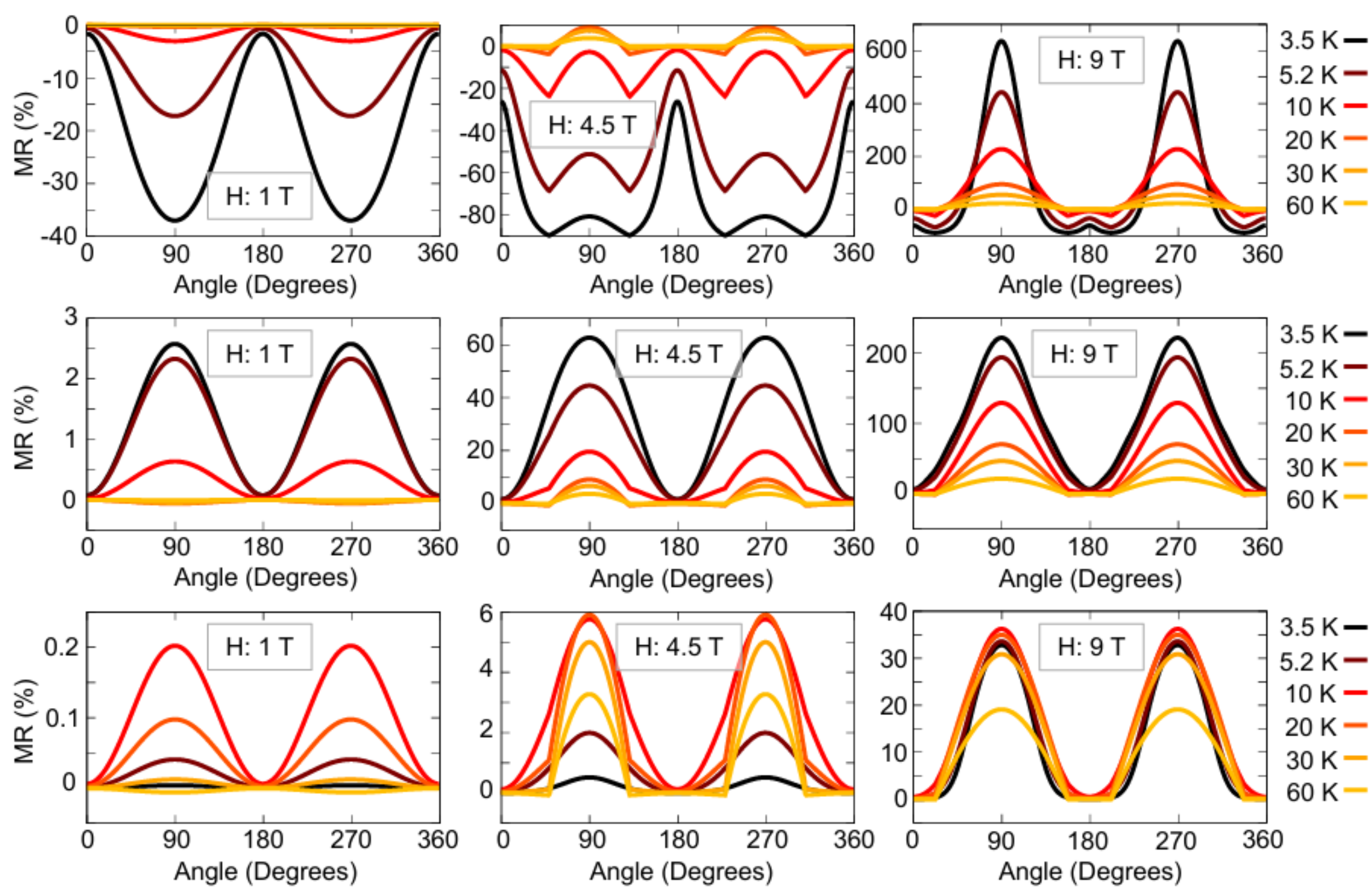

**Supplemental figure 5.1-4. Simulated MR (%) as a function of orientation angle, for multiple temperatures (depicted by color), across three different magnetic fields (horizontal) and three different positions of the Fermi level (vertical). The values of the Fermi level from top to bottom are 0, -1 meV, -2 meV. The four-fold periodic component is strongest when the CNT is at the charge neutrality point and at 4.5 T. For this field, it is weakened but survives small doping; at 1 T and 9 T in doped CNT the two-fold periodicity dominates.**

Averaging over an ensemble. In figure 5.1.5 below, we show the results of numerical simulations of a set of 24 uncoupled CNTs in parallel, with random curvature-induced band gaps and diameters. We set up four such ensembles, named by the ranges from which their gaps and diameters are sampled. Gaps are sampled from a wider range of 2-10 meV (L) or a narrow range of 2-4 meV (S). The diameters are sampled from a wider 1-2 nm (l) range or a narrower one 1.2-1.8 nm (s). In figure 5.1.5a we see the linear conductance at T=0.1 K (very close to the quantum mechanical transmission) of an "Ll" ensemble, in figure 5.1.5b that of an "Sl" ensemble. In both we can see that the gap closing does not occur at a specific value of the magnetic field, but is instead spread out. The reopening of the band gap is also clearly visible, far beyond the average gap at zero field. In figure 5.1.5c we show the MR curves for all four ensembles, with profiles similar to that in figure 2c in the main text. Note that the data in figure 2 in the main text are acquired from ribbons, which show a weaker fourfold periodic component in angle-resolved MR, and seem to have more disparate chiralities and diameters. This effect could therefore be an important contributor to the room temperature positive MR, most likely combined with the classical MR mechanism.

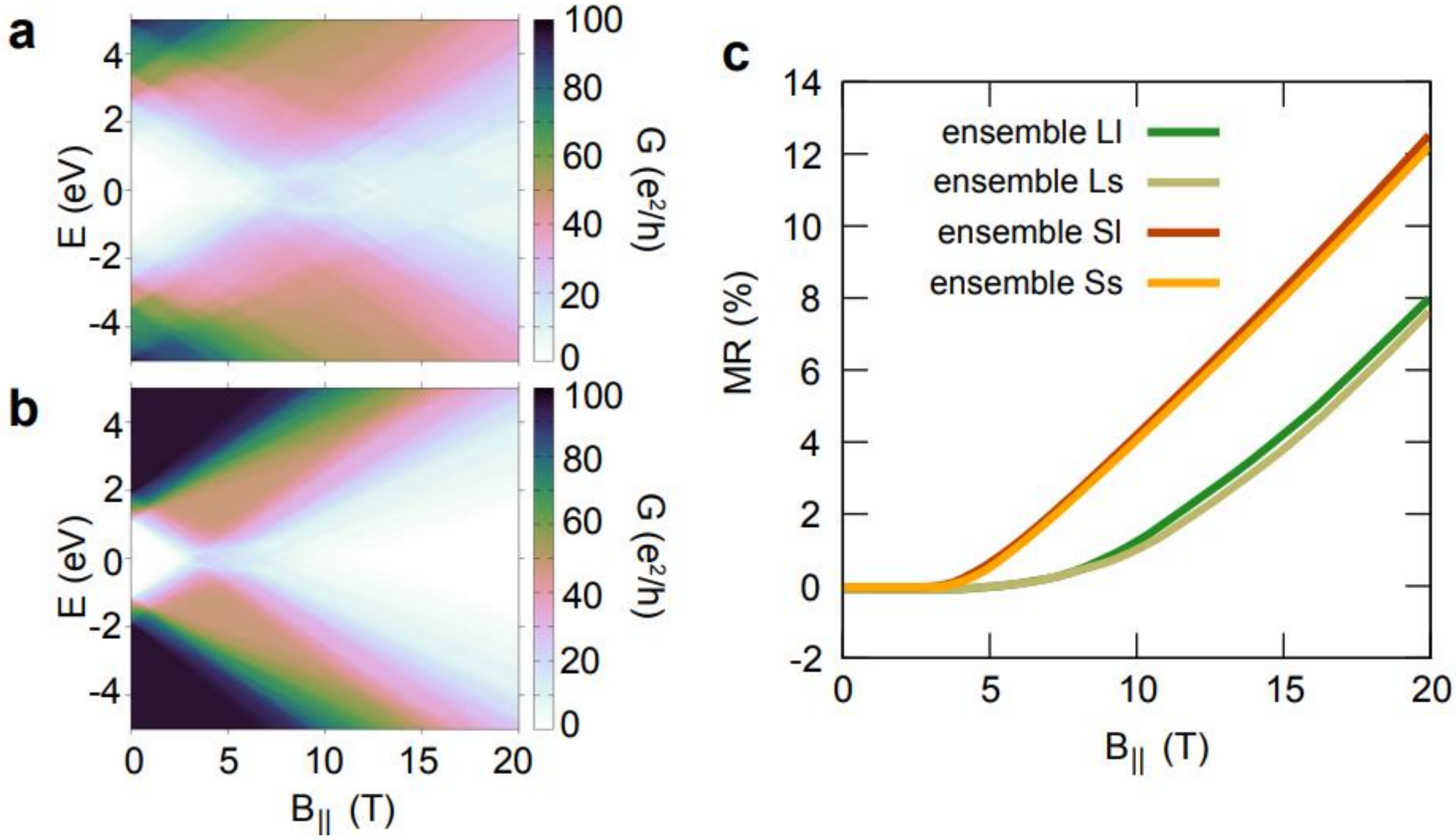

**Supplemental figure 5.1-5. Gradual closing and reopening of the band gap in a set of 24 uncoupled parallel CNTs with a random distribution of curvature induced gaps and diameters between 1 and 2 nm. a, linear conductance at T = 0.1 K of 24 CNTs with gaps from 2 to 10 meV; b, with gaps from 2 to 4 meV. c, the MR at T = 270 K for all four investigated sets of CNTs (details in the text).**

**Supplemental section 5.2 TB-NEGF modelling.** The spin-unpolarized tight-binding calculations of CNT bundles and junctions were performed using sisl python library[22] and TBtrans[23], a tight-binding code integrating the non-equilibrium Green's function formalism, with magnetic-field included in the Hamiltonian by Peierls substitution. In all structures, the carbon–carbon bond length was set to 1.42 Å. CNTs forming bundles and junctions composed of identical CNTs were placed at an inter-tube distance of 3.356 Å, while for mixed systems this spacing varied by ±0.1 Å. This value is based on the experimentally determined interlayer distance in graphite[24, 25].

CNT bundles and junctions were modelled as two-probe systems, with a central scattering region (C) positioned between semi-infinite left (L) and right (R) electrodes (figures 5a and 5c). Each electrode consisted of the same type of CNT as the one bridging the central region. Electrodes were typically one unit cell long, except for metallic (9,9) CNTs, where two units were used. The total length of the device—including electrodes and the central region—depended on the specific configuration. For metallic (9,9) bundles, the length was 11.31 nm; for semiconducting and mixed bundles composed of (n,0) nanotubes, including semiconducting (17,0), (19,0), (20,0) and metallic (18,0), (21,0) SWCNTs, the length was 11.08 nm. Bundles composed of metallic chiral (12,3) SWCNTs had a length of 11.71 nm. In the case of junctions of CNT bundles, the overlap region was 10.33 nm long.

To model the electronic structure of these systems, we used the tight-binding Hamiltonian with interactions up to third nearest neighbours[26, 27]:

$$\mathrm{H} = 2\sum_{ij} \left(t_{ij}\, a_i^{\dagger} a_j + h.c.\right) + U \tag{6}$$

where the first term is hopping with $a_{i\sigma}^{\dagger}$ $(a_{i\sigma})$ being creation (annihilation) operators, and $t_{ij}$ being a hopping integral from site $j$ to $i$. Hopping parameters are[26]: first nearest neighbor (distance from the initial atom smaller than 1.5 Å) $t_1 = -2.97$ eV, second nearest neighbor (distance between 1.5 Å and 2.5 Å) $t_2 = -0.073$ eV, third nearest neighbor (distance between 2.5 Å and 3.7 Å) $t_3 = -0.33$ eV. In this model if the distance between atoms is bigger than 3.7 Å they do not interact, and the interaction of atoms from different nanotubes is of a third nearest neighbor type. *U* is a constant that shifts energy so the Fermi level is at 0 eV.

In our calculations we use non-orthogonal atomic orbital basis set, so the overlap matrix $S$ is not an identity matrix. $S$ elements between basis states centered on different lattice sites written in local basis states, $|i>$, are $S_{ij} = <i|j>$, with values being [26]: first nearest neighbor $s_1 = 0.073$, second nearest neighbor $s_2 = 0.018$, third nearest neighbor $s_3 = 0.026$.

Both the Hamiltonian $H$ and the overlap matrix $S$ change when the magnetic-field is applied. Peierls substitution[27, 28, 29] allows us to include the magnetic-field as a phase factor that multiplies $S$ and the hopping term of $H$:

$$-\sum_{ij} t_{ij} a_i^{\dagger} a_j \rightarrow -\sum_{ij} t_{ij} a_i^{\dagger} \left(e^{i\frac{2\pi}{\Phi_0}\int_{r_j}^{r_i} \underline{A}\cdot\underline{dr}}\right) a_j \tag{7}$$

$$S = \sum_{ij} \quad S_{ij} \rightarrow S = \sum_{ij} \quad S_{ij}(e^{i\frac{2\pi}{\phi_0}\int_{\underline{r_j}}^{\underline{r_i}} \quad \underline{A}\cdot\underline{dr}}) \tag{8}$$

$\underline{r_i} = (x_i, y_i, z_i)$ are $i^{th}$ site coordinates, $\underline{A} = (-yB_z, 0, yB_x)$ is the magnetic vector potential defined up to the gauge that was chosen in such a way to ensure translational invariance in $z$ direction, and which gives:

$$i\int_{\underline{r_j}}^{r_i} \quad \underline{A}\cdot\underline{dr} = i/2\cdot(-B_z\Delta x + B_x\Delta z)(y_i + y_j) \tag{9}$$

For perpendicular magnetic field $B_z = 0$, while for parallel $B_x = 0$. As the inclusion of the Peierls substitution renders the Hamiltonian and overlap matrices complex, we modified version 0.15.1 of the open-source sisl Python library ourselves, as it did not originally support complex matrices. After defining the structure with its corresponding $H$ and $S$ matrices, we performed transport calculations.

For transport calculations, the energy window was set between −1.5 eV and 1.5 eV with a step of 0.001 eV. Since none of the structures were periodic, a 1×1×1 k-point grid was used, and the electronic temperature was set to the default value of 300 K. TBtrans transport code uses non-equilibrium Green's function method [30] [23], where the Green's function is computed by combining both Hamiltonian and overlap matrices. The Green function $G$ can be expressed in terms of $S$, $H$, and self-energy $\Sigma$ as:

$$G^{-1}(E) = S(E + i\eta) - H - \delta H - \sum_i \quad \Sigma_i - \delta\Sigma \tag{10}$$

While the open-source version of TBtrans permits user intervention, it is limited to the central region of the device's Hamiltonian. However, as we required the inclusion of the Peierls substitution in the Hamiltonians and overlap matrices of the electrodes as well as the entire device, we developed our own modified version of TBtrans—based on version 5.2.0-alpha of SIESTA code [31] [32]—to enable support for complex *H* and *S* matrices. As a result of the calculations we obtained transmission of an electron of energy $E$ between two different electrodes *L* and *R*, which was computed as:

$$T_{L\rightarrow R}(E) = Tr[\Gamma_R(E)G(E)\Gamma_L(E)G^\dagger(E)] \tag{11}$$

Here $\Gamma_L$ is a scattering matrix from *L*, and $E$ also corresponds to the chosen doping level. Apart from the transmission, we calculated bond currents, which are local currents flowing between two orbitals $\nu$ and $\mu$, and are associated with outgoing states from electrode *L*:

$$\tau_{L,\nu\mu} = \frac{e}{h} Im[A_{L,\nu\mu}(H_{\mu\nu} - ES_{\mu\nu}) - A_{L,\mu\nu}(H_{\nu\mu} - ES_{\nu\mu})] \tag{12}$$

$A_{L,\nu\mu}$ is a spectral function originating from *L*. To analyze the distribution of electronic-transport across individual CNTs within the bundle, we define the CNT-resolved transmission contribution, denoted as $\tau_{tube}$. This quantity is obtained by summing the bond currents—local currents flowing between pairs of orbitals—associated with each carbon atom in a specific CNT. For a given CNT, the CNT-resolved transmission contribution $\tau_{tube}$ is calculated as:

$$\tau_{tube} = \sum_{\nu\in tube} \quad \sum_\mu \quad \tau_{\nu\mu} \tag{13}$$

Here, $\tau_{\nu\mu}$ represents the bond current between orbitals $\nu$ and $\mu$, with $\nu$ belonging to chosen CNT and the summation over $\mu$ includes all orbitals connected to $\nu$. This approach allows us to quantify the individual contribution of each nanotube to the overall electron transport in the bundle. Note that inelastic processes, such as electron–phonon coupling, are not included in the TB-NEGF calculations.

Discrete perpendicular magnetic fields (applied along the x-axis, as depicted in supplemental figure 5.2-4 and table 5.2-1) of 4.5 T, 9 T, 30 T, and 60 T were applied to the smallest bundle composed of metallic (9,9) SWCNTs (supplemental figure 5.2-4e), as well as to the mixed junction comprising metallic (18,0) and semiconducting (19,0) and (17,0) CNTs (supplemental figure 5.2-4f). Additionally, the response of the (9,9) bundle was studied under magnetic-fields applied along the z-direction (parallel) and at 45°, as illustrated in supplemental figure 5.2-5.

Geometry optimization of nanotube bundles was carried out using the LAMMPS code [33]. The interactions between carbon nanotubes were described using the AIREBO potential [34] with the cutoff radius for the Lennard-Jones term of 17 Å. To avoid interaction of periodic images of atoms, the supercell of 1x1x8 was considered for structural relaxation with a fixed size of the bundle unit cell and 1x1x9 in the case when the length of the unit cell along the bundle axis was also optimized. The structure optimization was performed using the Polak-Ribière version [35] of the conjugate gradient algorithm [36]. The calculation was performed till the energy change in successive iterations divided by the energy magnitude became less than $10^{-15}$ or forces on all atoms got smaller than $10^{-15}$ eV/Å.

The effects of CNT arrangement, structural optimization, and magnetic field orientation on transmission and magnetoresistance in SWCNT Bundles are shown below.

**Supplemental Table 5.2-1 Tube-resolved transmission contributions $\tau_{tube}$ at selected energies (0.0 eV, −0.1 eV, −0.6 eV) for all studied systems, including relaxed and unrelaxed configurations, with and without applied magnetic fields. Values are given in units of $G_0$; numbers in brackets indicate the percentage contribution to the total transmission at the specified energy. The sum of tube-resolved contributions over all nanotubes in the bundle (e.g., $\tau_{tube0-6}$ for small bundles or $\tau_{tube0-18}$ for larger ones) corresponds to the total transmission. For larger bundles, additional grouped contributions are shown, including the middle layer ($\tau_{tube1-6}$), all outer-layer tubes ($\tau_{tube7-18}$), outer-layer tubes closer to the bundle center ($\tau_{tube7-12}$), and those further away ($\tau_{tube13-18}$). A negative sign indicates only the opposite direction of transmission and does not imply negative conductance. Cross-sectional views of the systems are shown in the first column. Metallic nanotubes are numbered using darker colors, semiconducting ones with lighter colors, and each chirality is represented by a distinct color.**

| | **tube-resolved transmission contribution ($\tau_{tube}$)** | | |
|---|---|---|---|
| **System** | **$\tau_{tube}$ [$G_0$] at 0.0 eV** | **$\tau_{tube}$ [$G_0$] at -0.1 eV** | **$\tau_{tube}$ [$G_0$] at -0.6 eV** |
| (9,9)<br>3 2<br>4 0 1<br>5 6<br>metallic | $\tau_{tube0}$=2.590x$10^{-4}$ [5.63%]<br>$\tau_{tube1}$=7.236x$10^{-4}$ [15.73%]<br>$\tau_{tube2}$=7.232x$10^{-4}$ [15.72%]<br>$\tau_{tube3}$=7.236x$10^{-4}$ [15.73%]<br>$\tau_{tube4}$=7.232x$10^{-4}$ [15.72%]<br>$\tau_{tube5}$=7.236x$10^{-4}$ [15.73%]<br>$\tau_{tube6}$=7.232x$10^{-4}$ [15.72%]<br>$\tau_{tube0-6}$=45.995x$10^{-4}$ [100%]<br>lower τ inside | $\tau_{tube0}$=1.370 [11.42%]<br>$\tau_{tube1}$=1.772 [14.77%]<br>$\tau_{tube2}$=1.771 [14.76%]<br>$\tau_{tube3}$=1.772 [14.77%]<br>$\tau_{tube4}$=1.771 [14.76%]<br>$\tau_{tube5}$=1.772 [14.77%]<br>$\tau_{tube6}$=1.771 [14.76%]<br>$\tau_{tube0-6}$=12.000 [100%]<br>lower τ inside | $\tau_{tube0}$=2.004 [14.31%]<br>$\tau_{tube1}$=1.998 [14.27%]<br>$\tau_{tube2}$=2.001 [14.29%]<br>$\tau_{tube3}$=1.998 [14.27%]<br>$\tau_{tube4}$=2.001 [14.29%]<br>$\tau_{tube5}$=1.998 [14.27%]<br>$\tau_{tube6}$=2.001 [14.29%]<br>$\tau_{tube0-6}$=14.000 [100%]<br>similar τ values |
| (21,0)<br>3 2<br>4 0 1<br>5 6<br>metallic | $\tau_{tube0}$=1.199 [17.71%]<br>$\tau_{tube1}$=0.926 [13.69%]<br>$\tau_{tube2}$=0.930 [13.74%]<br>$\tau_{tube3}$=0.926 [13.69%]<br>$\tau_{tube4}$=0.930 [13.74%]<br>$\tau_{tube5}$=0.926 [13.69%]<br>$\tau_{tube6}$=0.930 [13.74%]<br>$\tau_{tube0-6}$=6.767 [100%]<br>larger τ inside | $\tau_{tube0}$=1.250 [11.37%]<br>$\tau_{tube1}$=1.626 [14.78%]<br>$\tau_{tube2}$=1.624 [14.76%]<br>$\tau_{tube3}$=1.626 [14.78%]<br>$\tau_{tube4}$=1.624 [14.76%]<br>$\tau_{tube5}$=1.626 [14.78%]<br>$\tau_{tube6}$=1.624 [14.76%]<br>$\tau_{tube0-6}$=11.001 [100%]<br>lower τ inside | $\tau_{tube0}$=3.022 [14.39%]<br>$\tau_{tube1}$=3.002 [14.30%]<br>$\tau_{tube2}$=2.990 [14.24%]<br>$\tau_{tube3}$=3.002 [14.30%]<br>$\tau_{tube4}$=2.990 [14.24%]<br>$\tau_{tube5}$=3.002 [14.30%]<br>$\tau_{tube6}$=2.990 [14.24%]<br>$\tau_{tube0-6}$=20.999 [100%]<br>similar τ values |
| (12,3)<br>3 2<br>4 0 1<br>5 6<br>metallic | $\tau_{tube0}$=0.017 [0.33 %]<br>$\tau_{tube1}$=0.817 [16.49%]<br>$\tau_{tube2}$=0.829 [16.73%]<br>$\tau_{tube3}$=0.817 [16.49%]<br>$\tau_{tube4}$=0.829 [16.73%]<br>$\tau_{tube5}$=0.817 [16.49%]<br>$\tau_{tube6}$=0.829 [16.73%]<br>$\tau_{tube0-6}$=4.955 [100%]<br>lower τ inside | $\tau_{tube0}$=2.000 [14.29%]<br>$\tau_{tube1}$=1.999 [14.28%]<br>$\tau_{tube2}$=2.001 [14.29%]<br>$\tau_{tube3}$=1.999 [14.28%]<br>$\tau_{tube4}$=2.001 [14.29%]<br>$\tau_{tube5}$=1.999 [14.28%]<br>$\tau_{tube6}$=2.001 [14.29%]<br>$\tau_{tube0-6}$=14.000 [100%]<br>similar τ values | $\tau_{tube0}$=2.000 [14.29%]<br>$\tau_{tube1}$=1.999 [14.28%]<br>$\tau_{tube2}$=2.001 [14.29%]<br>$\tau_{tube3}$=1.999 [14.28%]<br>$\tau_{tube4}$=2.001 [14.29%]<br>$\tau_{tube5}$=1.999 [14.28%]<br>$\tau_{tube6}$=2.001 [14.29%]<br>$\tau_{tube0-6}$=14.000 [100%]<br>similar τ values |
| (20,0)<br>3 2<br>4 0 1<br>5 6<br>semiconducting | inside band gap | inside band gap | $\tau_{tube0}$=4.008 [14.31%]<br>$\tau_{tube1}$=3.997 [14.28%]<br>$\tau_{tube2}$=3.999 [14.28%]<br>$\tau_{tube3}$=3.997 [14.28%]<br>$\tau_{tube4}$=3.999 [14.28%]<br>$\tau_{tube5}$=3.997 [14.28%]<br>$\tau_{tube6}$=3.999 [14.28%]<br>$\tau_{tube0-6}$=28.000 [100%]<br>similar τ values |
| 3 (17,0) 2 (19,0)<br>4 (18,0) 0 (18,0) 1 (18,0)<br>5 (19,0) 6 (17,0)<br>mixed<br>(3 metallic &<br>4 semiconducting) | $\tau_{tube0}$=1.462 [29.89%]<br>$\tau_{tube1}$=1.707 [34.88%]<br>$\tau_{tube2}$=-0.018 [0.36%]<br>$\tau_{tube3}$=0.001 [0.03%]<br>$\tau_{tube4}$=1.743 [35.63%]<br>$\tau_{tube5}$=-0.001 [0.02%]<br>$\tau_{tube6}$=-0.003 [0.06%]<br>$\tau_{tube0-6}$=4.892 [100%]<br>lower τ inside | $\tau_{tube0}$=1.933 [32.22%]<br>$\tau_{tube1}$=1.959 [32.64%]<br>$\tau_{tube2}$=0.066 [1.10%]<br>$\tau_{tube3}$=0.009 [0.14%]<br>$\tau_{tube4}$=1.962 [32.70%]<br>$\tau_{tube5}$=0.061 [1.01%]<br>$\tau_{tube6}$=0.011 [0.19%]<br>$\tau_{tube0-6}$=6.000 [100%]<br>lower τ inside | $\tau_{tube0}$=2.626 [11.42%]<br>$\tau_{tube1}$=2.348 [10.21%]<br>$\tau_{tube2}$=4.002 [17.40%]<br>$\tau_{tube3}$=3.848 [16.73%<br>$\tau_{tube4}$=2.320 [10.09%]<br>$\tau_{tube5}$=4.012 [17.44%]<br>$\tau_{tube6}$=3.843 [16.71%]<br>$\tau_{tube0-6}$=23.000 [100%]<br>similar τ values |
| 3 (19,0) 2 (20,0)<br>4 (21,0) 0 (20,0) 1 (21,0)<br>5 (20,0) 6 (19,0)<br>mixed<br>(2 metallic &<br>5 semiconducting) | $\tau_{tube0}$=-0.015 [0.86%]<br>$\tau_{tube1}$=0.678 [39.85%]<br>$\tau_{tube2}$=-0.015 [0.86%]<br>$\tau_{tube3}$=-0.015 [0.89%]<br>$\tau_{tube4}$=1.064 [62.49%]<br>$\tau_{tube5}$=0.017 [1.01%]<br>$\tau_{tube6}$=-0.012 [0.73%]<br>$\tau_{tube0-6}$=1.703 [100%]<br>transmission dominated by metallic tube 4 | $\tau_{tube0}$=0.033 [0.82%]<br>$\tau_{tube1}$=1.991 [49.78%]<br>$\tau_{tube2}=-2.357\times10^{-4}$ [0.01%]<br>$\tau_{tube3}=-64.875\times10^{-4}$[0.16%]<br>$\tau_{tube4}$=1.986 [49.66%]<br>$\tau_{tube5}=78.102\times10^{-4}$[0.20%]<br>$\tau_{tube6}$=-0.012 [0.29%<br>$\tau_{tube0-6}$=4.000 [100%]<br>transmission dominated by metallic tubes | $\tau_{tube0}$=4.023 [14.37%]<br>$\tau_{tube1}$=3.986 [14.24%]<br>$\tau_{tube2}$=3.963 [14.16%]<br>$\tau_{tube3}$=4.004 [14.30%]<br>$\tau_{tube4}$=4.034 [14.41%]<br>$\tau_{tube5}$=3.996 [14.27%]<br>$\tau_{tube6}$=3.991 [14.25%]<br>$\tau_{tube0-6}$=27.997 [100%]<br>similar τ values |

| 3 (20,0) 2 (19,0)<br>4 (21,0) 0 (20,0) 1 (21,0)<br>5 (20,0) 6 (19,0)<br>mixed<br>(2 metallic &<br>5 semiconducting) | $\tau_{tube0}$=-0.015 [1.55%]<br>$\tau_{tube1}$=0.230 [23.22%]<br>$\tau_{tube2}$=-0.005 [0.50%]<br>$\tau_{tube3}$=-0.001 [0.09%]<br>$\tau_{tube4}$=0.786 [79.51%]<br>$\tau_{tube5}$=-0.001 [0.09%]<br>$\tau_{tube6}$=-0.005 [0.50%]<br>$\tau_{tube0\text{-}6}$=0.989 [100%]<br>transmission dominated by metallic tube 4 | $\tau_{tube0}$=0.022 [0.55%]<br>$\tau_{tube1}$=2.013 [50.32%]<br>$\tau_{tube2}$=-0.013 [0.32%]<br>$\tau_{tube3}$=0.008 [0.20%]<br>$\tau_{tube4}$=1.975 [49.39%]<br>$\tau_{tube5}$=0.008 [0.20%<br>$\tau_{tube6}$=-0.013 [0.32%]<br>$\tau_{tube0\text{-}6}$=4.000 [100%]<br>transmission dominated by metallic tubes | $\tau_{tube0}$=4.024 [14.38%]<br>$\tau_{tube1}$=3.997 [14.29%]<br>$\tau_{tube2}$=3.995 [14.28%]<br>$\tau_{tube3}$=3.996 [14.28%]<br>$\tau_{tube4}$=3.978 [14.22%]<br>$\tau_{tube5}$=3.996 [14.28%]<br>$\tau_{tube6}$=3.995 [14.28%]<br>$\tau_{tube0\text{-}6}$=27.980 [100%]<br>similar τ values |
|---|---|---|---|
| (9,9)<br>15 7 14<br>8 3 2 12<br>16 4 0 1 13<br>9 5 6 11<br>17 10 18<br>metallic | $\tau_{tube0}$=0.001 [0.25%]<br>$\tau_{tube1}$=0.024 [7.17%]<br>$\tau_{tube2}$=0.024 [7.17%]<br>$\tau_{tube3}$=0.024 [7.17%]<br>$\tau_{tube4}$=0.024 [7.17%]<br>$\tau_{tube5}$=0.024 [7.17%]<br>$\tau_{tube6}$=0.024 [7.17%]<br>$\tau_{tube7}$=0.011 [3.25%]<br>$\tau_{tube8}$=0.011 [3.27%]<br>$\tau_{tube9}$=0.011 [3.25%]<br>$\tau_{tube10}$=0.011 [3.27%]<br>$\tau_{tube11}$=0.011 [3.25%]<br>$\tau_{tube12}$=0.011 [3.27%]<br>$\tau_{tube13}$=0.021 [6.20%]<br>$\tau_{tube14}$=0.021 [6.20%]<br>$\tau_{tube15}$=0.021 [6.20%]<br>$\tau_{tube16}$=0.021 [6.20%]<br>$\tau_{tube17}$=0.021 [6.20%]<br>$\tau_{tube18}$=0.021 [6.20%]<br>$\tau_{tube0\text{-}18}$=0.337 [100%]<br>$\tau_{tube1-6}$=0.145 [43.02%]<br>$\tau_{tube7-12}$=0.066 [19.55%]<br>$\tau_{tube13-18}$=0.125 [37.18%]<br>$\tau_{tube8-18}$=0.191 [56.73%]<br>lower τ inside | $\tau_{tube0}$=1.708 [5.34%]<br>$\tau_{tube0}$=1.534 [4.79%]<br>$\tau_{tube1}$=1.534 [4.79%]<br>$\tau_{tube2}$=1.534 [4.79%]<br>$\tau_{tube3}$=1.534 [4.79%]<br>$\tau_{tube4}$=1.534 [4.79%]<br>$\tau_{tube5}$=1.534 [4.79%]<br>$\tau_{tube6}$=1.710 [5.34%]<br>$\tau_{tube7}$=1.709 [5.34%]<br>$\tau_{tube8}$=1.710 [5.34%]<br>$\tau_{tube9}$=1.709 [5.34%]<br>$\tau_{tube10}$=1.710 [5.34%]<br>$\tau_{tube11}$=1.709 [5.34%]<br>$\tau_{tube12}$=1.805 [5.64%]<br>$\tau_{tube13}$=1.805 [5.64%]<br>$\tau_{tube14}$=1.805 [5.64%]<br>$\tau_{tube15}$=1.805 [5.64%]<br>$\tau_{tube16}$=1.805 [5.64%]<br>$\tau_{tube17}$=1.805 [5.64%]<br>$\tau_{tube0\text{-}18}$=32.000 [100%]<br>$\tau_{tube1-6}$=9.205 [28.77%]<br>$\tau_{tube7-12}$=10.26 [32.05%]<br>$\tau_{tube13-18}$=10.829 [33.84%]<br>$\tau_{tube8-18}$=21.087 [65.90%]<br>lower τ inside | $\tau_{tube0}$=2.000 [6.25%]<br>$\tau_{tube1}$=2.002 [6.25%]<br>$\tau_{tube2}$=2.001 [6.25%]<br>$\tau_{tube3}$=2.002 [6.25%]<br>$\tau_{tube4}$=2.001 [6.25%]<br>$\tau_{tube5}$=2.002 [6.25%]<br>$\tau_{tube6}$=2.001 [6.25%]<br>$\tau_{tube7}$=2.000 [6.25%]<br>$\tau_{tube8}$=2.000 [6.25%]<br>$\tau_{tube9}$=2.000 [6.25%]<br>$\tau_{tube10}$=2.000 [6.25%]<br>$\tau_{tube11}$=2.000 [6.25%]<br>$\tau_{tube12}$=2.000 [6.25%]<br>$\tau_{tube13}$=1.997 [6.24%]<br>$\tau_{tube14}$=2.000 [6.25%]<br>$\tau_{tube15}$=1.997 [6.24%]<br>$\tau_{tube16}$=2.000 [6.25%]<br>$\tau_{tube17}$=1.997 [6.24%]<br>$\tau_{tube18}$=2.000 [6.25%]<br>$\tau_{tube0\text{-}18}$=38.000 [100%]<br>$\tau_{tube1-6}$=12.009 [37.53%]<br>$\tau_{tube7-12}$=10.26 [37.50%]<br>$\tau_{tube13-18}$=10.829 [37.47%]<br>$\tau_{tube8-18}$=21.087 [74.97%]<br>similar τ values |
| (9,9)<br>10 9<br>11 7 8<br>12 13<br>3 2<br>4 0 1<br>5 6<br>metallic | $\tau_{tube0}=2.131\times10^{-5}$ [3.22%]<br>$\tau_{tube1}=3.728\times10^{-5}$ [5.63%]<br>$\tau_{tube2}=1.982\times10^{-4}$ [29.91%]<br>$\tau_{tube3}=1.912\times10^{-4}$ [28.85%]<br>$\tau_{tube4}=4.190\times10^{-5}$ [6.32%]<br>$\tau_{tube5}=8.778\times10^{-5}$ [13.25%]<br>$\tau_{tube6}=8.496\times10^{-5}$ [12.82%]<br>$\tau_{tube7}=-2.131\times10^{-5}$ [3.22%]<br>$\tau_{tube8}=-4.190\times10^{-5}$ [6.32%]<br>$\tau_{tube9}=-8.778\times10^{-5}$ [13.25%]<br>$\tau_{tube10}=-8.496\times10^{-5}$ [12.82%]<br>$\tau_{tube11}=-3.728\times10^{-5}$ [5.63%]<br>$\tau_{tube12}=-1.982\times10^{-4}$ [29.91%]<br>$\tau_{tube13}=-1.912\times10^{-4}$ [3.22%]<br>$\tau_{tube0-6}=6.626\times10^{-4}$ [100%]<br>$\tau_{tube7-13}=-6.626\times10^{-4}$ [100%]<br>lower τ inside, transmission dominated by tubes 2, 3, 12, 13 | $\tau_{tube0}=0.034$ [2.59%]<br>$\tau_{tube1}=0.080$ [6.18%]<br>$\tau_{tube2}=0.454$ [35.06%]<br>$\tau_{tube3}=0.480$ [37.10%]<br>$\tau_{tube4}=0.101$ [7.83%]<br>$\tau_{tube5}=0.076$ [5.88%]<br>$\tau_{tube6}=0.069$ [5.36%]<br>$\tau_{tube7}=-0.034$ [2.59%]<br>$\tau_{tube8}=-0.101$ [7.83%]<br>$\tau_{tube9}=-0.076$ [5.88%]<br>$\tau_{tube10}=-0.069$ [5.36%]<br>$\tau_{tube11}=-0.080$ [6.18%]<br>$\tau_{tube12}=-0.454$ [35.06%]<br>$\tau_{tube13}=-0.480$ [37.10%]<br>$\tau_{tube0-6}=1.294$ [100%]<br>$\tau_{tube7-13}=-1.294$ [100%]<br>lower τ inside, transmission dominated by tubes 2, 3, 12, 13 | $\tau_{tube0}=0.011$ [3.44%]<br>$\tau_{tube1}=0.016$ [4.95%]<br>$\tau_{tube2}=0.120$ [36.97%]<br>$\tau_{tube3}=0.133$ [41.13%]<br>$\tau_{tube4}=0.026$ [7.94%]<br>$\tau_{tube5}=0.011$ [3.40%]<br>$\tau_{tube6}=0.007$ [2.17%]<br>$\tau_{tube7}=-0.011$ [3.44%]<br>$\tau_{tube8}=-0.026$ [7.94%]<br>$\tau_{tube9}=-0.011$ [3.40%]<br>$\tau_{tube10}= -0.007$ [2.17%]<br>$\tau_{tube11}=-0.016$ [4.95%]<br>$\tau_{tube12}=-0.120$ [36.97%]<br>$\tau_{tube13}=-0.133$ [41.13%]<br>$\tau_{tube0-6}=0.324$ [100%]<br>$\tau_{tube7-13}=-0.324$ [100%]<br>similar τ values, transmission dominated by tubes 2, 3, 12, 13 |
| (9,9)<br>3 2<br>4 0 1<br>5 6<br>metallic,<br>tube 0 rotated by 20°<br>around the z-axis | $\tau_{tube0}=1.5091$ [75.29%]<br>$\tau_{tube1}=0.0813$ [4.06%]<br>$\tau_{tube2}=0.0837$ [4.18%]<br>$\tau_{tube3}=0.0813$ [4.06%]<br>$\tau_{tube4}=0.0837$ [4.18%]<br>$\tau_{tube5}= 0.0813$ [4.06%]<br>$\tau_{tube6}=0.0837$ [4.18%]<br>$\tau_{tube0-6}=2.004$ [100%]<br>larger τ inside | $\tau_{tube0}=1.5049$ [12.54%]<br>$\tau_{tube1}=1.7497$ [14.58%]<br>$\tau_{tube2}= 1.7502$ [14.58%]<br>$\tau_{tube3}=1.7497$ [14.58%]<br>$\tau_{tube4}= 1.7502$ [14.58%]<br>$\tau_{tube5}=1.7497$ [14.58%]<br>$\tau_{tube6}=1.7502$ [14.58%]<br>$\tau_{tube0-6}=12.005$ [100%]<br>lower τ inside | $\tau_{tube0}=1.9944$ [14.25%]<br>$\tau_{tube1}=1.9981$ [14.27%]<br>$\tau_{tube2}= 2.0037$ [14.31%]<br>$\tau_{tube3}=1.9981$ [14.27%]<br>$\tau_{tube4}=2.0037$ [14.31%]<br>$\tau_{tube5}=1.9981$ [14.27%]<br>$\tau_{tube6}= 2.0037$ [14.31%]<br>$\tau_{tube0-6}=14.000$ [100%]<br>similar τ values |
| (9,9)<br>3 2<br>4 0 1<br>5 6<br>metallic,<br>optimized | $\tau_{tube0}=4.607\times10^{-4}$ [3.00%]<br>$\tau_{tube1}=2.491\times10^{-3}$ [16.24%]<br>$\tau_{tube2}=2.468\times10^{-3}$ [16.09%]<br>$\tau_{tube3}=2.491\times10^{-3}$ [16.24%]<br>$\tau_{tube4}=2.468\times10^{-3}$ [16.09%]<br>$\tau_{tube5}=2.491\times10^{-3}$ [16.24%]<br>$\tau_{tube6}=2.468\times10^{-3}$ [16.09%]<br>$\tau_{tube0-6}=1.534\times10^{-2}$ [100%]<br>lower τ inside | $\tau_{tube0}=1.416$ [11.80%]<br>$\tau_{tube1}= 1.764$ [14.70%]<br>$\tau_{tube2}=1.764$ [14.70%]<br>$\tau_{tube3}= 1.764$ [14.70%]<br>$\tau_{tube4}= 1.764$ [14.70%]<br>$\tau_{tube5}= 1.764$ [14.70%]<br>$\tau_{tube6}=1.764$ [14.70%]<br>$\tau_{tube0-6}=12.000$ [100%]<br>lower τ inside | $\tau_{tube0}= 2.003$ [14.31%]<br>$\tau_{tube1}= 1.997$ [14.26%]<br>$\tau_{tube2}= 2.002$ [14.30%]<br>$\tau_{tube3}= 1.997$ [14.26%]<br>$\tau_{tube4}= 2.002$ [14.30%]<br>$\tau_{tube5}=1.997$ [14.26%]<br>$\tau_{tube6}= 2.002$ [14.30%]<br>$\tau_{tube0-6}=14.000$ [100%]<br>similar τ values |
| (9,9)<br>metallic,<br>tube 0 rotated by 20°<br>around the z-axis<br>optimized | $\tau_{tube0}$=1.211 [60.11%]<br>$\tau_{tube1}$=0.138 [6.85%]<br>$\tau_{tube2}$=0.130 [6.44%]<br>$\tau_{tube3}$=0.138 [6.85%]<br>$\tau_{tube4}$=0.130 [6.44%]<br>$\tau_{tube5}$=0.138 [6.85%]<br>$\tau_{tube6}$=0.130 [6.44%]<br>$\tau_{tube0\text{-}6}$=2.015 [100%]<br>larger τ inside | $\tau_{tube0}$=1.428 [11.90%]<br>$\tau_{tube1}$=1.758 [14.65%]<br>$\tau_{tube2}$=1.766 [14.71%]<br>$\tau_{tube3}$=1.758 [14.65%]<br>$\tau_{tube4}$=1.766 [14.71%]<br>$\tau_{tube5}$=1.758 [14.65%]<br>$\tau_{tube6}$=1.766 [14.71%]<br>$\tau_{tube0\text{-}6}$=12.001 [100%]<br>lower τ inside | $\tau_{tube0}$=1.995 [14.25%]<br>$\tau_{tube1}$=1.996 [14.26%]<br>$\tau_{tube2}$=2.006 [14.33%]<br>$\tau_{tube3}$=1.996 [14.26%]<br>$\tau_{tube4}$=2.006 [14.33%]<br>$\tau_{tube5}$=1.996 [14.26%]<br>$\tau_{tube6}$=2.006 [14.33%]<br>$\tau_{tube0\text{-}6}$=14.000 [100%]<br>similar τ values |
| (21,0)<br>metallic,<br>optimized | $\tau_{tube0}$=0.375 [6.49%]<br>$\tau_{tube1}$=0.899 [15.59%]<br>$\tau_{tube2}$=0.898 [15.58%]<br>$\tau_{tube3}$=0.899 [15.59%]<br>$\tau_{tube4}$=0.898 [15.58%]<br>$\tau_{tube5}$=0.899 [15.59%]<br>$\tau_{tube6}$=0.898 [15.58%]<br>$\tau_{tube0\text{-}6}$=5.768 [100%]<br>lower τ inside | $\tau_{tube0}$=1.160 [10.55%]<br>$\tau_{tube1}$=1.645 [14.95%]<br>$\tau_{tube2}$=1.635 [14.87%]<br>$\tau_{tube3}$=1.645 [14.95%]<br>$\tau_{tube4}$=1.635 [14.87%]<br>$\tau_{tube5}$=1.645 [14.95%]<br>$\tau_{tube6}$=1.635 [14.87%]<br>$\tau_{tube0\text{-}6}$=11.000 [100%]<br>lower τ inside | $\tau_{tube0}$=4.112 [12.85%]<br>$\tau_{tube1}$=4.680 [14.62%]<br>$\tau_{tube2}$=4.621 [14.43%]<br>$\tau_{tube3}$=4.680 [14.62%]<br>$\tau_{tube4}$=4.621 [14.43%]<br>$\tau_{tube5}$=4.680 [14.62%]<br>$\tau_{tube6}$=4.621 [14.43%]<br>$\tau_{tube0\text{-}6}$=32.012 [100%]<br>similar τ values |
| (12,3)<br>metallic,<br>tube 0 rotated by 60°<br>around the z-axis<br>optimized | $\tau_{tube0}$=0.008 [0.16%]<br>$\tau_{tube1}$=0.812 [16.49%]<br>$\tau_{tube2}$= 0.827 [16.79%]<br>$\tau_{tube3}$=0.812 [16.49%]<br>$\tau_{tube4}$=0.827 [16.79%]<br>$\tau_{tube5}$=0.812 [16.49%]<br>$\tau_{tube6}$=0.827 [16.79%]<br>$\tau_{tube0\text{-}6}$=4.927 [100%]<br>lower τ inside | $\tau_{tube0}$=2.001 [14.29%]<br>$\tau_{tube1}$=1.998 [14.27%]<br>$\tau_{tube2}$=2.002 [14.30%]<br>$\tau_{tube3}$=1.998 [14.27%]<br>$\tau_{tube4}$=2.002 [14.30%]<br>$\tau_{tube5}$=1.998 [14.27%]<br>$\tau_{tube6}$=2.002 [14.30%]<br>$\tau_{tube0\text{-}6}$=14.000 [100%]<br>similar τ values | $\tau_{tube0}$=2.003 [14.31%]<br>$\tau_{tube1}$=1.997 [14.27%]<br>$\tau_{tube2}$=2.002 [14.30%]<br>$\tau_{tube3}$=1.997 [14.27%]<br>$\tau_{tube4}$=2.002 [14.30%]<br>$\tau_{tube5}$=1.997 [14.27%]<br>$\tau_{tube6}$=2.002 [14.30%]<br>$\tau_{tube0\text{-}6}$=14.000 [100%]<br>similar τ values |
| (20,0)<br>semiconducting, tube 0 rotated by 9° around the z-axis optimized | inside band gap | inside band gap | $\tau_{tube0}=4.028$ [14.38%]<br>$\tau_{tube1}=4.001$ [14.29%]<br>$\tau_{tube2}=3.992$ [14.26%]<br>$\tau_{tube3}=3.992$ [14.26%]<br>$\tau_{tube4}=4.001$ [14.29%]<br>$\tau_{tube5}=3.992$ [14.26%]<br>$\tau_{tube6}=3.992$ [14.26%]<br>$\tau_{tube0-6}=28.000$ [100%]<br>similar $\tau$ values |
| (9,9)<br>at $B_{\perp}$=4.5 T | $\tau_{tube0}=2.723\times10^{-4}$ [5.83%]<br>$\tau_{tube1}=5.445\times10^{-4}$ [11.66%]<br>$\tau_{tube2}=1.473\times10^{-3}$ [31.54%]<br>$\tau_{tube3}=1.692\times10^{-3}$ [36.22%]<br>$\tau_{tube4}=9.243\times10^{-4}$ [19.79%]<br>$\tau_{tube5}=-3.009\times10^{-5}$ [0.64%]<br>$\tau_{tube6}=-2.048\times10^{-4}$ [4.38%]<br>$\tau_{tube0-6}=4.671\times10^{-3}$ [100%] | $\tau_{tube0}=1.370$ [11.42%]<br>$\tau_{tube1}=1.772$ [14.77%]<br>$\tau_{tube2}=1.769$ [14.74%]<br>$\tau_{tube3}=1.769$ [14.74%]<br>$\tau_{tube4}=1.771$ [14.76%]<br>$\tau_{tube5}=1.774$ [14.79%]<br>$\tau_{tube6}=1.774$[14.78%]<br>$\tau_{tube0-6}=12.000$ [100%]<br>lower $\tau$ inside | $\tau_{tube0}=2.004$ [14.31%]<br>$\tau_{tube1}=1.998$ [14.27%]<br>$\tau_{tube2}=2.001$ [14.29%]<br>$\tau_{tube3}=1.998$ [14.27%]<br>$\tau_{tube4}=2.001$ [14.29%]<br>$\tau_{tube5}=1.998$ [14.27%]<br>$\tau_{tube6}=2.001$ [14.29%]<br>$\tau_{tube0-6}=14.000$ [100%]<br>similar $\tau$ values |
| (9,9)<br>at $B_{\perp}$=9 T | $\tau_{tube0}=2.779\times10^{-4}$ [5.77%]<br>$\tau_{tube1}=3.87\times10^{-4}$ [8.03%]<br>$\tau_{tube2}=2.25\times10^{-3}$ [46.84%]<br>$\tau_{tube3}=2.73\times10^{-3}$[56.63%]<br>$\tau_{tube4}=1.137\times10^{-3}$ [23.62%]<br>$\tau_{tube5}=-7.934\times10^{-4}$ [16.49%]<br>$\tau_{tube6}=-1.175\times10^{-3}$ [24.42%]<br>$\tau_{tube0-6}=4.812\times10^{-3}$[100%] | $\tau_{tube0}=1.369$ [11.41%]<br>$\tau_{tube1}= 1.772$ [14.76%]<br>$\tau_{tube2}=1.768$ [14.73%]<br>$\tau_{tube3}=1.768$ [14.73%]<br>$\tau_{tube4}=1.771$ [14.75%]<br>$\tau_{tube5}=1.777$ [14.80%]<br>$\tau_{tube6}=1.777$ [14.81%]<br>$\tau_{tube0-6}=12.000$ [100%] | $\tau_{tube0}=2.004$[14.31%]<br>$\tau_{tube1}=1.998$ [14.27%]<br>$\tau_{tube2}=2.001$ [14.29%]<br>$\tau_{tube3}=1.998$ [14.27%]<br>$\tau_{tube4}=2.001$ [14.29%]<br>$\tau_{tube5}=1.998$ [14.27%]<br>$\tau_{tube6}=2.001$ [14.29%]<br>$\tau_{tube0-6}= 14.00$ [100%] |
| (9,9)<br>at $B_{\perp}$=30 T | $\tau_{tube0}=4.198\times10^{-4}$ [1.23%]<br>$\tau_{tube1}= 5.394\times10^{-4}$ [1.58%]<br>$\tau_{tube2}=6.967\times10^{-3}$ [20.43%]<br>$\tau_{tube3}=1.085\times10^{-2}$ [31.81%]<br>$\tau_{tube4}=2.136\times10^{-3}$ [6.27%]<br>$\tau_{tube5}=-4.898\times10^{-3}$ [14.37%]<br>$\tau_{tube6}= -8.289\times10^{-3}$ [24.31%]<br>$\tau_{tube0-6}=3.410\times10^{-2}$ [100%] | $\tau_{tube0}=1.349$ [11.24%]<br>$\tau_{tube1}=1.766$ [14.72%]<br>$\tau_{tube2}=1.793$ [14.94%]<br>$\tau_{tube3}=1.790$ [14.92%]<br>$\tau_{tube4}=1.765$ [14.70%]<br>$\tau_{tube5}=1.765$ [14.71%]<br>$\tau_{tube6}=1.771$ [14.76%]<br>$\tau_{tube0-6}=12.000$ [100%] | $\tau_{tube0}=2.004$ [14.31%]<br>$\tau_{tube1}=1.998$ [14.27%]<br>$\tau_{tube2}=2.000$ [14.29%]<br>$\tau_{tube3}= 1.998$ [14.27%]<br>$\tau_{tube4}=2.001$ [14.29%]<br>$\tau_{tube5}=1.998$ [14.27%]<br>$\tau_{tube6}=2.001$ [14.29%]<br>$\tau_{tube0-6}=14.000$ [100%] |
| (9,9)<br>3 2<br>4 0 1<br>5 6<br>at $B_{\perp}$=60 T | $\tau_{tube0}$=2.043x10$^{-3}$ [0.23%]<br>$\tau_{tube1}$= 8.936x10$^{-2}$ [10.11%]<br>$\tau_{tube2}$=0.133 [15.07%]<br>$\tau_{tube3}$=0.291 [32.96%]<br>$\tau_{tube4}$=-1.674x10$^{-2}$ [1.89%]<br>$\tau_{tube5}$=-0.106 [12.01%]<br>$\tau_{tube6}$=-0.245 [27.73%]<br>$\tau_{tube0-6}$=0.884 [100%] | $\tau_{tube0}$=1.278 [10.66%]<br>$\tau_{tube1}$= 1.740 [14.50%]<br>$\tau_{tube2}$=2.192 [18.28%]<br>$\tau_{tube3}$=2.130 [17.76%]<br>$\tau_{tube4}$=1.757 [14.64%]<br>$\tau_{tube5}$=1.411 [11.77%]<br>$\tau_{tube6}$=1.487 [12.40%]<br>$\tau_{tube0-6}$=11.994 [100%] | $\tau_{tube0}$=2.003 [14.31%]<br>$\tau_{tube1}$=1.998 [14.27%]<br>$\tau_{tube2}$=2.000 [14.29%]<br>$\tau_{tube3}$= 1.998 [14.28%]<br>$\tau_{tube4}$=2.001 [14.29%]<br>$\tau_{tube5}$= 1.999 [14.28%]<br>$\tau_{tube6}$=2.001 [14.30%]<br>$\tau_{tube0-6}$=14.000 [100%] |
| (9,9)<br>3 2<br>4 0 1<br>5 6<br>at $B_{\angle=45o}$=9 T | $\tau_{tube0}$=3.107x10$^{-4}$ [6.59%]<br>$\tau_{tube1}$=4.655x10$^{-4}$ [9.88%]<br>$\tau_{tube2}$=1.781x10$^{-3}$ [37.80%]<br>$\tau_{tube3}$=2.094x10$^{-3}$ [44.44%]<br>$\tau_{tube4}$=1.010x10$^{-3}$ [21.43%]<br>$\tau_{tube5}$=-3.491x10$^{-4}$ [7.41%]<br>$\tau_{tube6}$=-5.998x10$^{-4}$ [12.73%]<br>$\tau_{tube0-6}$=4.712x10$^{-3}$ [100%] | $\tau_{tube0}$= 1.370 [11.41%]<br>$\tau_{tube1}$=1.772 [14.76%]<br>$\tau_{tube2}$=1.768 [14.74%]<br>$\tau_{tube3}$=1.769 [14.74%]<br>$\tau_{tube4}$=1.771 [14.76%]<br>$\tau_{tube5}$=1.775 [14.79%]<br>$\tau_{tube6}$= 1.775 [14.79%]<br>$\tau_{tube0-6}$=12.000 [100%] | $\tau_{tube0}$=2.004 [14.31%]<br>$\tau_{tube1}$=1.998 [14.27%]<br>$\tau_{tube2}$=2.001 [14.29%]<br>$\tau_{tube3}$=1.998 [14.27%]<br>$\tau_{tube4}$=2.001 [14.29%]<br>$\tau_{tube5}$=1.998 [14.27%]<br>$\tau_{tube6}$=2.001 [14.29%]<br>$\tau_{tube0-6}$=14.000 [100%] |
| (9,9)<br>3 2<br>4 0 1<br>5 6<br>at $B_{\parallel}$=9 T | $\tau_{tube0}$=3.194x10$^{-4}$ [6.92%]<br>$\tau_{tube1}$=7.124x10$^{-4}$[15.43%]<br>$\tau_{tube2}$=7.198x10$^{-4}$ [15.59%]<br>$\tau_{tube3}$=7.124x10$^{-4}$ [15.43%]<br>$\tau_{tube4}$=7.198x10$^{-4}$ [15.59%]<br>$\tau_{tube5}$=7.124x10$^{-4}$ [15.43%]<br>$\tau_{tube6}$=7.198x10$^{-4}$ [15.59%]<br>$\tau_{tube0-6}$=4.616x10$^{-3}$ [100%] | $\tau_{tube0}$= 1.371 [11.42%]<br>$\tau_{tube1}$= 1.772 [14.76%]<br>$\tau_{tube2}$=1.771[14.76%]<br>$\tau_{tube3}$=1.772 [14.76%]<br>$\tau_{tube4}$=1.771 [14.76%]<br>$\tau_{tube5}$= 1.772 [14.76%]<br>$\tau_{tube6}$=1.771 [14.76%]<br>$\tau_{tube0-6}$=12.000 [100%] | $\tau_{tube0}$=2.004 [14.31%]<br>$\tau_{tube1}$= 1.998 [14.27%]<br>$\tau_{tube2}$=2.001 [14.29%]<br>$\tau_{tube3}$=1.998 [14.27%]<br>$\tau_{tube4}$=2.0012 [14.29%]<br>$\tau_{tube5}$=1.998 [14.27%]<br>$\tau_{tube6}$=2.001 [14.29%]<br>$\tau_{tube0-6}$=14.000 [100%] |
| 3 (17,0) 2 (19,0)<br>4 (18,0) 0 (18,0) 1 (18,0)<br>5 (19,0) 6 (17,0)<br>mixed<br>(3 metallic &<br>4 semiconducting)<br>at $B_{\perp}$=4.5 T | $\tau_{tube0}$=1.464 [29.50%]<br>$\tau_{tube1}$=1.742 [35.11%]<br>$\tau_{tube2}$=-0.017 [0.34%]<br>$\tau_{tube3}$=0.001 [0.02%]<br>$\tau_{tube4}$=1.775 [35.77%]<br>$\tau_{tube5}$=-0.001 [0.01%]<br>$\tau_{tube6}$=-0.002 [0.04%]<br>$\tau_{tube0-6}$=4.962 [100%] | $\tau_{tube0}$=1.931 [32.19%]<br>$\tau_{tube1}$= 1.957 [32.62%]<br>$\tau_{tube2}$=0.068 [1.13%]<br>$\tau_{tube3}$=0.009 [0.15%]<br>$\tau_{tube4}$= 1.961 [32.69%]<br>$\tau_{tube5}$=0.062 [1.03%]<br>$\tau_{tube6}$=0.011 [0.19%]<br>$\tau_{tube0-6}$=6.000 [100%] | $\tau_{tube0}$=2.626 [11.42%]<br>$\tau_{tube1}$=2.348 [10.21%]<br>$\tau_{tube2}$=4.002 [17.40%]<br>$\tau_{tube3}$=3.847 [16.72%]<br>$\tau_{tube4}$=2.321 [10.09%]<br>$\tau_{tube5}$= 4.011 [17.44%]<br>$\tau_{tube6}$=3.844 [16.71%]<br>$\tau_{tube0-6}$=23.000 [100%] |
| 3 (17,0) 2 (19,0)<br>4 (18,0) 0 (18,0) 1 (18,0)<br>5 (19,0) 6 (17,0)<br>mixed<br>(3 metallic &<br>4 semiconducting)<br>at $B_{\perp}$=9 T | $\tau_{tube0}$=1.464 [29.50%]<br>$\tau_{tube1}$=1.745 [35.16%]<br>$\tau_{tube2}$=-0.017 [0.33%]<br>$\tau_{tube3}$= 0.001 [0.02%]<br>$\tau_{tube4}$=1.772 [35.70%]<br>$\tau_{tube5}$=-0.001 [0.01%]<br>$\tau_{tube6}$=-0.002 [0.04%]<br>$\tau_{tube0\text{-}6}$=4.962 [100%] | $\tau_{tube0}$=1.931 [32.18%]<br>$\tau_{tube1}$=1.958 [32.63%]<br>$\tau_{tube2}$=0.068 [1.13%]<br>$\tau_{tube3}$=0.009 [0.15%]<br>$\tau_{tube4}$=1.962 [32.69%]<br>$\tau_{tube5}$=0.062 [1.03%]<br>$\tau_{tube6}$=0.011 [0.18%]<br>$\tau_{tube0\text{-}6}$=6.000 [100%] | $\tau_{tube0}$=2.627 [11.42%]<br>$\tau_{tube1}$=2.347 [10.20%]<br>$\tau_{tube2}$=4.003 [17.40%]<br>$\tau_{tube3}$=3.845 [16.72%]<br>$\tau_{tube4}$=2.322 [10.10%]<br>$\tau_{tube5}$=4.010 [17.44%]<br>$\tau_{tube6}$=3.845 [16.72%]<br>$\tau_{tube0\text{-}6}$=23.000 [100%] |
| 3 (17,0) 2 (19,0)<br>4 (18,0) 0 (18,0) 1 (18,0)<br>5 (19,0) 6 (17,0)<br>mixed<br>(3 metallic &<br>4 semiconducting)<br>at $B_{\perp}$=30 T | $\tau_{tube0}$=1.465 [29.50%]<br>$\tau_{tube1}$=1.757 [35.39%]<br>$\tau_{tube2}$=-0.014 [0.28%]<br>$\tau_{tube3}$=0.001 [0.01%]<br>$\tau_{tube4}$=1.758 [35.41%]<br>$\tau_{tube5}$=-0.001 [0.01%]<br>$\tau_{tube6}$=-0.001 [0.01%]<br>$\tau_{tube0\text{-}6}$=4.965 [100%] | $\tau_{tube0}$=1.932 [32.21%]<br>$\tau_{tube1}$=1.961 [32.68%]<br>$\tau_{tube2}$=0.065 [1.08%]<br>$\tau_{tube3}$=0.010 [0.173%]<br>$\tau_{tube4}$= 1.966 [32.76%]<br>$\tau_{tube5}$=0.056 [0.93%]<br>$\tau_{tube6}$=0.011 [0.18%]<br>$\tau_{tube0\text{-}6}$=6.000 [100%] | $\tau_{tube0}$= 2.629 [11.43%]<br>$\tau_{tube1}$=2.346 [10.202%]<br>$\tau_{tube2}$=4.005 [17.41%]<br>$\tau_{tube3}$=3.839 [16.69%]<br>$\tau_{tube4}$=2.327 [10.12%]<br>$\tau_{tube5}$=4.005[17.41%]<br>$\tau_{tube6}$=3.849 [16.74%]<br>$\tau_{tube0\text{-}6}$=23.000 [100%] |
| 3 (17,0) 2 (19,0)<br>4 (18,0) 0 (18,0) 1 (18,0)<br>5 (19,0) 6 (17,0)<br>mixed<br>(3 metallic &<br>4 semiconducting)<br>at $B_{\perp}$=60 T | $\tau_{tube0}$=1.466 [29.48%]<br>$\tau_{tube1}$=1.772 [35.63%]<br>$\tau_{tube2}$=-0.009 [0.19%]<br>$\tau_{tube3}$=-3.245x10$^{-4}$ [0.01%]<br>$\tau_{tube4}$=1.744 [35.06%]<br>$\tau_{tube5}$=-0.001[0.01%]<br>$\tau_{tube6}$=0.002 [0.03%]<br>$\tau_{tube0\text{-}6}$=4.973 [100%] | $\tau_{tube0}$=1.944 [32.40%]<br>$\tau_{tube1}$=1.966 [32.76%]<br>$\tau_{tube2}$=0.056 [0.94%]<br>$\tau_{tube3}$=0.011 [0.18%]<br>$\tau_{tube4}$=1.981 [33.01%]<br>$\tau_{tube5}$=0.032 [0.54%]<br>$\tau_{tube6}$=0.010 [0.16%]<br>$\tau_{tube0\text{-}6}$=6.000 [100%] | $\tau_{tube0}$=2.634 [11.45%]<br>$\tau_{tube1}$=2.346[10.20%]<br>$\tau_{tube2}$=4.008 [17.43%]<br>$\tau_{tube3}$=3.830 [16.65%]<br>$\tau_{tube4}$= 2.337 [10.16%]<br>$\tau_{tube5}$=3.993 [17.36%]<br>$\tau_{tube6}$= 3.850 [16.74%]<br>$\tau_{tube0\text{-}6}$=23.000 [100%] |

**Supplemental Table 5.2-2. Total energies (in eV/atom) of small metallic SWCNT bundles before and after structural optimization using the AIREBO potential[34] with the cutoff radius for the Lennard-Jones term of 17 Å. Each bundle consists of seven nanotubes arranged in a flower-like configuration—one central tube surrounded by six outer tubes in a hexagonal layout—as seen in cross-sectional views provided in supplemental table 5.2-1.**

| system composition | Unrelaxed | optimized |
| --- | --- | --- |
| (21,0) | −7.402259228 | −7.418501742 |
| (9,9) | −7.391263561 | −7.405152387 |
| (9,9),<br>tube 0 rotated by 20° around the z-axis | −7.391275523 | −7.405137757 |

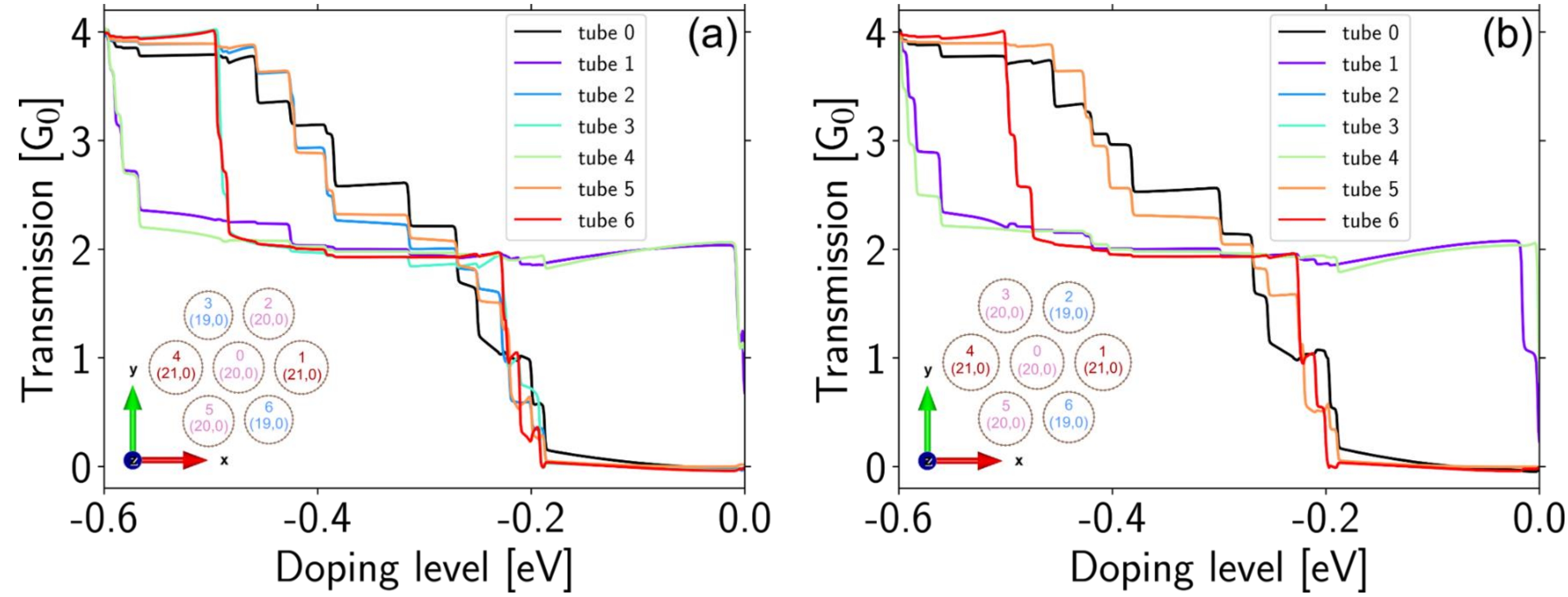

**Supplemental figure 5.2-1. Tube-resolved transmission contributions as a function of energy for two mixed SWCNT bundles at *H*=0 T. Transport occurs along the z-direction. Both systems contain two metallic (21,0) nanotubes located on the outer layer (marked with darker colors), and five semiconducting nanotubes—three (20,0) and two (19,0)—marked with lighter colors. The central tube is (20,0). The semiconducting (19,0) and (20,0) nanotubes are arranged in opposing pairs on the outer layer of the bundle. a, Configuration exhibiting only point symmetry. b, Configuration exhibiting axial symmetry, obtained by swapping the positions of the upper (19,0) and (20,0) tubes with their counterparts in (a), so that each semiconducting pair is aligned across the bundle (i.e., the top (19,0) lies above the bottom (19,0), and likewise for the (20,0) tubes). Each panel includes a cross-sectional visualization of the corresponding bundle. In the visualizations, nanotubes are labelled with colored numbers indicating their chirality and electronic character: semiconducting tubes are marked with lighter-colored numbers, while metallic tubes are labelled in darker colors. All nanotubes are plotted to scale, so differences in cross-sectional size reflect actual diameter differences.**

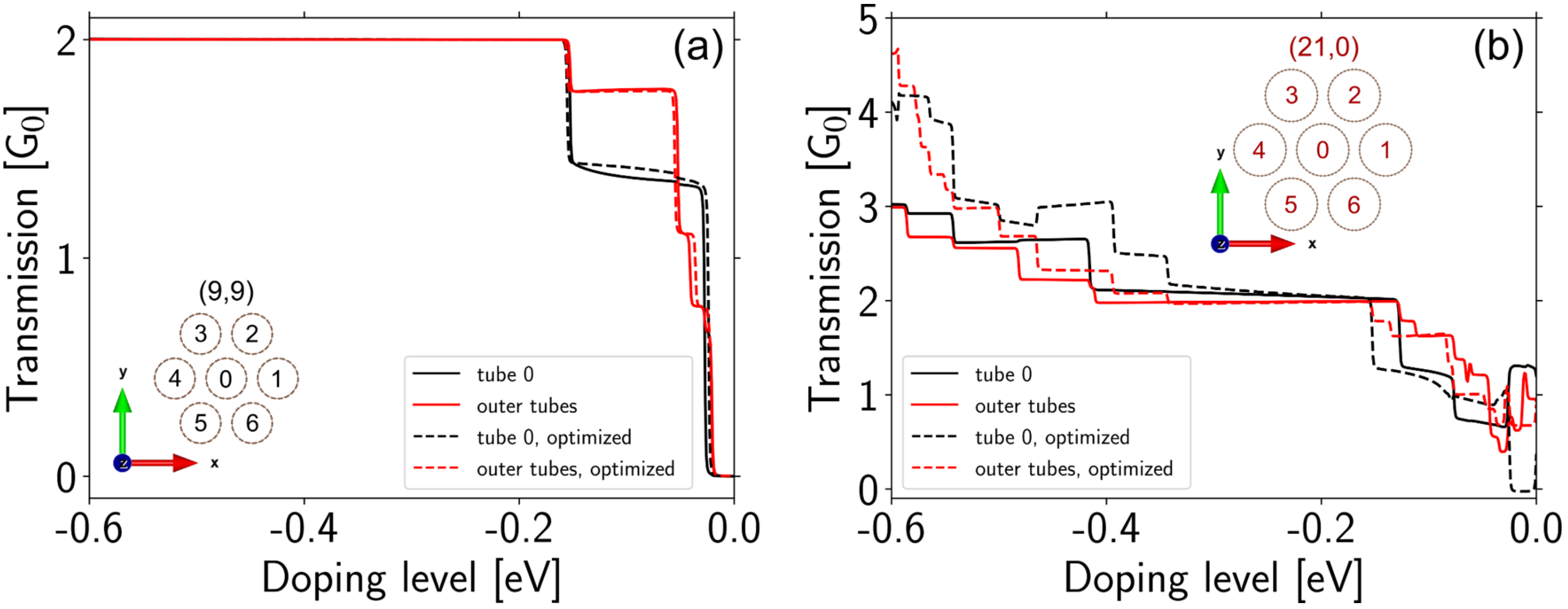

**Supplemental figure 5.2-2. Tube-resolved transmission contributions as a function of energy at *H*=0 T for metallic SWCNT bundles before and after structural optimization. a, Results for an armchair (9,9) bundle: solid lines correspond to the non-optimized structure, and dashed lines to the fully optimized system (atomic positions and lattice vectors). b, Results for a zigzag (21,0) bundle using the same optimization scheme. In both cases, transport is along the z-direction. Each panel includes a cross-sectional visualization of the corresponding system. In the visualizations, nanotubes are labelled with colored numbers indicating their chirality and electronic character, and are plotted to scale—so differences in cross-sectional size reflect real differences in tube diameters.**

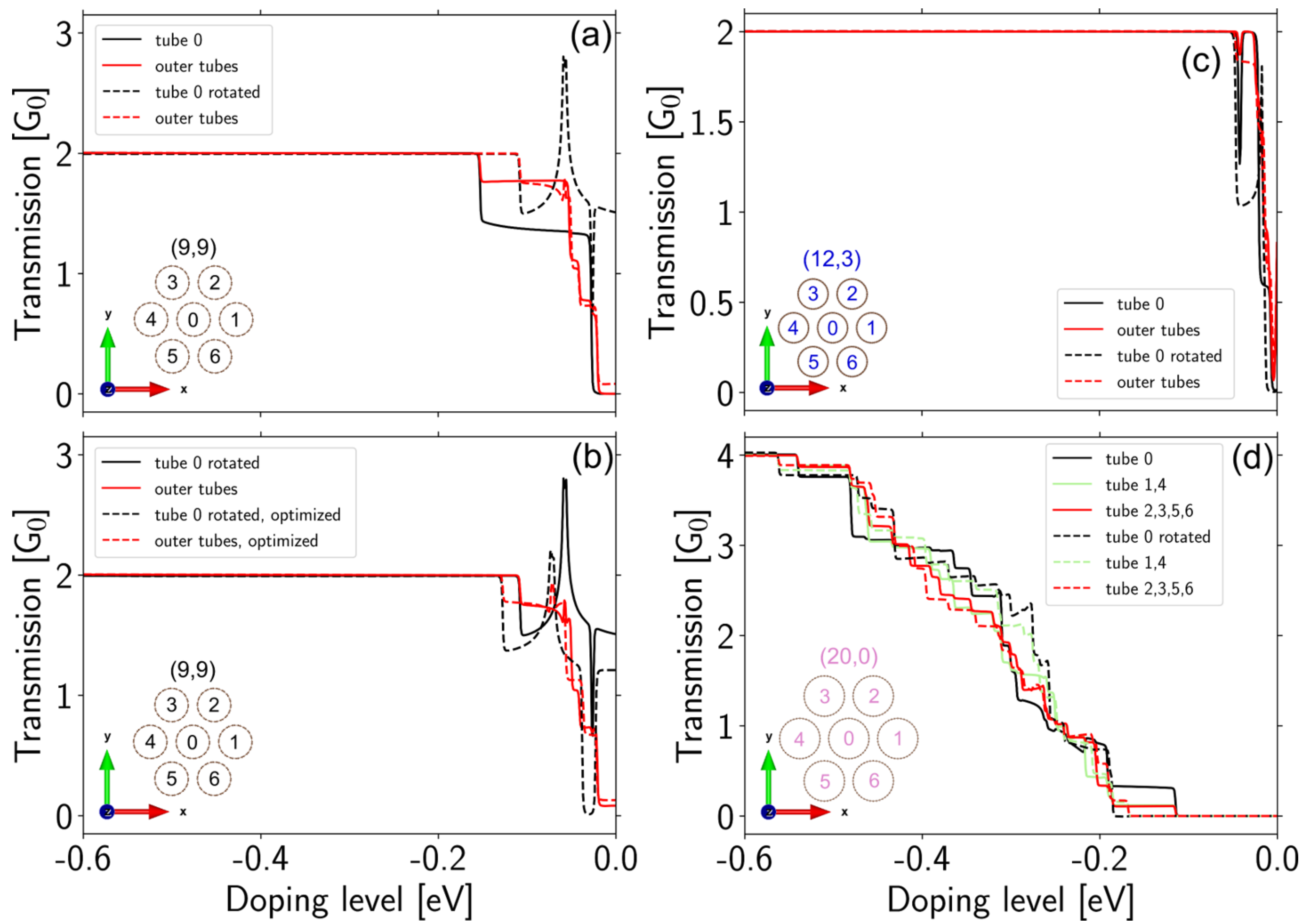

**Supplemental figure 5.2-3. Tube-resolved transmission contributions at *H*=0 T for SWCNT bundles composed of seven nanotubes, where tube 0 (central tube) is either unrotated or rotated around the bundle axis (z-axis). a, Transmission through a metallic armchair (9,9) bundle for the unrotated configuration (solid lines) and with tube 0 rotated by 20° around the z-axis (dashed lines). b, Transmission through a metallic (9,9) bundle with tube 0 rotated by 20° for the non-optimized system (solid lines) and after full optimization of atomic positions and lattice vectors (dashed lines). c, Transmission through a metallic chiral (12,3) bundle for the unrotated configuration (solid lines) and with tube 0 rotated by 60° (dashed lines). d, Transmission through a semiconducting (20,0) bundle for the unrotated configuration (solid lines) and with tube 0 rotated by 9° (dashed lines). Transport is along the z-direction. Each panel includes an inset showing a cross-sectional view of the corresponding system configuration. In the visualizations, nanotubes are labelled with colored numbers indicating their chirality and electronic character: semiconducting tubes are marked with lighter-colored numbers, while metallic tubes are labelled in darker colors. All nanotubes are plotted to scale, so differences in cross-sectional size reflect actual diameter differences.**

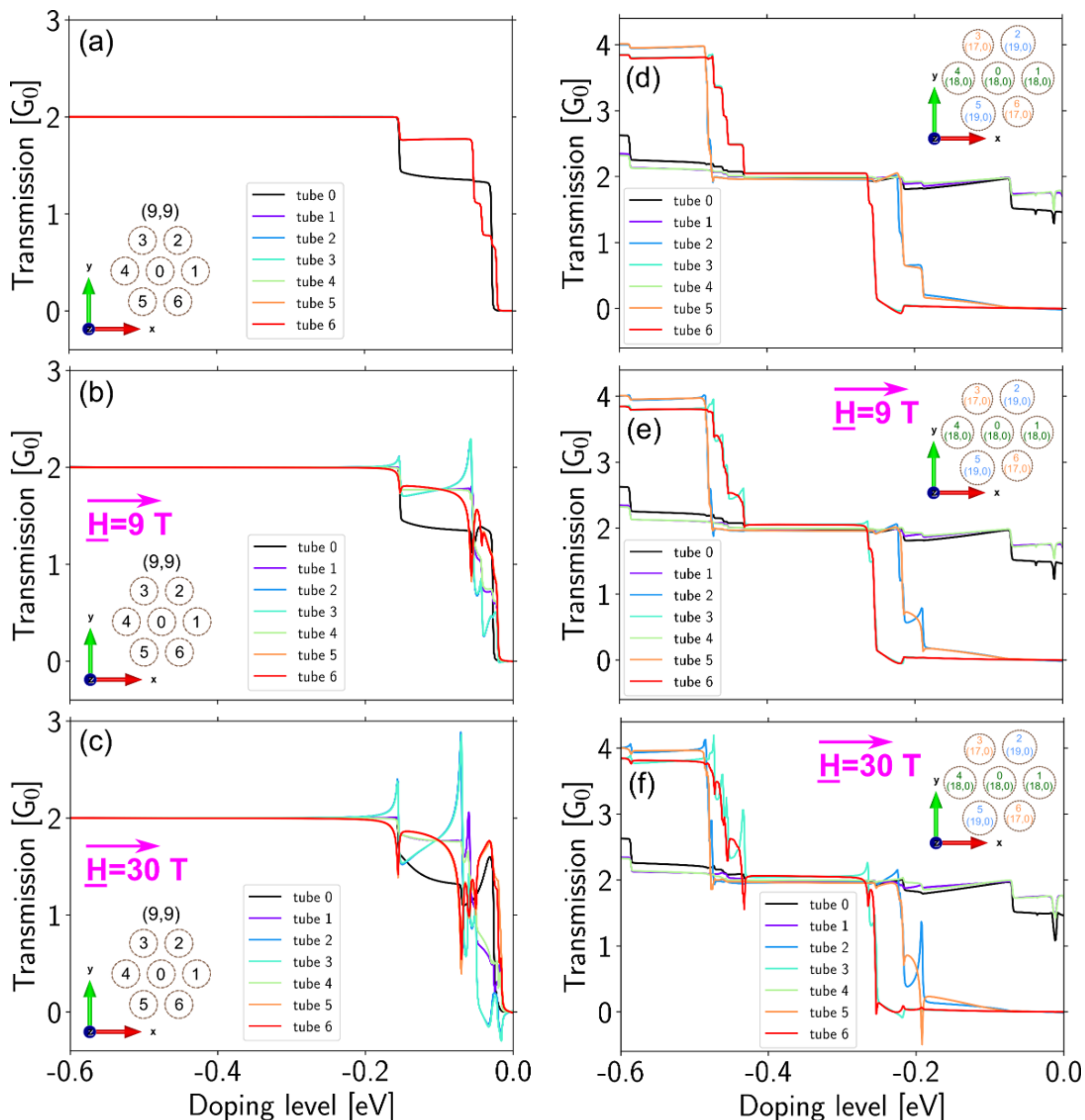

**Supplemental figure 5.2-4. Tube-resolved transmission contributions as a function of energy for two small SWCNT bundles under varying magnetic field conditions. Panels (a–c) show results for a metallic (9,9) bundle: a, in the absence of a magnetic field; b, under a perpendicular magnetic field of 9 T (along the x-axis); and, c, under a perpendicular magnetic field of 30 T. Panels (d–f) present analogous results for a mixed bundle composed of three metallic (18,0) SWCNTs and four semiconducting SWCNTs—two (17,0) and two (19,0). One metallic tube is positioned centrally, while the other two are on the outer layer along the same axis. The semiconducting tubes are arranged in opposing (17,0) and (19,0) pairs on opposite sides of the bundle, as shown in the insets. Each panel includes a cross-sectional visualization of the corresponding system. In the visualizations, individual nanotubes are labelled; semiconducting tubes are marked with lighter colors and metallic tubes with darker colors. All nanotubes are plotted to scale, so differences in cross-sectional size reflect actual diameter differences The direction of the applied magnetic field is indicated in panels (b), (c), (e), and (f).**

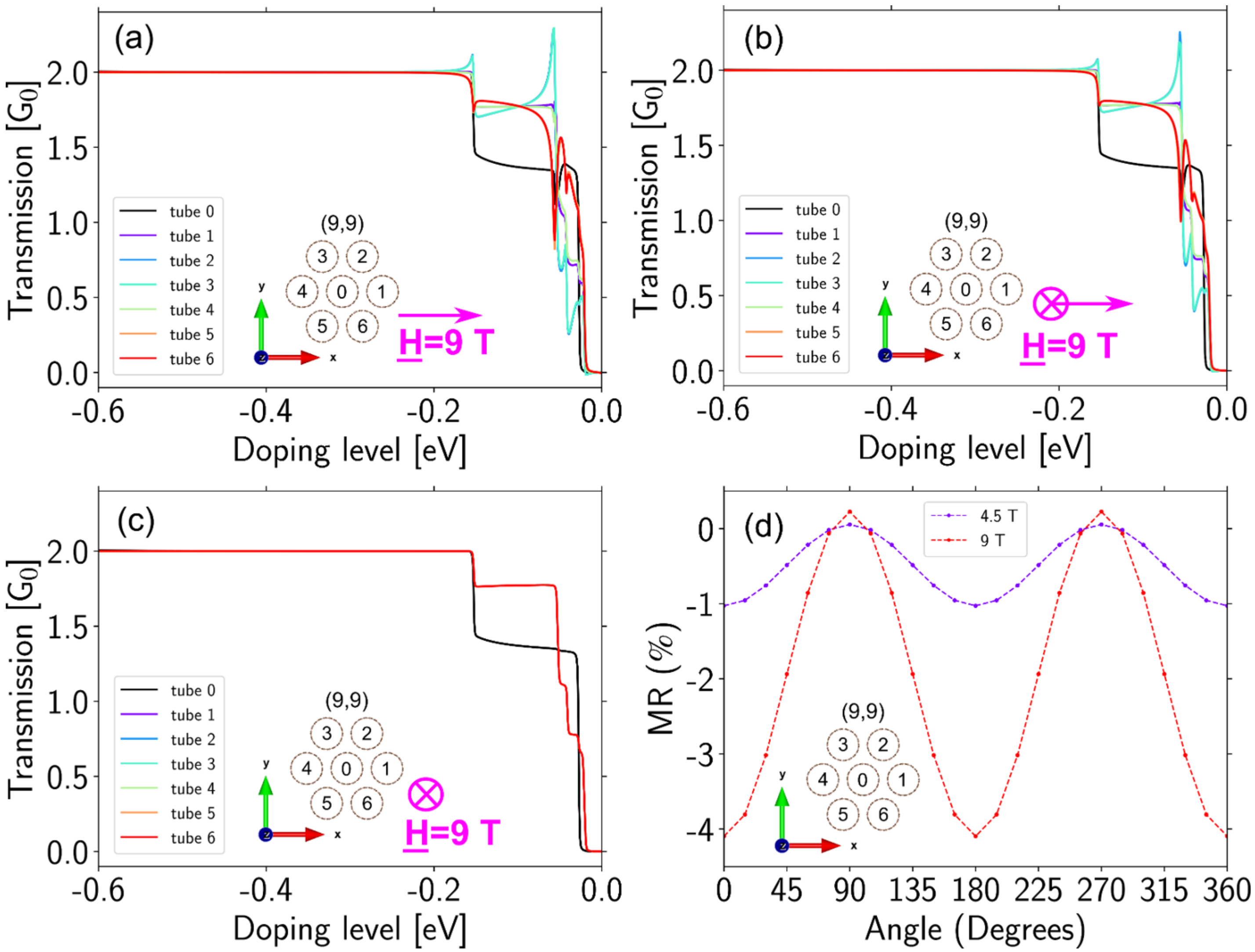

**Supplemental figure 5.2-5. Tube-resolved transmission contributions as a function of energy for a small metallic (9,9) SWCNT bundle under external magnetic fields of 4.5 T (purple line) and 9 T (red line). Panels show different magnetic field orientations: a, perpendicular to the bundle axis (along the x-axis); b, at 45°; and c, parallel to the bundle axis (along the z-axis). d, Magnetoresistance (MR) at $E_f$=0 eV as a function of the magnetic field angle for the de-doped bundle. 0° corresponds to the perpendicular case shown in the inset of panel (a), while 90° corresponds to the parallel case illustrated in the inset of panel (c). Each panel (a–d) includes a cross-sectional visualization of the bundle; the direction of the applied magnetic field is indicated in panels (a–c).**

2-fold Angle dependent MR with CNT Bundle. To complement the experimental angular MR measurements, we performed TB–NEGF simulations of a de-doped CNT bundle composed of seven (9,9) SWCNTs arranged in a flower-like geometry. As shown in Supplementary Figure 5.2-5, our calculations reproduce a clear angular dependence of the magnetoresistance, with a dominant two-fold symmetry as observed in the experiment. Increasing the magnetic field from 4.5 T to 9 T enhances the amplitude of MR oscillations by approximately a factor of four, consistent with the experimentally observed increase in MR contrast at higher fields.

It is important to emphasize that our simulations are performed at zero temperature and do not include temperature-dependent dephasing or classical transport effects. Moreover, our tight-binding Hamiltonian does not include curvature-induced modifications to the band structure. For this reason, we specifically chose armchair (9,9) nanotubes, which have a relatively large diameter and negligible curvature effects, allowing us to isolate the contribution of the quantum-coherent transport along a bundle. Under these idealized conditions, the observed MR modulation arises from magnetic phase effects encoded via the Peierls substitution and reflects the sensitivity of inter-tube coupling to the relative orientation of the magnetic field.

In addition to phase coherence, our results suggest that inter-tube interactions within the bundle—particularly those modulated by symmetry and relative positioning—play an important role in shaping the angular MR response. As the field orientation changes, interference between coupled transmission pathways across different tubes is altered, leading to measurable changes in tube-resolved transmission contributions. This highlights that the MR response of CNT bundles cannot be attributed solely to single-tube effects, but rather emerges from the collective electronic structure and geometry of the bundle.

While experimental MR features—such as the emergence of a positive component near 90°, four-fold symmetry, and strong temperature dependence—are influenced by a complex interplay of mechanisms including weak localization, Aharonov–Bohm band modulation, and possibly spin-orbit interaction, our modelling offers a complementary and geometry-resolved perspective. It supports a picture where surface-dominated conduction, inter-tube coupling, and quantum interference collectively govern the angular magnetoresistance behavior of CNT bundles under high magnetic fields.